\documentclass[a4paper,11pt]{article}
\usepackage[
bookmarks=true,bookmarksnumbered=true,
colorlinks = true,anchorcolor= blue, 
linkcolor = darkred, citecolor = darkgreen
,linktoc=page]{hyperref}
\usepackage{amsfonts,amsthm,amsmath,amssymb,graphicx}
\usepackage{tensor}
\usepackage[a4paper, textwidth=6.5in, textheight=9in]{geometry}
\usepackage{braket}
\usepackage{simpler-wick}
\usepackage{bm}
\usepackage{float} 
\usepackage{mathrsfs}
\usepackage{booktabs}
\usepackage{makeidx}
\usepackage{color} 
\usepackage{comment} 
\usepackage{tikz}
\usetikzlibrary{fit,shapes.geometric}
\numberwithin{equation}{section}
    \definecolor{darkgreen}{rgb}{0,0.5,0}
    \definecolor{darkblue}{rgb}{0,0,0.6}
    \definecolor{purple}{rgb}{0.4,0.2,0.7}
    \definecolor{darkred}{rgb}{0.7,0,0}
    \definecolor{airforceblue}{rgb}{0.36, 0.54, 0.66}
    \definecolor{cyan}{rgb}{0.0, 1.0, 1.0}
    \definecolor{cyan(process)}{rgb}{0.0, 0.72, 0.92}

\newcommand{\f}{\frac}
\newcommand{\s}{\sqrt}

\DeclareMathOperator{\sgn}{sgn} 
\DeclareMathOperator{\Tr}{Tr}
\DeclareMathOperator{\Sch}{Sch}
\DeclareMathOperator{\Pf}{Pf}

\newcommand{\arctanh}{\text{arctanh}}

\def\ba#1\ea{\begin{align}#1\end{align}}

\newcommand{\aff}[1]{${}^{#1}$}

\begin{document}

\begin{center}
\vspace*{0.5cm}
{\LARGE\bf
Effective Field Theory of Noncritical M-theory from Bosonization
}\\[1.2cm]
{\large  Patrick Jefferson\footnote{\tt pjefferson@umass.edu}}\aff{a}
 and 
{\large Tokiro Numasawa\footnote{\tt numasawa@issp.u-tokyo.ac.jp}}\aff{b} \\[0.8cm]
\aff{a} {\small
\it Department of Physics, University of Massachusetts, Amherst, MA 01003 USA
}\\
\aff{b} {\small
\it Institute for Solid State Physics, University of Tokyo, Kashiwa 277-8581, Japan
}\\
\abstract{We extend the coadjoint orbit approach to bosonization systematized by Delacr\'etaz-Du-Mehta-Son to double-scaled non-relativistic Fermi liquids with central inverted harmonic oscillator potential. In 1+1d, this reproduces the Das-Jevicki collective field theory describing quasiparticle excitations of the $c=1$ matrix model. In 2+1d, this yields an effective field theory for noncritical M-theory, namely the 2+1d Fermi liquid proposed by Ho\v{r}ava and Keeler as a unified framework for characterizing noncritical string vacua. With a suitable gauge choice this theory reduces to a continuous family of 1+1d chiral bosons resembling Das-Jevicki collective fields coupled by additional interactions. Utilizing the underlying integrability of noncritical M-theory, we compute select fermion density correlation functions order-by-order in perturbation theory, and provide evidence that they agree with their Fermi liquid counterparts in a suitable semiclassical limit. This represents a step toward an effective spacetime gravity description of noncritical M-theory, which we expect to shed light on both the landscape of 2d noncritical strings and 3d quantum gravity in general.}
\thispagestyle{empty}
\newpage
\end{center}

{\hypersetup{linkcolor= red, filecolor = magenta, urlcolor=magenta}}

\tableofcontents

\newpage

\section{Introduction}

\subsection{Background and motivation}

In the foundational papers \cite{Townsend:1995kk,Witten:1995ex,Horava:1995qa}, M-theory was proposed as the unifying framework underlying the web of dualities relating the five perturbative superstring theories \cite{Hull:1994ys}, and it has since become a central concept in the study of quantum gravity. Its low-energy effective description as eleven-dimensional supergravity has been especially important, serving as the starting point for exploring a broad range of phenomena, including compactifications, branes, dualities, black holes, and holography. A complete nonperturbative definition of M-theory itself, however, remains unknown. The known nonperturbative formulations are holographic, defining M-theory through a lower-dimensional non-gravitational theory rather than intrinsically in eleven dimensions, and each is moreover tied to a fixed asymptotic background. These include the BFSS matrix model, which defines M-theory in the flat-space light-cone frame and is dual to the near-horizon black D0-brane geometry \cite{Banks:1996vh,Itzhaki:1998dd}; its massive deformation, the BMN matrix model, dual to the maximally supersymmetric pp-wave \cite{Berenstein:2002jq}; the theory on M2-branes, dual to $AdS_4 \times S^7$ \cite{Aharony:2008ug}; and the six-dimensional $(2,0)$ theory on M5-branes, dual to $AdS_7 \times S^4$ \cite{Maldacena:1997re}. Two questions central to understanding M-theory remain largely unsettled, namely how precisely these bulk effective descriptions emerge from their microscopic definitions, and whether these constructions themselves arise as limits of a single background-independent formulation.

Low-dimensional solvable string theories provide an important laboratory for studying this problem.
In particular, in noncritical string theory \cite{Polyakov:1981rd,Polyakov:1981re} one can explicitly compare a closed-string worldsheet description with a nonperturbative definition based on solvable matrix models \cite{Douglas:1989ve,Klebanov:2003wg,Seiberg:2004ei}.
For example, it has been proposed that an exact definition of the $c=1$ string \cite{Takayanagi:2003sm,Douglas:2003up,McGreevy:2003kb} is given by the double-scaling limit of matrix quantum mechanics, for which the singlet sector can be formulated as a Fermi sea of noninteracting fermions moving in an inverted harmonic oscillator potential \cite{Gross:1990st} (for reviews, see \cite{Klebanov:1991qa,Polchinski:1994mb,Ginsparg:1993is, Martinec:2004td}). This formulation makes it possible to compare the exact matrix-model description with the worldsheet description through physical observables such as S-matrices, and indeed a great deal of evidence has been accumulated supporting the conjectured duality between the $c=1$ matrix model and perturbative $c=1$ string worldsheet theory; see, e.g., \cite{Bershadsky:1990gs, Gross:1990md, Moore:1991ir, DiFrancesco:1991daf, Moore:1991sf, Klebanov:2003km, Balthazar:2017mxh, Balthazar:2019rnh, Balthazar:2019ypi, Sen:2019qqg, Sen:2020oqr, Sen:2020eck, Sen:2021qdk, Alexandrov:2023fvb, Collier:2026pxi}.
Moreover, the tachyon field in the closed-string theory is identified with the collective excitations of the matrix model Fermi surface.
This correspondence is deeply related to the bosonization of fermionic systems familiar from condensed matter physics.
The fermions in the matrix model may be interpreted as D-brane-like degrees of freedom, while closed-string degrees of freedom arise as their collective excitations \cite{Klebanov:2003km}.
In this sense, noncritical string theory provides a controlled setting in which one can analytically study how a bulk effective theory emerges from a microscopic definition.

From this viewpoint, it is natural to consider the analogue of M-theory in noncritical string theory.
A candidate for such a noncritical version of M-theory was proposed by Ho\v{r}ava and Keeler in \cite{Horava:2005tt,Horava:2005wm,Horava:2007ds}.
Their noncritical M-theory is exactly defined as a double-scaled (2+1)-dimensional nonrelativistic free fermion system, or equivalently as a Fermi liquid. The vacuum states of the type 0A and type 0B two-dimensional noncritical string theories can be obtained from the canonical vacuum of this (2+1)-dimensional Fermi liquid by imposing appropriate constraints on the Fermi sea.
Thus, analogous to how ``critical'' (i.e., eleven-dimensional) M-theory unifies various perturbative superstring theories, noncritical M-theory may be viewed as a lower-dimensional model unifying the landscape of two-dimensional noncritical string theories.

However, noncritical M-theory is in a situation complementary to that of critical M-theory.
In critical M-theory, a powerful low-energy bulk effective theory is known, namely eleven-dimensional supergravity, while its complete microscopic definition is not known in general.
By contrast, noncritical M-theory has an exact microscopic definition in terms of free fermions, but the corresponding spacetime effective theory, in particular a three-dimensional gravitational bulk description in the conventional sense, is not yet well understood.
Taking inspiration from the relationship between the open and closed string descriptions of the $c=1$ string and other noncritical string theories, it is therefore natural to construct an effective theory for noncritical M-theory and compare it with the exact Fermi liquid description proposed by Ho\v{r}ava and Keeler.
This in principle provides a solvable setting in which to study the relationship between a microscopic definition and a bulk effective theory. Such a study may serve as a lower-dimensional analog of the critical superstring duality web, and by doing so may thus offer insights into the broader question of how bulk effective theories such as eleven-dimensional supergravity emerge from microscopic degrees of freedom in M-theory.

The purpose of this paper is to start from the exact Fermi liquid description of noncritical M-theory and construct the low-energy effective theory of its Fermi surface.
In a free fermion system, the low-energy degrees of freedom are described by deformations of the Fermi surface \cite{2005cond.mat..5529H}.
Since these deformations are generated by canonical transformations in phase space, they are naturally organized as coadjoint orbits of the algebra of canonical transformations.
Coadjoint-orbit methods have already appeared in the study of the collective field theory of the $c=1$ string---see, e.g., \cite{Das:1991uta}.
More recently, this idea has been developed into a systematic construction of effective field theories for Fermi surface bosonization \cite{Delacretaz:2022ocm,Huang:2024uap,Chen:2025jdv}.
In this paper, we apply this coadjoint-orbit method to noncritical M-theory and construct an effective theory in terms of bosonic fields living on the Fermi surface.

\subsection{Summary of results}

We first apply the coadjoint-orbit method to the $c=1$ matrix model and confirm that it reproduces an effective action equivalent to the known Das--Jevicki collective field theory. This discussion also serves as preparation for the generalization to noncritical M-theory. We then perform the bosonization of the Fermi surface of noncritical M-theory, which is defined as a (2+1)-dimensional Fermi liquid. Since the Fermi surface is higher-dimensional, unlike in the ordinary $c=1$ string, it contains residual continuous degrees of freedom. As a result, the effective theory is not a single two-dimensional scalar field, but rather a continuous family of chiral bosons labeled by conserved quantities that parametrize classical trajectories.

We further test the validity of the resulting bosonized effective theory by comparing its density correlation functions with those computed in the exact Fermi liquid description. In particular, we consider two types of density operators. The first is the eigenvalue density operator defined in terms of the ordinary rectilinear coordinate basis, while the second is a density operator defined in a basis of light-cone coordinates of action-angle type. For the eigenvalue density operators, we show that the structures of the one-, two-, and three-point functions obtained from the bosonized theory agree with the semiclassical limits of the exact Fermi liquid computations. For the two-point function, this agreement is fully established by comparing with the saddle-point approximation of the resolvent. For the three-point function, the bosonized theory reproduces the feature that the correlation function has support only when the three insertion points lie on a common classical trajectory. On the other hand, for the light-cone density operators, the integrability manifest in action-angle coordinates drastically simplifies the correlation functions. We show in both the exact Fermi liquid description and the bosonized description that connected correlation functions with three or more insertions vanish.

These results indicate that the effective theory obtained by coadjoint-orbit bosonization correctly captures, at least in the semiclassical regime, the physics of the exact Fermi liquid definition of noncritical M-theory. From a broader perspective, this work is a first step toward clarifying the relation between an exact microscopic definition and a bulk-like effective description in noncritical M-theory. If the connection between this effective theory and a three-dimensional gravitational description, or a geometric understanding of the noncritical string landscape, can be made more precise, it may provide useful insight into the more general problem of the emergence of spacetime effective theories in M-theory.

\subsection{Organization of the paper}

The organization of this paper is as follows.
In Section \ref{sec:reviewstring}, we review the $c=1$ noncritical string, its matrix model description, and the collective field theory of its Fermi surface.
In Section \ref{sec:FermiLiquidM}, we review the exact Fermi liquid description of noncritical M-theory and introduce the density operators and correlation functions that will later serve as our main observables.
In Section \ref{COM}, we explain the coadjoint-orbit method for Fermi surface bosonization and first apply it to the $c=1$ matrix model, thereby rederiving the known collective field theory.
In Section \ref{sec:bosonizationM}, we apply the coadjoint orbit method to noncritical M-theory and derive the bosonized effective action in different gauge choices.
In Section \ref{bosonicdensitycorr}, we compute density correlation functions in the bosonized theory and compare them with the exact Fermi liquid description.
Finally, in Section \ref{sec:conclusion}, we summarize our results and discuss future directions, including the spacetime effective description of noncritical M-theory and its possible applications to the noncritical string landscape. Various mathematical results along with the details of some of the calculations described in the previous Sections are collected in the Appendices.

\section{Review of $c=1$ String Theory}

\label{sec:reviewstring}

We begin with a brief overview of noncritical strings in two dimensions, focusing specifically on the $c=1$ string.
In Section \ref{noncritoverview}, we review noncritical strings and their dual descriptions in terms of matrix quantum mechanics. In Section \ref{EFTFermi}, we review various early approaches to deriving the effective field theory (EFT) description of the collective excitations of the $c=1$ matrix model.

\subsection{Overview of noncritical strings}
\label{noncritoverview}

    Open/closed string duality has been most concretely realized for noncritical string theories---namely, $c<1$ minimal strings, the $c = 1$ string, and type 0 strings (for reviews of two-dimensional gravity and noncritical strings see, e.g., \cite{Klebanov:1991qa, Ginsparg:1993is, Jevicki:1993qn, Polchinski:1994mb, Alexandrov:2003ut, Martinec:2004td, Nakayama:2004vk}). Among noncritical string theories, the $c=1$ string is particularly appealing because it is the simplest example of a string theory with a spacetime interpretation.  The worldsheet description of the $c=1$ string corresponds to $c=25$ Liouville theory coupled to a timelike boson and the $b,c$ ghost system, and the perturbative dynamics of this system is captured by a massless scalar field (i.e., the ``tachyon'') propagating in $1+1$ dimensions.

\begin{figure}
\begin{center}
    \includegraphics[scale=.4]{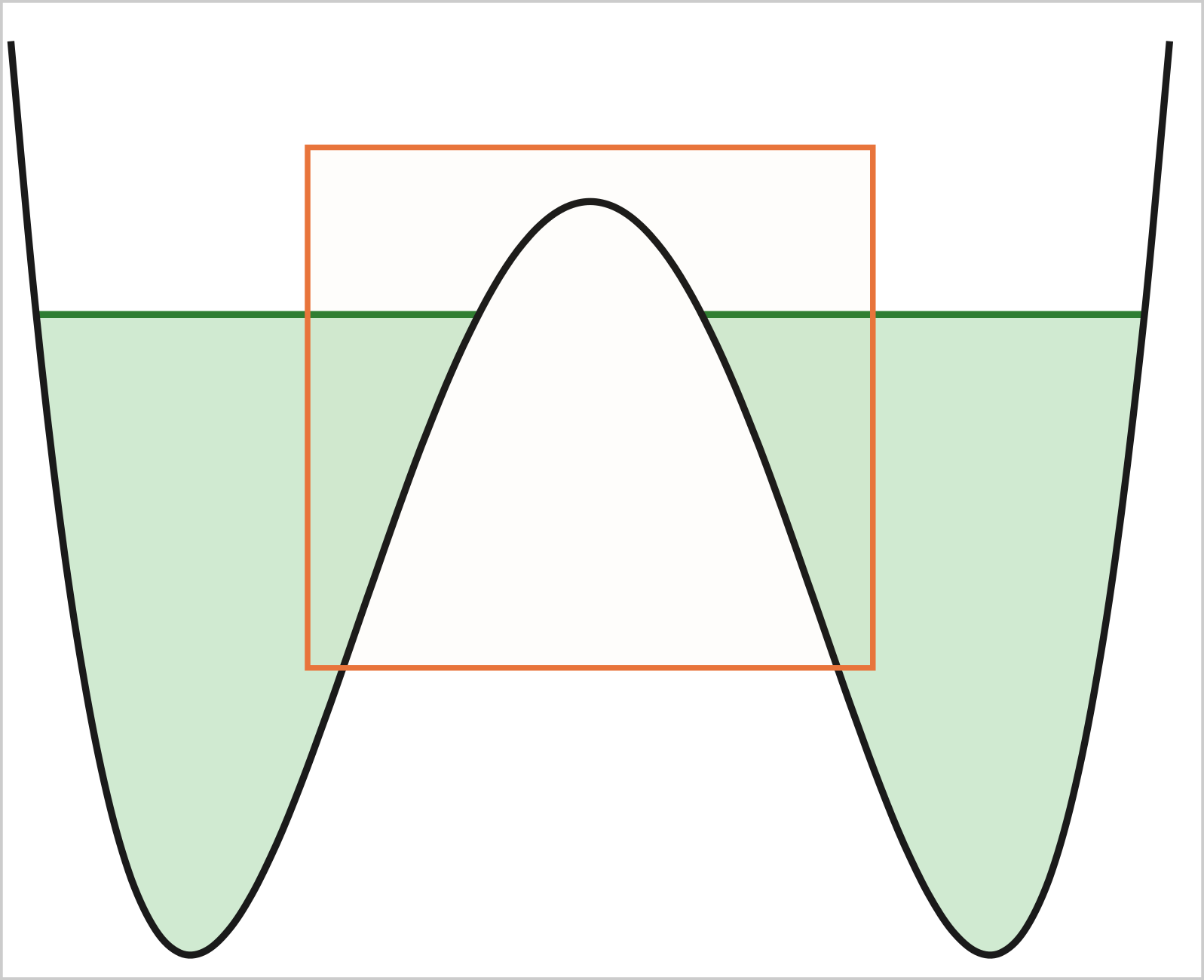}
    ~~~~~~~~~~~~~
     \includegraphics[scale=.4]{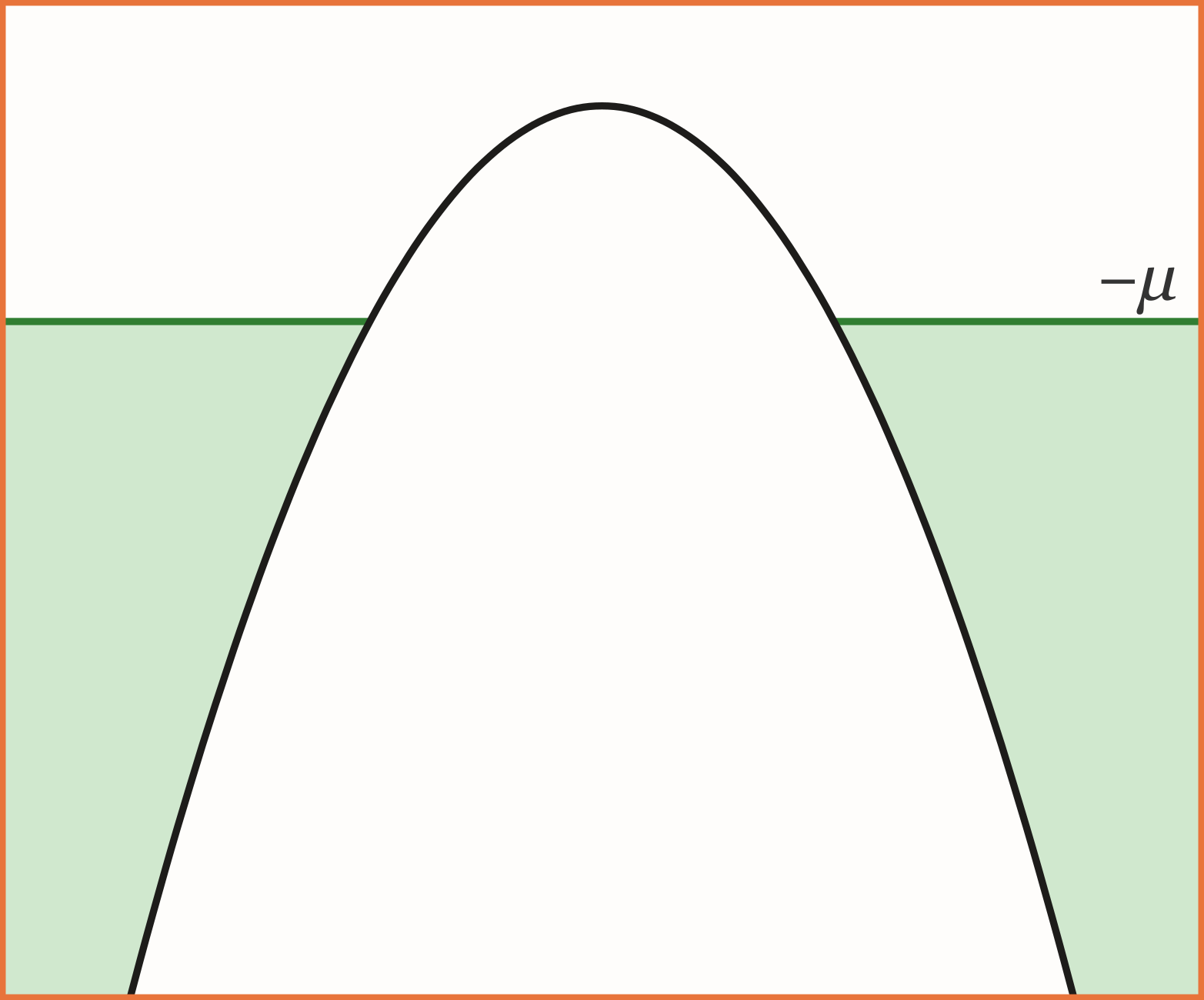}
\end{center}
\caption{The double scaling limit \eqref{eq:c=1DSlimit} effectively ``zooms in'' on the quadratic maximum of the potential in \eqref{MQM}, in the process scaling away other non-universal features  while holding fixed the renormalized Fermi level $-\mu$. The figure on the right depicts the inverted harmonic oscillator potential, which is the universal potential that remains after taking the double-scaling limit. In both figures, the shaded region is the filled Fermi sea, with its boundary corresponding to the static Fermi surface \eqref{eq:c=1FS}.}
\label{fig:c=1potential}
\end{figure}

   The closed string worldsheet description of the $c=1$ string is conjectured to have a dual open string description in terms of the $c=1$ matrix model. The $c=1$ matrix model is a double-scaling limit of gauged $U(N)$ matrix quantum mechanics with Euclidean action
    	\begin{align}
        \label{MQM}
		S[M]&=   N \int \mathrm dt \, \Tr\left( \frac{1}{2} (D_t M)^2 + V(M) \right),~~~~ V(M) = - \frac{1}{2} \omega^2 M^2 + \cdots,
	\end{align}
where $M$ is an $N\times N$ Hermitian matrix and the ellipses in $V(M)$ denote non-universal terms that vanish in the double scaling limit.   One can show that the ground state of (\ref{MQM}) corresponds to $N$ free fermions filling the lowest energy levels up to the Fermi energy $\varepsilon_{F} =-\bar \mu$. The double scaling limit is then
	\begin{align}
    \label{eq:c=1DSlimit}
		N \to \infty, ~~~~ \bar\mu \to 0,~~~~ \mu \equiv  N \bar \mu \text{ fixed}, 
	\end{align}
where $-\mu$ denotes the renormalized Fermi energy.  In the double scaling limit, the single-particle dispersion for the Fermi sea is given by
\begin{equation}
h(  \lambda,  p) = \frac{1}{2}  p^2 - \frac{1}{2} \omega^2  \lambda^2,
\end{equation}
with classical Fermi surface (see Figure \ref{fig:c=1potential})
    \begin{align}
    \label{eq:c=1FS}
        h( \lambda,p) = - \mu.
    \end{align}
The full system is described by one-dimensional non-relativistic fermions with the following Hamiltonian density
\begin{align}
\label{Hc=1}
\mathcal H_{c=1} = \frac{1}{2}  \partial_{  \lambda} \Psi^{\dagger} \partial_{ \lambda} \Psi + V( \lambda) \Psi^{\dagger}\Psi ,~~~~
        V( \lambda) = - \frac{1}{2} \omega^2  \lambda^2 + \cdots.
    \end{align}
This theory is perturbatively stable; a possible non-perturbative completion of the $c=1$ matrix model that has been studied extensively is the type 0B matrix model,  which corresponds to a double scaling limit of the gauged matrix quantum mechanics (\ref{MQM}) where the Fermi sea is filled on both sides of the quadratic maximum. It has been proposed that the type 0B matrix model can be interpreted as the worldline theory of $N$ unstable D0 branes and thus provides a dual description of type 0B string theory \cite{Douglas:2003up,Takayanagi:2003sm}.

Open/closed string duality identifies the asymptotic states of the $c=1$ string with quasi-particles in the $c=1$ matrix model, with the  quasi-particles corresponding to collective excitations of the Fermi surface \cite{Das:1990kaa, Gross:1990md,Gross:1990st,Polchinski:1991uq}. Bosonization provides an effective description of these collective excitations in terms of a massless scalar field living on the Fermi surface that becomes free at large distances from the classical turning point. This scalar field is related to the spacetime tachyon of the $c=1$ string by a nonlocal field redefinition in which the asymptotic modes of the scalar living on the Fermi surface are multiplied by a frequency-dependent ``leg-pole factor'' \cite{Gross:1990md,DiFrancesco:1991daf}. The nonlocal nature of this field redefinition is responsible for converting the
local interactions of the $c=1$ matrix model collective field into the nonlocal interactions of the $c=1$ string theory tachyon---see, e.g., \cite{Natsuume:1994sp,Dhar:1995nq} for early discussions of this point and \cite{Balthazar:2017mxh} for a more recent clarification of the origins of the leg pole factor in the $c=1$ string description.

\subsection{Effective description of the Fermi surface dynamics}

\label{EFTFermi}

A handful of equivalent effective descriptions of the collective excitations of the $c=1$ matrix model Fermi surface in terms of a scalar field were developed in rapid succession in the nineties. The first effective description of the $c=1$ matrix model is due to Das and Jevicki, who in  \cite{Das:1990kaa} introduced the gauge-invariant collective density
    \begin{align}
    \label{eq:density}
        \phi(t,\bar \lambda) = \Tr\,\delta(\bar \lambda I - M(t))
    \end{align} and rewrote the $c=1$ matrix model path integral in terms of this new field variable, leading to the Lagrangian
        \begin{align}
        \label{eq:DJaction}
                L = \int  \mathrm d \bar \lambda \, \left[ \frac{1}{2} \frac{\partial_{\bar \lambda}^{-1} \dot \phi \partial_{\bar \lambda}^{-1} \dot \phi }{\phi} - \frac{\pi^2}{6} \phi^3 + (-\bar \mu + \frac{\bar \lambda^2}{2} ) \phi \right].
        \end{align}
    The resulting theory captures the dynamics of a non-relativistic scalar field, with interactions arising from the field redefinition \eqref{eq:density}.

   In \cite{Gross:1990st}, Gross and Klebanov subsequently arrived at an equivalent description by splitting the non-relativistic fermions in (\ref{Hc=1}) into chiral components,
        \begin{align}
            \Psi &= \frac{e^{ i N\bar \mu t}}{\sqrt{2 \bar v_{ F}}} \left[ e^{\left(-i N  \int^{\bar \lambda} \mathrm d\bar \lambda' \, \bar v_{ F}(\bar \lambda') + i \frac{\pi}{4} \right)} \Psi_{ L}+ e^{\left(iN \int^{\bar \lambda} \mathrm d\bar \lambda' \, \bar v_{ F}(\bar \lambda') - i \frac{\pi}{4} \right)} \Psi_{ R}\right],
        \end{align}
    (above $\bar v_{ F} = \mathrm d\bar \lambda/ \mathrm d\tau = \sqrt{\bar \lambda^2-2\bar \mu}$ is the Fermi velocity) and then applying the standard bosonization procedure for two-dimensional massless Dirac fermions to the chiral components $\Psi_{L,R}$:
        \begin{align}
            \begin{split}
                \Psi_{ L} &= \frac{1}{\sqrt{2\pi}} : \exp\left(i \sqrt{\pi} \int^\tau \mathrm d\tau' \,(\Pi_S-S') \right) :\\
                \Psi_{ R} &=\frac{1}{\sqrt{2\pi}} : \exp\left(i \sqrt{\pi} \int^\tau \mathrm d\tau' \, (\Pi_S + S')\right) :.
            \end{split}
        \end{align}
    Above, $S$ is a massless two-dimensional periodic scalar and $\Pi_S$ is its canonically conjugate momentum. Substituting the above field redefinitions into \eqref{Hc=1} leads to an effective bosonic Hamiltonian that describes the dynamics of the collective excitations of the Fermi surface; after discarding some total derivative terms, this Hamiltonian takes the form \cite{Gross:1990st}
\begin{equation}
\label{eq:GKHam}
H= \frac{1}{2} \int _0 ^{\frac{T}{2}} \mathrm d\tau \left[  \Pi_S^2 + (S')^2 - \frac{\sqrt{\pi}}{N \bar v_{ F}^2} \left( \Pi_S S' \Pi_S + \frac{1}{3} (S')^3 \right) - \frac{1}{2N \bar v_{ F}^2 \sqrt{\pi} }S' \Sch ( \bar \lambda , \tau) \right],
\end{equation}
where
$\Sch(f,z) = \frac{f'''(z)}{f'(z)} - \frac{3}{2} \frac{f''(z)^2}{f'(z)^2}$ is the Schwarzian derivative. 
Introducing the redefinition
\begin{align}
        \phi = \frac{1}{\pi} (\bar v_F - \frac{\sqrt{\pi}}{\bar v_F} \partial_\tau S)
    \end{align}
and substituting it into the Das-Jevicki action  \eqref{eq:DJaction} recovers the action defined by the above Hamiltonian, up to the tadpole term.\footnote{The tadpole term is a result of the normal ordering prescription used in \cite{Gross:1990st}.}

\begin{figure}
\begin{center}
    \includegraphics[scale=.4]{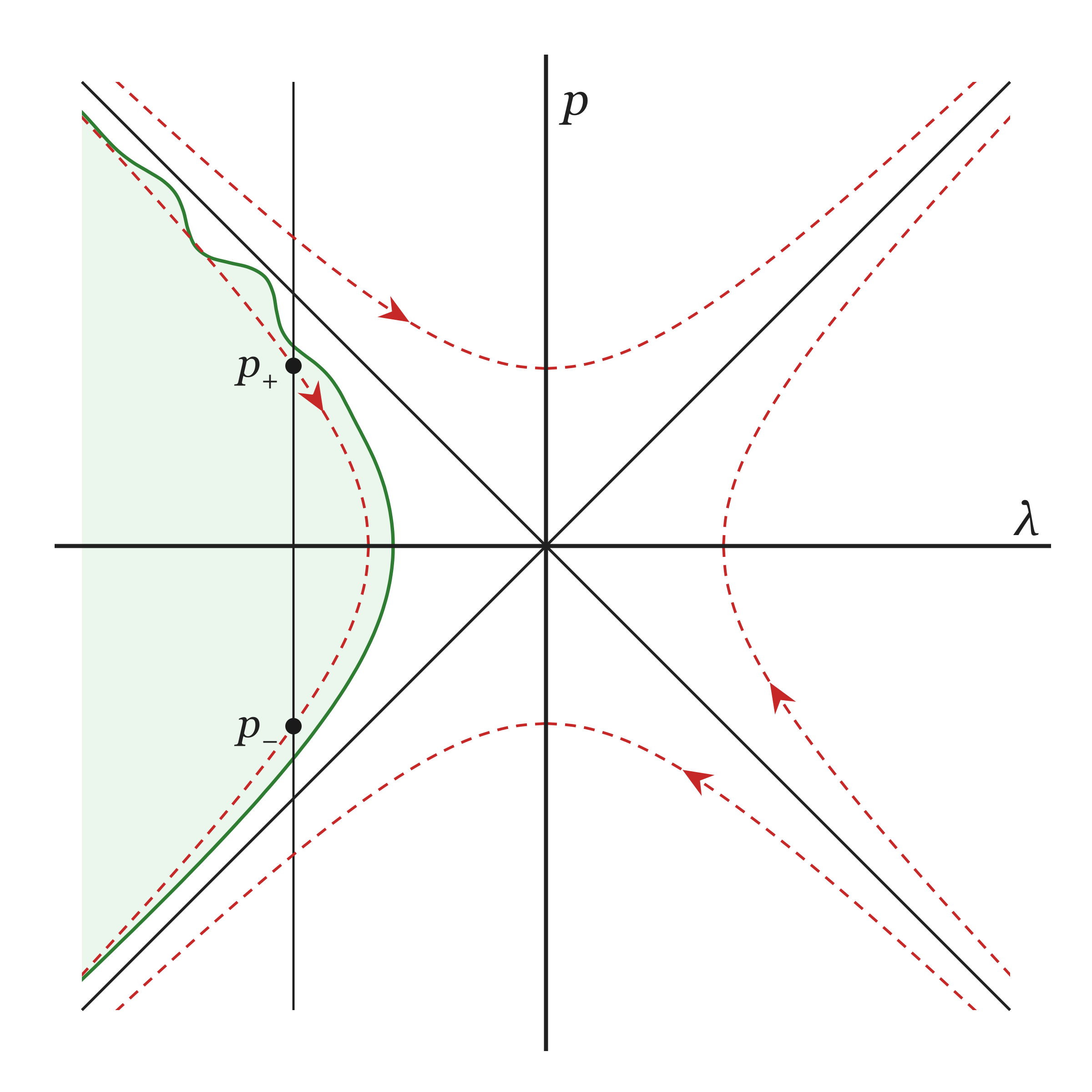}
\end{center}
\caption{The shaded region is the filled $c=1$ string Fermi sea. We can label the state of the Fermi sea in terms of the upper and lower surfaces $p_{\pm{}}(t,\lambda)$, which are indicated for a definite value of $\lambda$ as an illustration. This is essentially a duplicate of Figure 25 in \cite{Polchinski:1994mb}.}
\label{fig:c=1FS}
\end{figure}

    Shortly thereafter, Polchinski \cite{Polchinski:1991uq}  reinterpreted the double-scaling limit of the Das--Jevicki theory as the phase-space hydrodynamics of the static Fermi surface $p^2 = \lambda^2 - 2\mu$, by parametrizing its fluctuations in terms of the upper and lower Fermi momenta
        \begin{align}
            p_\pm( \lambda,t) = \mp{}  \lambda \pm{} \frac{\sqrt{\pi}}{ \lambda} \left( \pm{} \Pi_{ S}(q,t) - \partial_q  S(q,t) \right),~~~~ \lambda = - e^{- q},
        \end{align}
where the unbarred notation above indicates double-scaled quantities.
In doing so, he showed that the classical theory decouples into chiral sectors corresponding to upper and lower ``branches'' of the Fermi surface---see Figure \ref{fig:c=1FS}. These chiral sectors are captured by the dynamics of a single scalar field $S$; substituting the above expansion into the classical Hamiltonian leads to the effective Hamiltonian
    \begin{align}
        H =  \int_{FS} \frac{\mathrm d\lambda \mathrm dp}{2\pi} \, \frac{1}{2} (p^2 - \lambda^2) = \frac{1}{2} \int \mathrm d q \, \left[ \Pi_S^2 + (\partial_q S)^2 + e^{2q} \mathcal O(S^3) \right],
    \end{align}
which is again equivalent to the Gross-Klebanov Hamiltonian up to coordinate redefinitions.

The picture that has emerged from the above and subsequent work is that the spacetime effective description of the $c=1$ string can be captured by bosonizing the shape dynamics of the Fermi surface, with these and other approaches to bosonization essentially leading to equivalent results \cite{Das:1991uta}.  In fact, we will show in Section \ref{COM} that an effective action equivalent to the ones described above can be derived by means of a systematic approach to bosonization involving the coadjoint orbit method applied to canonical transformations in phase space \cite{Delacretaz:2022ocm}.\footnote{We stress that this is not the first time the coadjoint orbit method has been used to study the collective dynamics of the $c=1$ matrix model---see, e.g., \cite{Dhar:1992hr}.} Since this approach to bosonization can be readily generalized to higher-dimensional Fermi surfaces, the coadjoint orbit method will also form the basis of our derivation of the effective action for noncritical M-theory, which proceeds by a completely analogous calculation and is the main result of this paper.

\section{Exact Fermi Liquid Description of Noncritical M-Theory}

\label{sec:FermiLiquidM}

One of the major advantages of the Fermi liquid description of noncritical strings is that it provides an exact, solvable definition of the dual worldsheet theory. For example, the full non-perturbative S-matrix of the $c=1$ string was computed in \cite{Moore:1991zv} using the $c=1$ matrix model description, whereas the perturbative worldsheet S-matrix is still a topic of ongoing investigation---see, e.g., \cite{Balthazar:2017mxh} and more recently \cite{Collier:2026pxi}. Another advantage, perhaps less appreciated, is that the Fermi liquid description provides a simple and convenient way to characterize the landscape of two-dimensional noncritical strings in terms of their vacuum states. In \cite{Horava:2005tt} (see also \cite{Horava:2005wm,Horava:2007ds}) 
it was shown that the vacuum states of various noncritical string theories could be obtained from a more general many body vacuum state corresponding to a $(2+1)$-dimensional non-relativistic Fermi liquid, by restricting the occupancy of the Fermi sea in a suitable manner. It was proposed that this vacuum state is the canonical vacuum of a ``noncritical'' $(2+1)$-dimensional version of M-theory. An intriguing question to ask is whether or not there exists a dual description of noncritical M-theory as an effective theory of three-dimensional quantum gravity, and this question largely motivates the work done in this paper, namely developing an effective spacetime description of noncritical M-theory via bosonization.

In this section, we review the definition of noncritical M-theory and summarize some of its physical properties that we will revisit in the context of its bosonized EFT description in Section \ref{Mtheoryboson}, namely the correlation functions of its density operators, whose excitations roughly correspond to deformations of the classical Fermi surface. In Section \ref{defMFermi}, we introduce the Fermi liquid description of noncritical M-theory. In Section \ref{sec:cancoord}, we define the coordinate systems that we will use when comparing the physics of the Fermi liquid formulation of noncritical M-theory to its bosonized effective description. In Section \ref{MFermidensity} we introduce a general set of density operators whose Euclidean correlation functions will be the main observables of interest. Then, in Sections \ref{Mrectdensity} and  \ref{exactLCFermi}, respectively, we study the correlation functions of the eigenvalue density operators $\rho(x,\boldsymbol \lambda)$ and ``light-cone'' density operators $\rho(x,\tau_\sigma, \varphi_\sigma) =\rho(u_\sigma,\varphi_\sigma)$ (note that these are distinct physical observables). We find that the connected $n$-point correlation functions of the light-cone density operators vanish identically for $n\geq 3$, which signals a free theory.

\subsection{Basic definition of noncritical M-theory}
\label{defMFermi}

\begin{figure}
    \begin{center}
        \includegraphics[scale=.125]{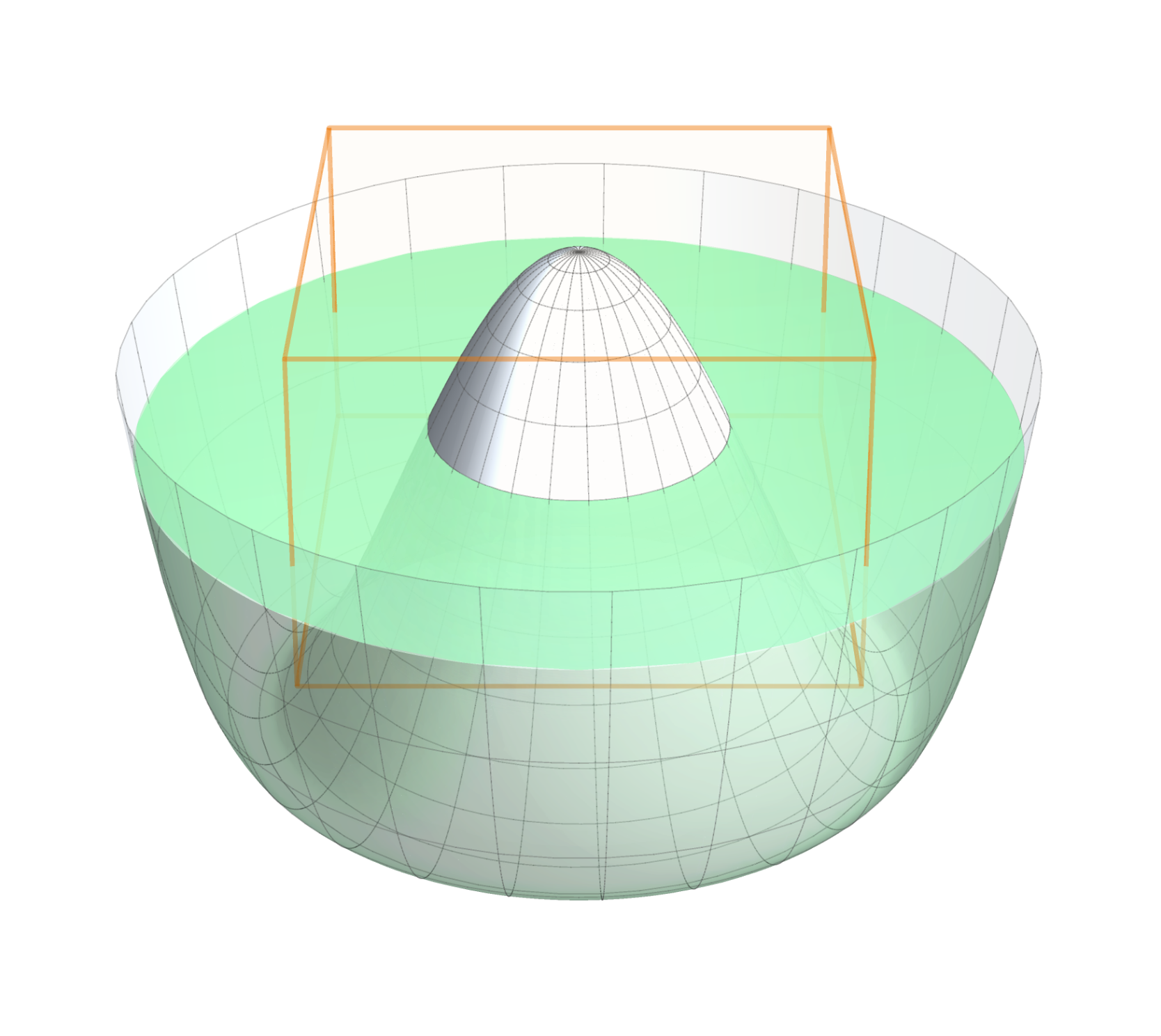}~~~~~~\includegraphics[scale=.125]{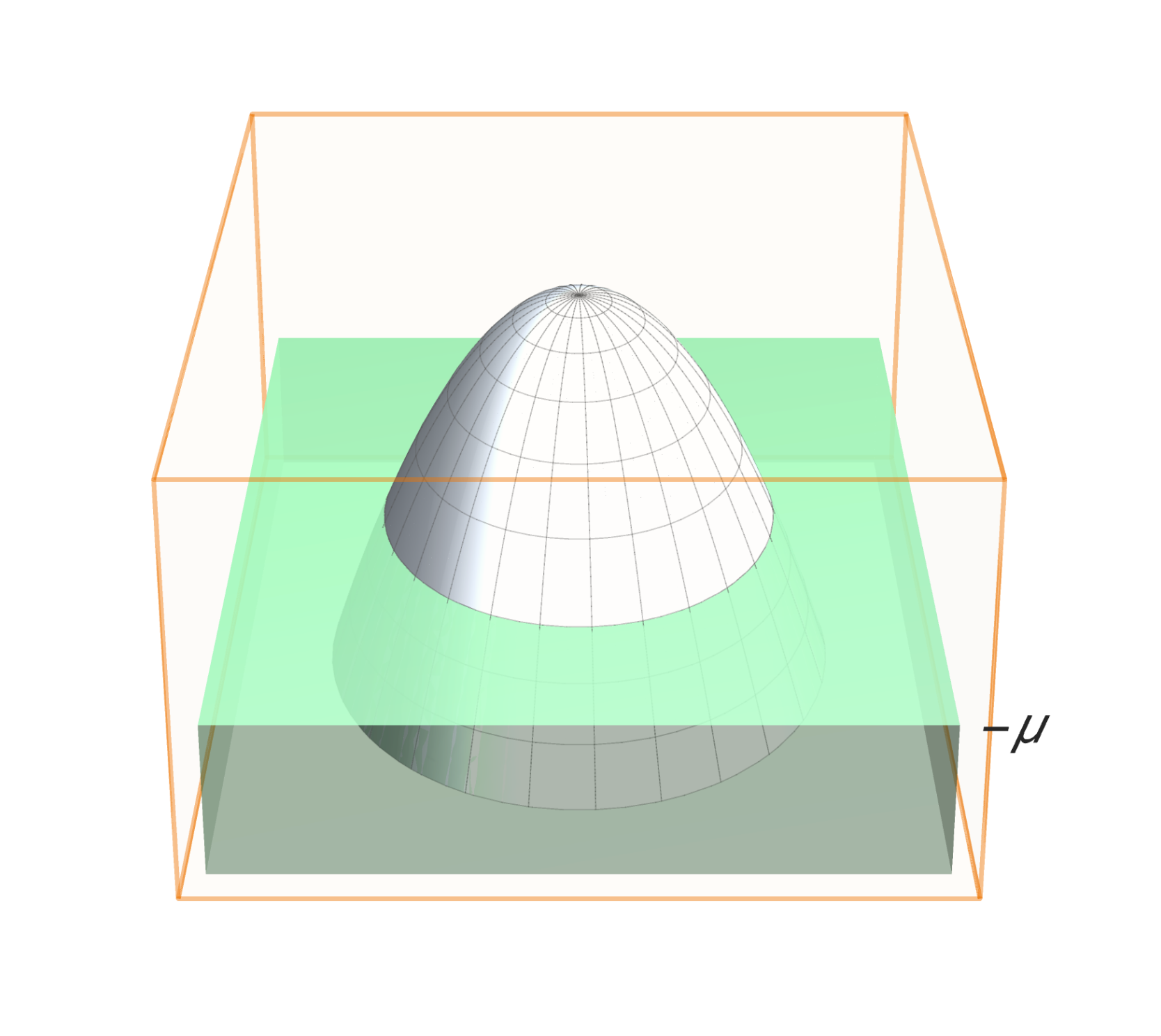}
    \end{center}
    \caption{The double scaling limit of noncritical M-theory is analogous to the double scaling limit of the $c=1$ matrix model (compare to Figure \ref{fig:c=1potential}). The Fermi sea is again filled up to a Fermi level $-\mu$, but in this case the single particle dispersion is given by a two-dimensional rotationally-symmetric inverted harmonic oscillator potential. In the above images, the shaded region is the filled Fermi sea, with the image on the right illustrating the universal potential that survives the double-scaling limit.}
    \label{fig:Mpotential}
\end{figure}

Noncritical M-theory was introduced in \cite{Horava:2005tt,Horava:2005wm} as a double-scaled non-relativistic Fermi liquid in $2+1$ dimensions.
The single-particle Hamiltonian is given by
\begin{equation}
h(\boldsymbol \lambda, \boldsymbol p) = \frac{1}{2} \boldsymbol p^2 - \frac{1}{2} \omega^2\boldsymbol \lambda^2, ~~~~ \boldsymbol \lambda = (\lambda_1, \lambda_2), ~~~~ \boldsymbol p = (p_1,p_2),
\end{equation}
and the theory is defined with respect to the classical Fermi surface (see Figure \ref{fig:Mpotential})
    \begin{align}
        h(\boldsymbol \lambda, \boldsymbol p ) = -\mu.
    \end{align}
The double-scaled field theory for these non-relativistic fermions is defined by the following Hamiltonian density
\begin{align}
\mathcal H = \frac{1}{2}  \partial_{\boldsymbol \lambda} \Psi^{\dagger}\cdot \partial_{\boldsymbol \lambda} \Psi + V(\boldsymbol \lambda) \Psi^{\dagger}\Psi ,~~~~
        V(\boldsymbol \lambda) = - \frac{1}{2} \omega^2 \boldsymbol \lambda^2 + \cdots,
    \end{align}
where throughout the paper we set $\omega = 1$.
Notice that the above Hamiltonian is very much a 2+1d generalization of the $c=1$ matrix model double-scaled Hamiltonian density given in (\ref{Hc=1}).

It is sometimes convenient to represent the Hamiltonian in a (spatial) coordinate-free manner as
    \begin{align}
    \label{eq:coordfreeHFermi}
       \mathcal H =\Psi^\dagger \left( -\frac{1}{2}  \boldsymbol \partial_{} \cdot \boldsymbol \partial_{} + V  \right)\Psi.
    \end{align}
Since this system is free, we can introduce the following schematic mode expansion:
\begin{align}
\label{Psialpha}
\Psi = \sum_{\boldsymbol n}a_{\boldsymbol n} \psi_{\boldsymbol n},~~~~ \{ a_{\boldsymbol n} , a^\dagger_{\boldsymbol n'}\} = \delta_{\boldsymbol n, \boldsymbol n'}.
\end{align}
In the above expression, $\boldsymbol n$ collectively denotes a complete set of quantum numbers, including the single-particle energy $E$. Time dependence is incorporated into this description in the usual way, namely by inserting phase factors $e^{-i Et}$ in the above sum.

The standard vacuum state $\ket{M,\mu}$ of noncritical M-theory in the double scaling limit corresponds to a Fermi sea filled up to negative Fermi energy $-\mu$:
    \begin{align}
    \begin{split}
        a_{\boldsymbol n}(E) \ket{M, \mu} &= 0,~~~~  E > -\mu\\
        a^\dagger_{\boldsymbol n}(E) \ket{M, \mu} &=0 ,~~~~ E < -\mu ,
    \end{split}
    \end{align}
for all values of the other quantum numbers $\boldsymbol n$.
It was proposed in \cite{Horava:2005tt}
 that the vacuum state $\ket{M,\mu}$ represents the M-theory lift of the linear dilaton vacua of the type 0A and 0B string theories. Other noncritical string vacua can be recovered from the noncritical M-theory vacuum by ``draining'' parts of the Fermi sea in a specific manner, as we demonstrate below.

Within noncritical M-theory, the quantum numbers used to define the type 0A string vacuum are the energy and angular momentum, $\boldsymbol n = ( E, q)$, and the quantum numbers used to define the type 0B string are the energies and parities of two independent $1+1$d inverted harmonic oscillators, $\boldsymbol n = (E_1,E_2,s_1,s_2)$:\footnote{Note that other mode expansions are also possible as there are many choices of observables that commute with the single particle Hamiltonian.}
	\begin{align}
	\begin{split}
    \label{type0modes}
	\text{0A} ~~~~:&~~~~	\Psi = \int \mathrm dE\,   \sum_{q \in \mathbb Z} a_{q}(E) e^{-i E t} \psi_q(E)\\
	\text{0B}~~~~:&~~~~	\Psi= \int \mathrm dE_1 \mathrm dE_2 \,  \sum_{s_1,s_2 = \pm{}} a_{s_1,s_2}(E_1,E_2)  e^{-i (E_1 +E_2) t} \psi_{s_1}(E_1) \psi_{s_2}(E_2).
	\end{split}
	\end{align}
 The type 0A vacuum can be recovered from noncritical M-theory by restricting the angular momentum $q$ to a fixed value:
\begin{align}
\begin{split}
    a_q(E) \ket{0A, q, \mu} &=  0 ,~~~~ E >-\mu\\
    a^\dagger_q(E) \ket{0A, q,\mu} &=0,~~~~ E < -\mu\\
    a_{q'}(E) \ket{0A,q,\mu}& = 0,~~~~ ~ \text{any $E$},~~~~q' \ne q.
\end{split}
\end{align}
For reference, the single particle eigenstates that appear in the type 0A mode expansion, which diagonalize the Hamiltonian $h =  (p_r^2 +  p_\theta^2/ \lambda_r^2 - \lambda_r^2 )/2$, are given by \cite{Demeterfi:1993cm}
    \begin{align}
    \begin{split}
    \label{0Afun}
        \psi_{q}(E;\lambda_r,\theta) &= \frac{1}{\sqrt{2\pi}} e^{i q\theta} \psi_{q}(E;\lambda_r)\\
        \psi_q(E;\lambda_r) &= \frac{1}{\sqrt{2\pi} \lambda_r} e^{- i \frac{\pi}{2}\left( \frac{|q|}{2} + \frac{1}{2} \right)} e^{\frac{\pi}{4} E } \frac{\left|\Gamma\left(\frac{1}{2}- \frac{i E}{2} + \frac{|q|}{2} \right)\right|}{ \Gamma( |q|+1) } M_{-\frac{iE}{2},\frac{|q|}{2}} (i \lambda_r^2).
    \end{split}
    \end{align}
    Likewise, the type 0B vacuum can be recovered by restricting, e.g., the energy $E_2$ and parity $s_2$ to fixed values:
    \begin{align}
        \begin{split}
            a_{s_1,s_2}(E_1,E_2)\ket{0B,\mu} &=0,~~~~E_1 >-\mu,~~~~E_2 = \bar E_2, ~~~~s_2 =\bar s_2\\
            a^\dagger_{s_1,s_2}(E_1,E_2) \ket{0B,\mu} &=0,~~~~ E_1 < -\mu,~~~~ E_2 = \bar E_2,~~~~ s_2 = \bar s_2\\
            a_{s_1,s_2}(E_1,E_2) \ket{0B,\mu} &=0,~~~~~\text{any $E_1$},~~~~~ (E_2,s_2) \ne (\bar E_2,\bar s_2),
        \end{split}
    \end{align}
and the single particle eigenstates appearing in the type 0B mode expansion, which diagonalize the  Hamiltonian $h = h_1 + h_2, h_i =  ( p_i^2 - \lambda_i^2 )/2$, are given by
    \begin{align}
        \begin{split}
            \psi_{s_1,s_2} (E_1,E_2;\lambda_1,\lambda_2) &= \psi_{s_1}(E_1;\lambda_1) \psi_{s_2}(E_2;\lambda_2) \\
            \psi_{s}(E;\lambda) &= \sgn(\lambda)^{\frac{1-s}{2}}\frac{ e^{i \frac{\pi}{2}\left(\frac{s}{4}-\frac{1}{2}\right) }}{2^{\frac{1+s}{2}} \pi } e^{\frac{\pi}{4} E} \frac{\left| \Gamma\left( \frac{1}{2} - \frac{i E}{2}-\frac{s}{4}\right)\right| }{\sqrt{|\lambda|}}M_{- \frac{i E}{2},-\frac{s}{4}}(i \lambda^2).
        \end{split}
    \end{align}
In the above expressions $M_{\mu,\nu}(z)$ are Whittaker functions; these wavefunctions are described in more detail in, e.g., \cite{Moore:1991zv}.

\subsection{Coordinate systems}

\label{sec:cancoord}

In Section \ref{defMFermi}, we  presented the Fermi liquid formulation of noncritical M-theory in a coordinate-free manner. An advantage of the coordinate-free presentation is that it gives us the freedom to choose different (typically canonical) coordinate systems, depending on what features of the theory we wish to highlight. In particular, we will see that the integrable structure of the theory is made manifest by working in terms of so-called action-angle coordinates, which we define below. In this section, we summarize the various coordinate systems that we will employ when studying the correlation functions of density operators in noncritical M-theory, in Sections \ref{Mrectdensity} and \ref{exactLCFermi}.  More details about canonical coordinates and other properties of the classical phase space of noncritical M-theory are collected in Appendix \ref{classPS}.

The first set of coordinates that we will use is (non-canonical) doubly-polar coordinates, defined implicitly by
    \begin{align}
    \label{dub}
        \boldsymbol \lambda = (\lambda_r \cos \theta, \lambda_r \sin \theta),~~~~\boldsymbol p = (p \cos \xi, p \sin \xi).
    \end{align}
We sometimes write  
    \begin{align}
        \boldsymbol p = p \boldsymbol n_\xi, ~~~~ \boldsymbol t_{\xi} \equiv \partial_\xi \boldsymbol n_\xi~~\implies ~~ \boldsymbol p \cdot \boldsymbol t_\xi = 0.
    \end{align}
The single particle Hamiltonian in these coordinates is given by
    \begin{align}
        h(\lambda_r, p) = \frac{1}{2} p^2- \frac{1}{2} \lambda_r^2.
    \end{align}
We parametrize the Fermi surface by writing
    \begin{align}
        p_{\text F}(\lambda_r) = \sqrt{\lambda_r^2 - 2 \mu},~~~~\lambda_r \geq \sqrt{2\mu}.
    \end{align}

Another class of canonical coordinates that we will use is action-angle coordinates $\tau, h, \varphi, m$, defined by
    \begin{align}
    \label{genaa}
        \{ \tau, h\} =\{ \varphi, m\} =1
    \end{align}
with all other Poisson brackets vanishing, and where $h$ is the Hamiltonian. Note that this coordinate system is not unique, as there are many choices of canonical coordinates satisfying the above properties. Notice in particular that the coordinates $\varphi,m$ are constants of motion, since they Poisson-commute with the Hamiltonian:
    \begin{align}
       \{ h, \varphi\} = \{ h, m\} =0.
    \end{align}
The Fermi surface is parametrized by
    \begin{align}
            h = - \mu.
    \end{align}
One specific set of action-angle coordinates that we use in Section \ref{Mrectdensity} is defined by
    \begin{alignat}{2}
        \tau &= \tau_1 +\tau_2 & \qquad h&= h_1 +h_2 \\
        \varphi &= \tau_1-\tau_2&\qquad m&= h_1-h_2,
    \end{alignat}
where
    \begin{align}
        \tau_i = \frac{1}{2} \arctanh\left( \frac{p_i}{\lambda_i} \right),~~~~h_i =\frac{1}{2} (p_i^2 - \lambda_i^2).
    \end{align}
This is a natural set of coordinates in which to describe the type 0B string vacuum, as $\hat h_1,\hat h_2$ are diagonalized in this basis.

In Section \ref{exactLCFermi}, we use another set of action-angle coordinates, given by the polar representation of the light-cone coordinates $z_\sigma = [ p_1 + \sigma \lambda_1 + i (p_2 + \sigma \lambda_2)]/2 = e^{\tau_\sigma + i \varphi_\sigma} / \sqrt{2}$, namely
\begin{alignat}{2}
\label{eq:LCcoord1}
    \tau_\sigma &= \log\left(\sqrt{2 z_\sigma z_\sigma^*}\right) &\qquad
    h_\sigma &= \sigma \left(z_\sigma z_{-\sigma}^* + z_\sigma^* z_{-\sigma}\right) \equiv \sigma h \\
    \varphi_\sigma &= -i \arctanh\left(\frac{z_\sigma - z_\sigma^*}{z_\sigma + z_\sigma^*}\right) &\qquad
    \ell &= \sigma i \left(z_\sigma z_{-\sigma}^* - z_\sigma^* z_{-\sigma}\right),
\end{alignat}
where $\sigma =\pm{}1$ distinguishes ``left-moving'' from ``right-moving'' coordinates.\footnote{Although $\ell$ is the same angular momentum that appears in polar coordinates, note that $\varphi_\sigma$ is not the same as the polar angle $\theta$.} By abuse of terminology we refer to these coordinates as ``light-cone coordinates'' when there is little chance for confusion from the context. The non-vanishing Poisson brackets are
    \begin{align}
        \{ \tau_\sigma, h_\sigma\} =  \{\varphi_\sigma, \ell\} = 1      .
    \end{align}
It follows that $\varphi_\sigma, \ell$ Poisson-commute with $h = \sigma h_\sigma$. This is a natural system of coordinates in which to describe the type 0A string vacuum, as $\hat h, \hat \ell$ are diagonalized in this system.

\subsection{Density operators}
\label{MFermidensity}

In this section, we give a general description of the main observables of interest in noncritical M-theory, namely the  Euclidean time-ordered correlation functions of the eigenvalue density operators
    \begin{align}
\label{rhot}
\rho \equiv \Psi^\dagger \Psi = \sum_{\boldsymbol n,\boldsymbol m}a_{\boldsymbol n}^{\dagger} a_{\boldsymbol m}\psi^*_{\boldsymbol n}  \psi_{\boldsymbol m}.
\end{align}
These eigenvalue density operators are the natural M-theory analog of the eigenvalue density operators defined in the context of the $c=1$ string. Note that we are free to project the above fermion bilinear operator onto any basis $\ket{\boldsymbol \lambda}$:
\begin{align}
\rho(t,  \boldsymbol \lambda)  = \sum_{\boldsymbol n,\boldsymbol m}a_{\boldsymbol n}^{\dagger} a_{\boldsymbol m} \psi_{\boldsymbol n}(t,{\boldsymbol \lambda})^* \psi_{\boldsymbol m}(t,{\boldsymbol \lambda}),
\end{align}
where we stress here that two density operators $ \rho(t,\boldsymbol \lambda), \rho(t,\tilde{\boldsymbol{ \lambda}})$ defined in terms of two different coordinate bases $\ket{\boldsymbol \lambda}, \ket{\tilde{\boldsymbol{\lambda}}}$ are not related by a unitary map, and they correspond to distinct physical observables. We will see, in particular, that comparing the correlation functions of different Fermi liquid density operators provides complementary sets of physical observables to compare with their counterparts in the bosonized theory, as discussed in Section \ref{bosonicdensitycorr}.

After defining the Euclidean propagator below, we give a general expression for the $n$-point correlation functions of the operators, which follows from a standard application of Wick's theorem.

\subsubsection{Single particle propagator}
\label{MFermiprop}

The basic building block of the correlation functions (\ref{Mcorrelators}) is the Euclidean time-ordered two-point function
\begin{align}
        \begin{split}
        \label{Fermiprop}
        S(i,j) =\braket{ M,\mu | T (\Psi^\dagger(i) \Psi(j))| M,\mu}.
        \end{split}
	\end{align}
Given a mode expansion \eqref{Psialpha} of the fermion field $\Psi$ in terms of some basis of eigenfunctions, the time-ordered two-point function can be expressed more explicitly as
	\begin{align}
		\begin{split}
			S(i,j)
			& =\int \mathrm dE\, \sum_{\boldsymbol n} \psi^*_{\boldsymbol n}(E,i) \psi_{\boldsymbol n}(E,j) \biggl[ \theta(-E-\mu) \theta(\Delta t )  - \theta(E + \mu) \theta(-\Delta t) \bigg]  e^{iE \Delta t},
		\end{split}
	\end{align}
where $  \Delta t \equiv t_i - t_j$. We analytically continue to Euclidean time $t = -i x$ to obtain the Euclidean propagator, which by abuse of notation we denote with the same symbol:
    \begin{align}
        \begin{split}
        \label{Sij}
        S(i,j) 	& = \int \mathrm dE\, \sum_{\boldsymbol n} \psi^*_{\boldsymbol n}(E,i) \psi_{\boldsymbol n}(E,j) \biggl[ \theta(-E-\mu) \theta(\Delta x )  - \theta(E + \mu) \theta(-\Delta x) \bigg]  e^{E \Delta x}\\
            &= \int \mathrm dE \,e^{-\mu  \Delta x} \int \frac{\mathrm dp}{2\pi} \frac{-i}{p + i (E+\mu)} e^{i p  \Delta x} \sum_{\boldsymbol n}  \psi^*_{\boldsymbol n}(E,i) \psi_{\boldsymbol n}(E,j)\\
            &=-i e^{-\mu  \Delta x} \int \frac{\mathrm dp}{2\pi} e^{i p  \Delta x} \int_0^{\sgn(p) \infty} \mathrm ds\, e^{-(p+i\mu)s} \int dE\,\sum_{\boldsymbol n}\psi^*_{\boldsymbol n}(E,i) \psi_{\boldsymbol n}(E,j) e^{-i E s}\\
            &=-i e^{-\mu \Delta x} \int \frac{\mathrm dp}{2\pi} e^{i p \Delta x} \int_0^{\sgn(p) \infty} \mathrm ds\, e^{-(p+i\mu)s} \braket{i | e^{-i h s} | j}.
        \end{split}
    \end{align}
In going to the second line above, we introduced the Fourier integral representation of the step function; in going to the third line, we replaced the resulting integrand by its Schwinger parametrization; and in going to the fourth line, we expressed the sum over $E,\boldsymbol n$ as the single particle propagator.

In terms of the type 0B mode basis given in the second line of (\ref{type0modes}), it is easy to calculate the Euclidean propagator because the eigenmodes factorize:
\begin{align}
\begin{split}
\braket{\boldsymbol \lambda_f | e^{-  i h \Delta x } |\boldsymbol \lambda_i} &= \braket{\lambda_f^1|e^{-i h_1 \Delta x}|\lambda_i^1}\braket{\lambda_f^2|e^{-i h_2 \Delta x}|\lambda_i^2}  \\
&= \frac{1}{2\pi i \sinh \Delta x} \exp \Bigg( \frac{i}{2\sinh \Delta x} \Big[ (\bm{\lambda}_i^2 + \bm{\lambda}_f^2 )\cosh \Delta x  - 2 \bm{\lambda}_i \cdot \bm{\lambda}_f \Big] \Bigg),
\end{split}
\end{align}
where $\Delta x = x_f - x_i$.

\subsubsection{Resolvent and WKB approximation}
In the $\ket{\boldsymbol \lambda}$ basis, the noncritical M-theory resolvent is given by
\begin{align}
I(p,\bm{\lambda}_f,\bm{\lambda}_i) = \braket{\bm{\lambda}_f|\frac{1}{h + \mu - i p}|\bm{\lambda}_i} = i \int  _0^{ \sgn(p) \infty} \mathrm ds \braket{\bm{\lambda}_f|e^{-ih s }|\bm{\lambda}_i}e^{-i\mu s} e^{-p s}.
\end{align}
Using the single particle propagator for the inverted harmonic oscillator, the integral on the right hand side of the above expression becomes
\begin{align}
I(p,\bm{\lambda}_f,\bm{\lambda}_i)  &= i \int  _0^{ \sgn(p) \infty}   \frac{\mathrm ds}{2\pi i \sinh s}\exp \Big[ \frac{i}{2} \frac{1}{\tanh s}(\bm{\lambda}_i^2 + \bm{\lambda}_f^2)- i\frac{1}{\sinh s} \bm{\lambda}_i \cdot \bm{\lambda}_f - i\mu s - ps\Big]  \notag \\
&= \frac{1}{2\pi} \int_0^{\sgn(p)1} \frac{\mathrm du}{u} \left(\frac{1-u}{1+u} \right)^{ p + i\mu} \exp \Big[\frac{i}{4}  \left( u +\frac{1}{u} \right)(\bm{\lambda}_i^2 + \bm{\lambda}_f^2) -\frac{i}{2}  \bm{\lambda}_i \cdot \bm{\lambda}_f \left( \frac{1}{u} -u\right) \Big] \notag \\
&= \frac{1}{2\pi} \int_0^{\sgn(p)1} \frac{\mathrm du}{u} \left(\frac{1-u}{1+u} \right)^{ p + i\mu} \exp \Big[\frac{i}{4u}  (\bm{\lambda}_i - \bm{\lambda}_f)^2  + i\frac{u}{4} (\bm{\lambda}_i + \bm{\lambda}_f)^2 \Big]. \label{eq:MresInt}
\end{align}
In going to the second line above, we have changed the integration variable to $u = \tanh \frac{s}{2}$.\footnote{Both $u=0$ and $u=1$ are essential singularities and, in particular, the line $u = \pm i t, t \in \mathbb{R}_+$ is not a good contour.}
For $p>0$, we can change the integration contour to $u(\theta) = \frac{1-e^{i\theta}}{2}, \theta \in (0,\pi)$ so that the integrand has better damping behavior at $u=0$.
\begin{figure}[t]
\begin{center}
\includegraphics[width=6.5cm]{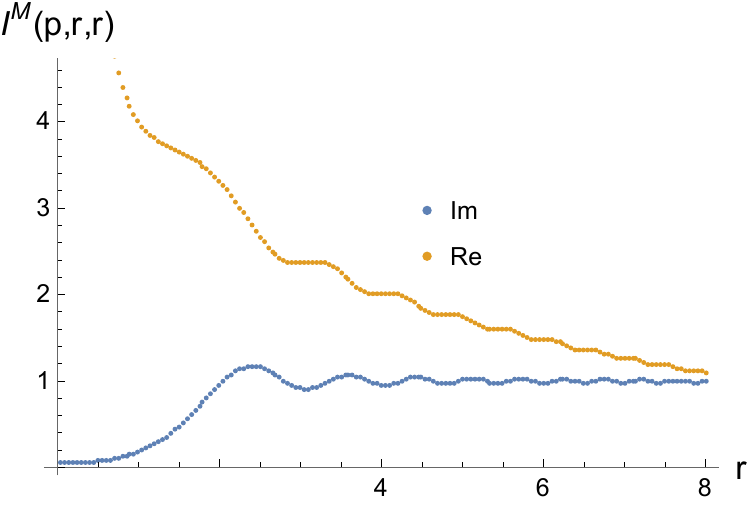}
\includegraphics[width=6.5cm]{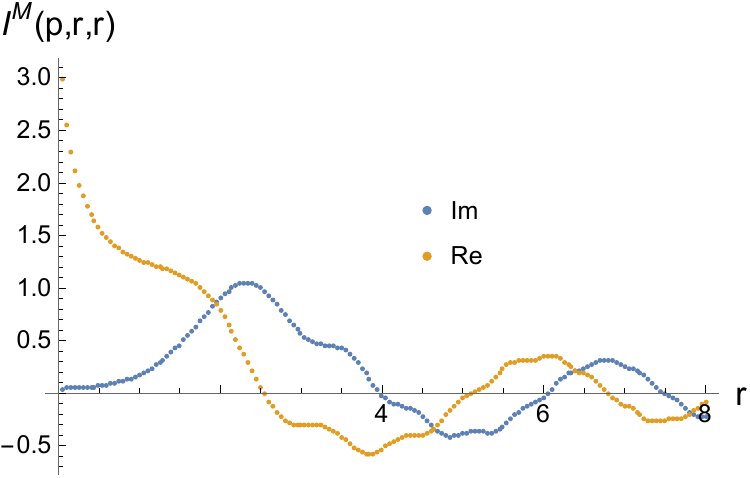}
\caption{Plot of the M-theory resolvent. Here $\theta$ denotes the angle between $\bm{\lambda}_i$ and $\bm{\lambda}_f$, with $|\bm{\lambda}_i| = |\bm{\lambda}_f| = \lambda_r$ in the figures.
For both plots we choose $\mu = 1$.
In the left panel we take $p = 0.001, \theta = 0$.
In the right panel we take $p = 0.001, \theta = \pi/20$.
}
\label{fig:Resolventp0001}
\end{center}
\end{figure}
\begin{figure}[t]
\begin{center}
\includegraphics[width=6.5cm]{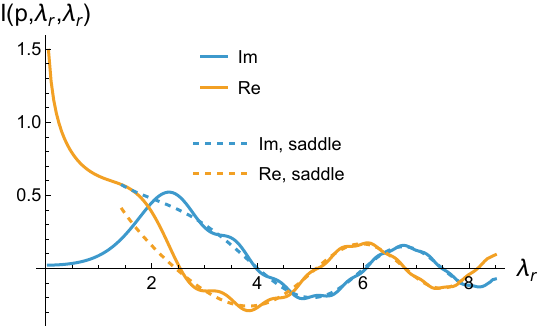}
\includegraphics[width=6.5cm]{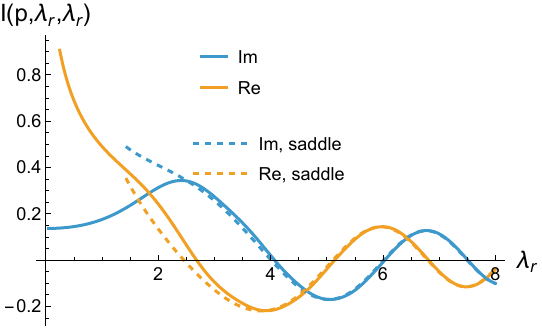}
\caption{
Comparison between the exact resolvent and its saddle-point approximation.
The left panel shows the case $p=0.001$, while the right panel corresponds to $p=1$.
The parameters are chosen as $\theta=\pi/20$ and $\mu=1$.
The qualitative behavior remains unchanged for sufficiently small $|\theta|$ and arbitrary $\mu$.
}\label{fig:Resolventp001}
\end{center}
\end{figure}
It follows that the M-theory resolvent can be computed using a  saddle point approximation, which we describe. First, let us write
\begin{align}
I(p,\bm{\lambda}_f,\bm{\lambda}_i) 
&= \frac{1}{2\pi} \int_0^{\sgn(p)1} \mathrm du \exp \left( i S(u) \right),
\end{align}
where
\begin{align}
\label{eq:saddleS}
S(u) = \frac{1}{4u} (\bm{\lambda}_i -  \bm{\lambda}_f)^2 +  \frac{u}{4} (\bm{\lambda}_i + \bm{\lambda}_f)^2  - i (p + i\mu) \log  \left( \frac{1-u}{1+u}\right) +i \log u.
\end{align}
Next, consider the limit $\mu \to \infty, |\boldsymbol \lambda_i| \to \infty, |\boldsymbol \lambda_f| \to \infty$, where all three quantities are scaled at the same rate. Using the expression for the phase $S(u)$ given above, we find that the saddle-point equations are
\begin{align}
\begin{split}
S'(u) &= - \frac{1}{4u^2}(\bm{\lambda}_i -  \bm{\lambda}_f)^2 +  \frac{1}{4} (\bm{\lambda}_i + \bm{\lambda}_f)^2 +i (p + i \mu) \left(\frac{1}{1-u} + \frac{1}{1+u} \right)  + \frac{i}{u}\\
S''(u) &= \frac{1}{2u^3} (\bm{\lambda}_i -  \bm{\lambda}_f)^2+i (p + i \mu) \left(\frac{1}{(1-u)^2} - \frac{1}{(1+u)^2} \right)  - \frac{i}{u^2},
\end{split}
\end{align}
hence the saddle point approximation for the resolvent integral is
\begin{align}
I(p,\bm{\lambda}_f,\bm{\lambda}_i) \approx \frac{1}{2\pi} \sum_{u_0} \sqrt{\frac{2\pi i}{S''(u_0)}} e^{i S(u_0)}.
\end{align}
For large $|\bm{\lambda}_i|$ and $|\bm{\lambda}_f|$, the last two terms in the action \eqref{eq:saddleS} are subleading and may be neglected, and in this limit 
the saddle-point equation reduces to
\begin{align}
S'(u)\approx - \frac{1}{4u^2}(\bm{\lambda}_i -  \bm{\lambda}_f)^2 +  \frac{1}{4} (\bm{\lambda}_i + \bm{\lambda}_f)^2 = 0 ~~\implies ~~ u_0 = \pm
\sqrt{\frac{(\bm{\lambda}_i -  \bm{\lambda}_f)^2}{(\bm{\lambda}_i + \bm{\lambda}_f)^2}}.
\end{align}
For $|\theta|<\pi/2$, the saddle
\begin{align}
u_0=
\sqrt{\frac{(\bm{\lambda}_i-\bm{\lambda}_f)^2}
{(\bm{\lambda}_i+\bm{\lambda}_f)^2}}
\end{align}
lies on the original integration contour.
Evaluating the action at this saddle yields
\begin{align}
\begin{split}
S(u_0) &\approx \frac{1}{2}\sqrt{(\bm{\lambda}_i -  \bm{\lambda}_f)^2( \bm{\lambda}_i +  \bm{\lambda}_f)^2} \\
&- i (p+ i\mu) \log \frac{\sqrt{(\bm{\lambda}_i +  \bm{\lambda}_f)^2}-\sqrt{(\bm{\lambda}_i -  \bm{\lambda}_f)^2}}{\sqrt{(\bm{\lambda}_i +  \bm{\lambda}_f)^2}+\sqrt{(\bm{\lambda}_i -  \bm{\lambda}_f)^2}}+ i \log \sqrt{\frac{(\bm{\lambda}_i -  \bm{\lambda}_f)^2}{(\bm{\lambda}_i + \bm{\lambda}_f)^2}},
\end{split}
\end{align}
and the one-loop contribution is given by
\begin{align}
S''(u_0)  \approx \frac{1}{2} \sqrt{\frac{(\bm{\lambda}_i + \bm{\lambda}_f)^2}{(\bm{\lambda}_i -\bm{\lambda}_f)^2}}(\bm{\lambda}_i + \bm{\lambda}_f)^2.
\end{align}
Substituting these results into the saddle-point formula, we obtain
\begin{align}
&I(p,\bm{\lambda}_f,\bm{\lambda}_i) \notag \\
&\approx  \frac{1}{2\pi}\sqrt{\frac{2\pi i}{{S''(u_0)}}} e^{i S(u_0) } \notag \\
&= \frac{1}{2\pi}\sqrt{\frac{4\pi i}{\sqrt{(\bm{\lambda}_i - \bm{\lambda}_f)^2(\bm{\lambda}_i + \bm{\lambda}_f)^2}} } \Bigg ( \frac{\sqrt{(\bm{\lambda}_i+\bm{\lambda}_f)^2}-\sqrt{(\bm{\lambda}_i-\bm{\lambda}_f)^2}}{\sqrt{(\bm{\lambda}_i+\bm{\lambda}_f)^2}+\sqrt{(\bm{\lambda}_i-\bm{\lambda}_f)^2}} \Bigg)^{p+i\mu} e^{\frac{i}{2} \sqrt{(\bm{\lambda}_i+\bm{\lambda}_f)^2(\bm{\lambda}_i-\bm{\lambda}_f)^2}} \label{eq:RNMsaddle1}.
\end{align}
We compare the exact resolvent with its saddle-point approximation in Figure~\ref{fig:Resolventp001}.
The agreement is excellent for
$|\bm{\lambda}_i|\gg1$
and
$|\bm{\lambda}_f|\gg1$,
apart from small rapid oscillations around the saddle-point result \eqref{eq:RNMsaddle1}.
These oscillations are analogous to the brane effects in noncritical string theory \cite{Saad:2019lba}.
They can also be reproduced by including the contribution from the second saddle point, as discussed in Appendix~\ref{sec:resolvent2nd}.

\subsubsection{Euclidean time-ordered correlation functions}

\label{Mdensitycorr}

It is straightforward to use the expansion (\ref{rhot}) to compute time-ordered correlation functions of the fermion bilinears $\rho$ in the M-theory vacuum state $\ket{M,\mu}$ that are valid for any coordinate representation, namely
    \begin{align}
    \label{Mcorrelators}
          G^{(n)}(1,\dots,n)=  \braket{M,\mu|T (\rho(1) \cdots \rho(n)) |M,\mu}.
    \end{align}
A standard application of Wick's theorem for fermions (see Appendix \ref{Wick} for a review) shows that the time-ordered $n$-point functions are given by
	\begin{align}
    \begin{split}
    \label{FermiG}
		G^{(n)}(1,\dots, n) &=\det(S)
    \end{split}
    \end{align}
where the components $S(i,j)$ of the matrix $S$ are the time-ordered fermion propagators (\ref{Fermiprop}).

\subsection{Eigenvalue density operators}
\label{Mrectdensity}

Below, we collect some results for the connected correlation functions of the Euclidean time eigenvalue density operators
    \begin{align}
        \begin{split}\rho(x,\boldsymbol \lambda ) &= \Psi^\dagger(x, \boldsymbol \lambda) \Psi(x,\boldsymbol \lambda) =\sum_{\boldsymbol n, \boldsymbol m} a_{\boldsymbol n}^\dagger a_{\boldsymbol m} \psi_{\boldsymbol n}^*(x,\boldsymbol \lambda) \psi_{\boldsymbol m}(x, \boldsymbol \lambda),
        \end{split}
    \end{align}
namely the density operators expressed in the coordinate basis $\ket{\boldsymbol \lambda}$. Physically, the bilinear operator $\rho(x, \boldsymbol \lambda)$ corresponds to a perturbation of the Fermi sea expressed in the standard basis of rectilinear coordinates $\boldsymbol \lambda = (\lambda_1,\lambda_2)$.

\subsubsection{One-point function}
\begin{figure}[t]
\begin{center}
\includegraphics[width=7cm]{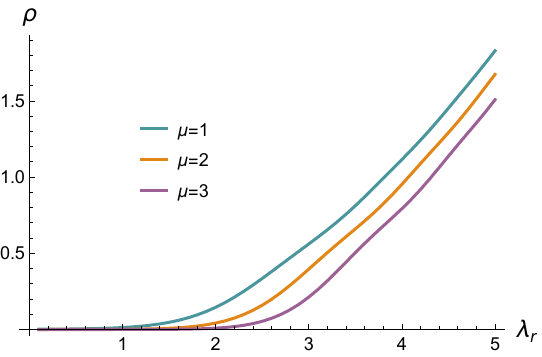}
\includegraphics[width=7cm]{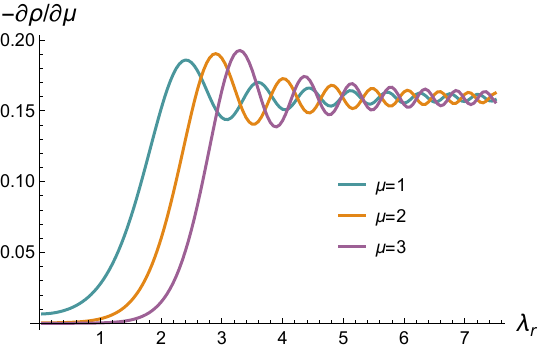}
\end{center}
\caption{Representative plots of the exact density one-point function.
The left panel shows the density $\rho$, while the right panel shows its derivative with respect to the Fermi energy, $-\partial \rho / \partial \mu$.
}
\label{fig:MtheoryOnePointMuD}
\end{figure}

The one-point function is given by
    \begin{align}
     \braket{ \rho} &= \sum_{\{\boldsymbol n  | E \leq - \mu\}} | \psi_{\boldsymbol n}|^2.
    \end{align}
In the type 0A basis, the one-point function of $\rho(x,\boldsymbol \lambda)$ is given in polar coordinates $\boldsymbol \lambda = (\lambda_r, \theta)$ by (see (\ref{0Afun}))
    \begin{align}
    \begin{split}
\label{0A1pt}
\braket{\rho(x,\boldsymbol\lambda)}
&= \int_{-\infty}^{-\mu} \mathrm dE \, \sum_{q \in \mathbb Z} |\psi_{q}(E;\lambda_r ,\theta) |^2= \frac{1}{2\pi} \int_{-\infty}^{-\mu} \mathrm dE \, \sum_{q \in \mathbb Z} |\psi_{q}(E;\lambda_r ) |^2
    \end{split}
    \end{align}
and its derivative with respect to the Fermi level is given by
\begin{equation}
 \frac{\partial}{\partial (-\mu)} \braket{\rho(x,\boldsymbol\lambda)}  = \frac{1}{2\pi} \sum_{q \in \mathbb{Z}} |\psi_q(-\mu,\lambda_r)|^2 .
\end{equation}
Thus, we see from \eqref{0A1pt} and below that the exact one-point function can be computed in terms of the type 0A eigenfunctions.
Several representative plots are displayed in Figure~\ref{fig:MtheoryOnePointMuD}.

We can also write the one-point function of $\rho(x, \boldsymbol \lambda)$ in the type 0B basis:
    \begin{align}
        \begin{split}
            \braket{\rho(x,\boldsymbol \lambda)} &= \int \mathrm dE_1 \mathrm dE_2 \, \theta(-E_1-E_2-\mu) \sum_{s_1} \left| \psi_{s_1}(E_1;\lambda_1)\right|^2 \sum_{s_2} \left| \psi_{s_2}(E_2;\lambda_2)\right|^2.
        \end{split}
    \end{align}
From (\ref{0A1pt}) and below, we can see that the one-point function is rotationally symmetric. Thus, we are free to write the above expression in polar coordinates and fix the angle to any convenient choice, e.g., $\theta = \pi/2$, so that
  its derivative with respect to the Fermi level is given by
\begin{equation}
\frac{\partial}{\partial (-\mu)} \braket{\rho(x,\boldsymbol\lambda)} = \int \mathrm dE\,  \psi_{+}(E;0)^2 \sum_{s}|\psi_{s}(-E-\mu;\lambda_r)|^2,
\end{equation}
where $\lambda_r =| \boldsymbol \lambda |$.

\subsubsection{Two-point function}

The connected two-point function of the density operator is defined by
\begin{align}
&\braket{T(\rho(x_1, \bm{\lambda}_1)\rho(x_2, \bm{\lambda}_2))} _c
\notag \\
&\equiv
\braket{M,\mu|
T(\Psi^{\dagger}(x_1,\bm{\lambda}_1)\Psi(x_1,\bm{\lambda}_1)
\Psi^{\dagger}(x_2,\bm{\lambda}_2)\Psi(x_2,\bm{\lambda}_2))
|M,\mu}  \notag \\
& -  \braket{M,\mu|\Psi^{\dagger}(x_1,\bm{\lambda}_1)\Psi(x_1,\bm{\lambda}_1)|M,\mu}\braket{M,\mu|\Psi^{\dagger}(x_2,\bm{\lambda}_2)\Psi(x_2,\bm{\lambda}_2)|M,\mu}.
\end{align}
The fermion two-point function \eqref{Fermiprop} in the eigenvalue basis is
\begin{align}
S(1,2)
&=\braket{ M,\mu|T (\Psi^{\dagger}(x_1, \bm{\lambda}_1)\Psi(x_2, \bm{\lambda}_2)) |M,\mu} \notag \\
&= \theta(\Delta x) S^h(1,2) -\theta(-\Delta x)S^p(2,1) \notag \\
&= \int \mathrm dE \Big[ \theta(-E-\mu)\theta(\Delta x) - \theta( E + \mu)\theta(-\Delta x) \Big] e^{E \Delta x} \sum_{q \in \mathbb{Z}} \psi_q(E,\lambda_r^1)\psi_q(E,\lambda_r^2)\frac{e^{iq \Delta \theta}}{2\pi}.
\end{align}
Here $\Delta x \equiv x_1 - x_2$, etc.
For large Euclidean time separation, $\Delta x \to \infty$, the fermion propagator behaves as
\begin{align}
S(1,2) \approx  \frac{e^{-\mu \Delta x}}{\Delta x} \frac{1}{2\pi}
\sum_{q\in\mathbb Z}
\psi_q(-\mu,r_1)\psi_q(-\mu,r_2)
e^{-iq \Delta \theta}.
\end{align}
Here $r_i \equiv \lambda_{ri}$ denotes the radial coordinate of the point $\bm{\lambda}_i$.
Using Wick's theorem, we find that the connected density two point function is given by
\begin{align}
\braket{T(\rho(x_1, \bm{\lambda}_1)\rho(x_2, \bm{\lambda}_2))} _c = -S(1,2) S(2,1),
\end{align}
where $\Delta x\equiv x_1-x_2$.
Therefore, for large Euclidean time separation, it follows that the density two-point correlator behaves as
\begin{align}
G^{(2)}(x_f, \bm{\lambda}_f,x_i, \bm{\lambda}_i) &=\braket{\rho(x_f, \bm{\lambda}_f)\rho(x_i, \bm{\lambda}_i)}_c \notag \\
&\approx\frac{1}{|\Delta x|^2}
\left(
\frac{1}{2\pi}
\sum_{q\in\mathbb Z}
\psi_q(-\mu,\lambda_{rf})\psi_q(-\mu,\lambda_{ri})
e^{-iq\Delta \theta}
\right)^2 .
\end{align}
The quantity inside the parentheses in the last line of the above equation can be expressed in terms of the resolvent as
\begin{align}
\lim_{p\to0}
\frac{1}{\pi}
\operatorname{Im}
I(p,\bm{\lambda}_f,\bm{\lambda}_i)
=
\frac{1}{2\pi}
\sum_{q\in\mathbb Z}
\psi_q(-\mu,\lambda_{rf})\psi_q(-\mu,\lambda_{ri})
e^{-iq \Delta \theta} .
\end{align}
For $|\bm\lambda_i|,|\bm\lambda_f|\gg \mu$, the resolvent may be evaluated using the saddle-point approximation in \eqref{eq:RNMsaddle1}.
Substituting the saddle-point approximation of the resolvent into the above expression, we obtain
\begin{align}
& I(p,\bm{\lambda}_f,\bm{\lambda}_i) \notag \\
& \approx \frac{1}{2\pi}\sqrt{\frac{2\pi i}{{S''(u_0)}}} e^{i S(u_0) } \notag \\
&= \frac{1}{2\pi}\sqrt{\frac{4\pi i}{\sqrt{(\bm{\lambda}_i - \bm{\lambda}_f)^2(\bm{\lambda}_i + \bm{\lambda}_f)^2}} } \Bigg ( \frac{\sqrt{(\bm{\lambda}_i+\bm{\lambda}_f)^2}-\sqrt{(\bm{\lambda}_i-\bm{\lambda}_f)^2}}{\sqrt{(\bm{\lambda}_i+\bm{\lambda}_f)^2}+\sqrt{(\bm{\lambda}_i-\bm{\lambda}_f)^2}} \Bigg)^{p+i\mu} e^{\frac{i}{2} \sqrt{(\bm{\lambda}_i+\bm{\lambda}_f)^2(\bm{\lambda}_i-\bm{\lambda}_f)^2}} \notag \\
&=  \frac{1}{\sqrt{\pi}} \sqrt{\frac{1}{\sqrt{(\bm{\lambda}_i - \bm{\lambda}_f)^2(\bm{\lambda}_i + \bm{\lambda}_f)^2} }} e^{ i \Theta(\mu-ip,\bm{\lambda}_i,\bm{\lambda}_f)},
\end{align}
where the phase $\Theta(\mu-ip,\bm{\lambda}_i,\bm{\lambda}_f)$ is defined by
\begin{align}
  &\Theta(\mu-ip,\bm{\lambda}_i,\bm{\lambda}_f) \notag \\
  &= \frac{1}{2} \sqrt{(\bm{\lambda}_i+\bm{\lambda}_f)^2(\bm{\lambda}_i-\bm{\lambda}_f)^2} + (\mu - ip) \log  \Bigg ( \frac{\sqrt{(\bm{\lambda}_i+\bm{\lambda}_f)^2}-\sqrt{(\bm{\lambda}_i-\bm{\lambda}_f)^2}}{\sqrt{(\bm{\lambda}_i+\bm{\lambda}_f)^2}+\sqrt{(\bm{\lambda}_i-\bm{\lambda}_f)^2}} \Bigg) + \frac{\pi}{4}.
\end{align}
In the limit $p\to 0$, $\Theta(\mu,\bm{\lambda}_i,\bm{\lambda}_f)$ is purely real.
Hence, the imaginary part of the resolvent leads to the following contribution,
\begin{align}
\begin{split}
\frac{1}{2\pi}\sum_{q \in \mathbb{Z}} \psi_q(-\mu,\lambda_{ri})  \psi_q(-\mu,\lambda_{rf})e^{-iq \Delta \theta}
&\approx \frac{1}{\pi^{\frac{3}{2}}} \sqrt{\frac{1}{\sqrt{(\bm{\lambda}_i - \bm{\lambda}_f)^2(\bm{\lambda}_i + \bm{\lambda}_f)^2} } } \sin \Theta(\mu,\bm{\lambda}_i,\bm{\lambda}_f),
\end{split}
\end{align}
where its square can be written as
\begin{align}
\begin{split}
\bigg( \frac{1}{2\pi}\sum_{q \in \mathbb{Z}} \psi_q(-\mu,\lambda_{r_i})  \psi_q(-\mu,\lambda_{r_f})e^{-iq \Delta \theta}   \bigg)^2 
\approx \frac{1}{2\pi^3} \frac{ 1 - \cos 2 \Theta(\mu,\bm{\lambda}_i,\bm{\lambda}_f)}{\sqrt{(\bm{\lambda}_i - \bm{\lambda}_f)^2(\bm{\lambda}_i + \bm{\lambda}_f)^2}}  .
\end{split}
\label{eq:SepAv}
\end{align}
For later comparison with the bosonized description in Section \ref{bosonicdensitycorr}, it is useful to rewrite the result in terms of action-angle variables.
Using the identities\footnote{Note that $\bm{\lambda}(\tau + i \pi,m,\varphi) = -\bm{\lambda}(\tau ,m,\varphi)$.}
\begin{align}
(\bm{\lambda}_f - \bm{\lambda}_i )^2 &= 4 \sinh^2 \frac{\tau_f -\tau_i}{2} \left ( -\mu + \mu \cosh (\tau_i + \tau_f) \cosh (2\varphi) - m \sinh (\tau_i + \tau_f) \sinh (2\varphi)\right) \notag \\
(\bm{\lambda}_f + \bm{\lambda}_i )^2 &= 4 \cosh^2 \frac{\tau_f -\tau_i}{2} \left ( \mu + \mu \cosh (\tau_i + \tau_f) \cosh (2\varphi) - m \sinh (\tau_i + \tau_f) \sinh (2\varphi)\right), \label{eq:sqAA}
\end{align}
we find that in the asymptotic region $\tau_i+\tau_f\gg1$, irrespective of the magnitude of $\tau_f-\tau_i$, the dominant contribution is given by
\begin{align}
&\sqrt{(\bm{\lambda}_f - \bm{\lambda}_i )^2(\bm{\lambda}_f + \bm{\lambda}_i )^2}  \notag \\
& \approx   4 \sinh  \frac{|\tau_f -\tau_i|}{2} \cosh  \frac{\tau_f -\tau_i}{2}\left (  \mu \cosh (\tau_i + \tau_f) \cosh (2\varphi) - m \sinh (\tau_i + \tau_f) \sinh (2\varphi)\right)  \notag \\
 &= 2 \sinh |\tau_f - \tau_i| \left(  \mu \cosh (\tau_i + \tau_f) \cosh (2\varphi) - m \sinh (\tau_i + \tau_f) \sinh (2\varphi)\right).
\end{align}
Using the above expansion and neglecting rapidly oscillating terms after making the stationary-phase approximation, the two-point function reduces to
\begin{align}
& \braket{\rho(x_1, \bm{\lambda}_1)\rho(x_2, \bm{\lambda}_2)}_c \notag \\
& \approx \frac{1}{2\pi^3 |\Delta x|^2} \frac{1}{\sqrt{(\bm{\lambda}_i - \bm{\lambda}_f)^2(\bm{\lambda}_i + \bm{\lambda}_f)^2}} \notag \\
&= \frac{1}{4\pi^3 |\Delta x|^2} \frac{1}{|\sinh (\tau_f - \tau_i)|} \frac{1}{\sqrt{ -\mu^2  + \big[ \mu \cosh (\tau_i + \tau_f) \cosh (2\varphi) - m \sinh (\tau_i + \tau_f) \sinh (2\varphi)\big]^2}}. \label{eq:twoptSemiAve}
\end{align}
\begin{figure}[t]
\begin{center}
\includegraphics[width=8cm]{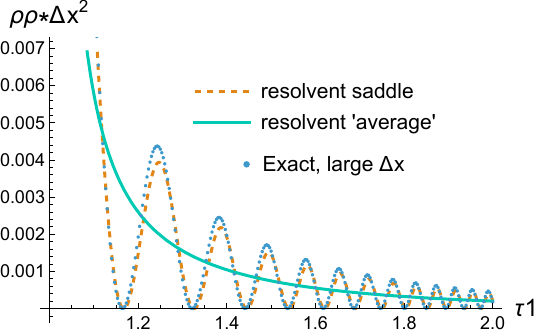}
\end{center}
\caption{Comparison of the density two-point function in the $\Delta x\to\infty$ limit.
The exact result is compared with the saddle-point approximation of the resolvent and with its averaged form obtained by dropping the rapidly oscillating term proportional to $\cos 2\Theta$.
}
\label{fig:MtheoryTwoPointMuD}
\end{figure}
We compare the exact result, the saddle-point approximation, and the averaged saddle-point result obtained by dropping the rapid oscillations proportional to $\cos 2\Theta$ in Figure~\ref{fig:MtheoryTwoPointMuD}.

\subsubsection{Three-point function \label{sec:3ptFermi}}

We consider the connected density three-point function,
\begin{align}
G^{(3)}(x_1,\bm{\lambda}_1,x_2,\bm{\lambda}_2,x_3,\bm{\lambda}_3)
\equiv
\braket{
T\!\left(
\rho(x_1,\bm{\lambda}_1)
\rho(x_2,\bm{\lambda}_2)
\rho(x_3,\bm{\lambda}_3)
\right)
}_c .
\end{align}
Using Wick's theorem (see Appendix~\ref{Wick} for details), the connected three-point function can be expressed in terms of fermionic propagators as
\begin{align}
G^{(3)}(1,2,3)
=
S(1,2)S(2,3)S(3,1)
+
S(1,3)S(3,2)S(2,1).
\end{align}
To evaluate this expression, we consider the asymptotic regime
\begin{align}
x_1-x_2\gg1,
\qquad
x_2-x_3\gg1.
\end{align}
In this limit, the three-point function is approximated by
\begin{align}
&G^{(3)}(x_1,\bm{\lambda}_1,x_2,\bm{\lambda}_2,x_3,\bm{\lambda}_3)
\notag\\
&\approx
\frac{1}{x_{12}x_{23}x_{31}}
\left(
\frac{1}{2\pi}
\sum_{q\in\mathbb Z}
\psi_q(-\mu,r_1)\psi_q(-\mu,r_2)
e^{-iq\theta_{12}}
\right)
\left(
\frac{1}{2\pi}
\sum_{q\in\mathbb Z}
\psi_q(-\mu,r_2)\psi_q(-\mu,r_3)
e^{-iq \theta_{23}}
\right)
\notag\\
&\qquad\times
\left(
\frac{1}{2\pi}
\sum_{q\in\mathbb Z}
\psi_q(-\mu,r_3)\psi_q(-\mu,r_1)
e^{-iq\theta_{31}}
\right)
+(132)
\notag\\
&=
\frac{1}{x_{12}x_{23}x_{31}} \frac{1}{\pi^3}
\lim_{p\to0}
\Big[
\operatorname{Im}I(p,\bm{\lambda}_1,\bm{\lambda}_2)\,
\operatorname{Im}I(p,\bm{\lambda}_2,\bm{\lambda}_3)\,
\operatorname{Im}I(p,\bm{\lambda}_3,\bm{\lambda}_1)
\Big]
+(132),
\end{align}
where above we have defined $x_{ij}\equiv x_i-x_j$ and $\theta_{ij} \equiv \theta_i - \theta_j$.
The notation $(132)$ denotes the second Wick contraction, obtained by exchanging $\lambda_2\leftrightarrow\lambda_3$, which corresponds to the term
$S(1,3)S(3,2)S(2,1)$.
Using the saddle-point approximation \eqref{eq:RNMsaddle1} for each resolvent, we obtain
\begin{align}
&G^{(3)}(x_1,\bm{\lambda}_1,x_2,\bm{\lambda}_2,x_3,\bm{\lambda}_3) \notag \\
& =\frac{1}{ x_{12}x_{23}x_{31}}\frac{1}{\pi^\frac{9}{2}}   \sqrt{\frac{1}{\sqrt{(\bm{\lambda}_1 - \bm{\lambda}_2)^2(\bm{\lambda}_1 + \bm{\lambda}_2)^2} } } \sqrt{\frac{1}{\sqrt{(\bm{\lambda}_2 - \bm{\lambda}_3)^2(\bm{\lambda}_2 + \bm{\lambda}_3)^2} } }\sqrt{\frac{1}{\sqrt{(\bm{\lambda}_3 - \bm{\lambda}_1)^2(\bm{\lambda}_3 + \bm{\lambda}_1)^2} } } \notag \\
& \times \sin \Theta(\mu,\bm{\lambda}_1,\bm{\lambda}_2)\sin \Theta(\mu,\bm{\lambda}_2,\bm{\lambda}_3)\sin \Theta(\mu,\bm{\lambda}_3,\bm{\lambda}_1) + (132).
\end{align}
For the two-point functions, the non-oscillating contribution can be isolated using the identity in \eqref{eq:SepAv}.
For three-point functions, however, the phase factors $\Theta(\mu, \boldsymbol \lambda_i,\boldsymbol \lambda_j)$ do not cancel in general.
To determine when a phase factor cancels,  we note that the phase factor $\Theta(\mu,\bm{\lambda}_f, \bm{\lambda}_i)$ in the regime $\tau_i + \tau_f \gg1$ is nothing but Hamilton's characteristic function:
\begin{align}
\Theta(\mu,\bm{\lambda}_f,\bm{\lambda}_i)& \approx W (\bm{\lambda}_f, \bm{\lambda}_i; -\mu) + \frac{\pi}{4}, \notag \\
 W (\bm{\lambda}_f, \bm{\lambda}_i; -\mu) &\equiv \int_{\bm{\lambda}_i}^{\bm{\lambda}_f} \bm{p}\cdot \mathrm d\bm{\lambda} = \int_{\tau_i}^{\tau_f} \bm{p}\cdot \frac{\mathrm d\bm{\lambda}}{\mathrm d\tau}\mathrm  d\tau.
\end{align}
To see this we first express the phase factor $\Theta(\mu,\bm{\lambda}_f,\bm{\lambda}_i)$ in this limit using \eqref{eq:sqAA}:
\begin{align}
&\Theta(\mu,\bm{\lambda}_f,\bm{\lambda}_i) \notag \\
&\approx \sinh(\tau_f -\tau_i) (\mu \cosh (\tau_i + \tau_f) \cosh (2\varphi)- m \sinh(\tau_i + \tau_f)\sinh(2\varphi)) - \mu (\tau_f - \tau_i) + \frac{\pi}{4} .
\end{align}
On the other hand, Hamilton's characteristic function is calculated directly and becomes
\begin{align}
W  &= \int _{\tau_i}^{\tau_f} \bm{p} \cdot \frac{\mathrm d\bm{\lambda}}{\mathrm d\tau}\mathrm d\tau  \notag \\
&= -\mu (\tau_f - \tau_i) + \sinh(\tau_f - \tau_i) [ \mu \cosh(\tau_f + \tau_i ) \cosh(2 \varphi) - m \sinh(\tau_f + \tau_i) \sinh(2 \varphi) ].
\end{align}
(See Appendix~\ref{on-shell} for a summary of relevant classical mechanics background.) It is now straightforward to see the agreement directly. This relation is expected on general grounds from the semiclassical form of the single-particle propagator \cite{Gutzwiller:1990}.
The phase can cancel when the three points $\boldsymbol \lambda_1, \boldsymbol\lambda_2,\boldsymbol\lambda_3$ lie on a common classical trajectory.

To make this statement explicit, we parametrize the six coordinates of the three insertion points $\bm{\lambda}_1,\bm{\lambda}_2,\bm{\lambda}_3$ by $(\tau_1,\tau_2,\tau_3,m,\varphi,s)$ as
\begin{align}
\bm{\lambda}_1 &= \bm{\lambda} (\tau_1,m,\varphi) \notag \\
\bm{\lambda}_2 &= \bm{\lambda} (\tau_2,m,\varphi) \notag \\
\bm{\lambda}_3 &= \bm{\lambda} (\tau_3,m,\varphi) + s  \bm{n} (\tau_3,m,\varphi), \label{eq:sixparam3pt}
\end{align}
where
\begin{align}
&\bm{\lambda} (\tau,m,\varphi) =
\begin{pmatrix}
\sqrt{\mu - m} \cosh (\tau + \varphi) \\
\sqrt{\mu + m} \cosh (\tau - \varphi)
\end{pmatrix}, \notag \\
&\bm{n}(\tau,m,\varphi) =
\begin{pmatrix}
\lambda_2(\tau,m,\varphi) \\
-\lambda_1(\tau,m,\varphi)
\end{pmatrix} =
\begin{pmatrix}
\sqrt{\mu + m} \cosh (\tau - \varphi) \\
-\sqrt{\mu - m} \cosh (\tau + \varphi)
\end{pmatrix}.
\end{align}
Here $\bm\lambda(\tau,m,\varphi)$ gives the position at time $\tau$ on the classical trajectory labeled by the conserved quantities $m$ and $\varphi$, while $s$ parametrizes the deviation of the third point from this trajectory.
We then focus on the phase factor
\begin{align}
F_{123}(\bm{\lambda}_1,\bm{\lambda}_2,\bm{\lambda}_3) & = \Theta(\mu,\bm{\lambda}_1,\bm{\lambda}_2) + \Theta(\mu,\bm{\lambda}_2,\bm{\lambda}_3) + \Theta(\mu,\bm{\lambda}_3,\bm{\lambda}_1)   \notag \\
&\approx W (\bm{\lambda}_1, \bm{\lambda}_2; -\mu) + W (\bm{\lambda}_2, \bm{\lambda}_3; -\mu) - W (\bm{\lambda}_1, \bm{\lambda}_3; -\mu) + \frac{3 \pi}{4}.
\end{align}
We can also study the other phase factors like $\pm \Theta(\mu,\bm{\lambda}_1,\bm{\lambda}_2) \pm \Theta(\mu,\bm{\lambda}_2,\bm{\lambda}_3) \pm \Theta(\mu,\bm{\lambda}_3,\bm{\lambda}_1)$ appearing from the expansion of $\sin \Theta(\mu,\bm{\lambda}_1,\bm{\lambda}_2)\sin \Theta(\mu,\bm{\lambda}_2,\bm{\lambda}_3)\sin \Theta(\mu,\bm{\lambda}_3,\bm{\lambda}_1)$ in the same manner.
The first derivative with respect to the transverse displacement $s$ vanishes at $s=0$:
\begin{align}
\frac{\partial F_{123}}{\partial s}\bigg |_{s=0} &= \frac{\partial}{\partial \bm{\lambda}_3}  W (\bm{\lambda}_2, \bm{\lambda}_3; -\mu) \cdot \frac{\partial \bm{\lambda}_3}{\partial s } - \frac{\partial}{\partial \bm{\lambda}_3 } W (\bm{\lambda}_1, \bm{\lambda}_3; -\mu)\cdot \frac{\partial \bm{\lambda}_3}{\partial s }   \notag \\
& =-\bm{p}_3 \cdot \frac{\partial \bm{\lambda}_3}{\partial s }  +\bm{p}_3\cdot \frac{\partial \bm{\lambda}_3}{\partial s }  = 0,
\end{align}
where we used the fact that derivatives of Hamilton's characteristic function with respect to the endpoint coordinates give the corresponding initial and final momenta:
\begin{align}
\frac{\partial}{\partial \bm{\lambda}_f}  W (\bm{\lambda}_f, \bm{\lambda}_i; E) = \bm{p}_f(\bm{\lambda}_f,\bm{\lambda}_i;E) ,\quad \frac{\partial}{\partial \bm{\lambda}_i}  W (\bm{\lambda}_f, \bm{\lambda}_i; E) = -\bm{p}_i(\bm{\lambda}_f,\bm{\lambda}_i;E).
\end{align}
Thus, near $s=0$, the phase factor takes the Gaussian form
\begin{align}
e^{i F_{123} - \frac{3}{4}\pi i} \approx \exp \left[ \frac{i}{2} \frac{\partial^2 F_{123}}{\partial s^2}\bigg|_{s=0} s^2 \right],
\end{align}
and may be approximated by a delta function localized at $s=0$:
\begin{align}
\exp \left[ \frac{i}{2} \frac{\partial^2 F_{123}}{\partial s^2}\bigg|_{s=0} s^2 \right] = \sqrt{2\pi i \left(\frac{\partial^2 F_{123}}{\partial s^2}\bigg|_{s=0} \right)^{-1}} \delta(s). \label{eq:Caus3Fermi}
\end{align}
We therefore find that, in the semiclassical limit, the density three-point function has support only when the three insertion points lie on the same classical trajectory.
We will see in Section~\ref{3ptfn} that the bosonized description reproduces precisely the same delta-function support.

\subsection{Light-cone density operators}
\label{exactLCFermi}

In this section, we consider the correlation functions of the Euclidean time ``light-cone'' density operators
    \begin{align}
    \begin{split}
        \rho(x,\tau_\sigma, \varphi_\sigma) &= \sum_{\boldsymbol n,\boldsymbol m} a_{\boldsymbol n}^\dagger a_{\boldsymbol m} \psi_{\boldsymbol n}(x,\tau_\sigma, \varphi_\sigma)^* \psi_{\boldsymbol m}(x,\tau_\sigma, \varphi_\sigma).
    \end{split}
    \end{align}
The above bilinear operators correspond to perturbations of the Fermi surface in the basis
    \begin{align}
        \ket{\tau_\sigma, \varphi_\sigma}.
    \end{align}
The coordinates $\tau_\sigma, \varphi_\sigma$ are related to light-cone coordinates  \eqref{eq:LCcoord1}
by $z_{\sigma} = e^{\tau_\sigma + i \varphi_\sigma}/\sqrt{2}$, where $\sigma = \pm{}1$ distinguishes between ``left-moving'' and ``right-moving'' coordinates. The relation between the ``position space'' eigenkets is
\begin{align}
\ket{\boldsymbol z_\sigma} = e^{-2\tau_\sigma} \ket{\tau_\sigma, \varphi_\sigma}.
\end{align}
Note that the operators $\rho(x, \tau_\sigma, \varphi_\sigma)$ are not unitarily-equivalent to the operators $\rho(x,\boldsymbol \lambda)$ considered in the previous subsection, and thus represent distinct observables.

The operators $\rho(x,\tau_\sigma, \varphi_\sigma)$ are  appealing to study because working in action-angle coordinates makes the underlying integrability of the theory manifest---the  coordinate system $\tau_\sigma,h_\sigma, \varphi_\sigma, \ell$ (where $h_\sigma \equiv \sigma h$) can be regarded as a set of action-angle coordinates (see \eqref{eq:LCcoord1}), with $\varphi_\sigma$ and $\ell$ being constants of motion.
As a result, it turns out that the correlation functions of the operators $\rho(x,\tau_\sigma, \varphi_\sigma)$ simplify dramatically and only depend non-trivially on the chiral coordinates\footnote{We have analytically continued to Euclidean time $u_\sigma =  i x + \sigma \tau_\sigma$.} $u_\sigma = ix + \sigma \tau_\sigma$, i.e.,
    \begin{align}
\rho(x,\tau_\sigma, \varphi_\sigma) = \rho( u_\sigma, \varphi_\sigma).
    \end{align}

To see how this works, expand the operators $\rho(x,\tau_\sigma, \varphi_\sigma)$ in a basis of ``type 0A'' eigenfunctions
    \begin{align}
        \psi_q(E;x,\tau_\sigma,\varphi_\sigma) &= \frac{1}{2\pi} e^{ i ( E u_\sigma +  q  \varphi_\sigma)},~~~~ u_\sigma \equiv i x + \sigma \tau_\sigma,
    \end{align}
which diagonalize the single particle Hamiltonian and angular momentum expressed in polar light-cone coordinates, namely
    \begin{align}
       \hat  h =- i \sigma \partial_{\tau_\sigma},~~~~\hat \ell = -i  \partial_{\varphi_{\sigma}}.
    \end{align}
The density operators expressed in this basis are
    \begin{align}
    \label{LCFermirho}
      \rho(u_\sigma, \varphi_\sigma)  &= \frac{1}{(2\pi)^2} \int \mathrm dE \mathrm dE' \sum_{q,q'} e^{i [(E-E') u_\sigma +  (q-q') \varphi_\sigma]} a_{q'}^\dagger(E') a_q(E).
    \end{align}
The one-point function is given by the formal divergent expression
    \begin{align}
    \label{LCFermi1pt}
        \braket{\rho(u_\sigma, \varphi_\sigma)} &= \frac{1}{(2\pi)^2} \sum_{q} \int_{-\infty}^{-\mu} \mathrm dE.
    \end{align}
Furthermore, using the integral formulae (\ref{Eint}) in conjunction with the general expression (\ref{Sij}) for the Euclidean propagator, we find that in light-cone coordinates the propagator is given by
    \begin{align}
    \begin{split}
   S(u_{\sigma f}, u_{\sigma i})
    &=\frac{1}{2\pi i}  \delta(\Delta \varphi_{\sigma})  \frac{e^{ i \mu  \Delta u_{\sigma }}}{\Delta u_{\sigma} }.
    \end{split}
    \end{align}
Using the above two expressions, it is straightforward to calculate the time-ordered $n$-point functions of the densities $\rho(u_\sigma, \varphi_\sigma)$ using the rules described in (\ref{FermiG}), for arbitrary $n$.

The connected two-point function is given by
		\begin{align}
			\begin{split}
            \label{LC2ptFermi}
			G^{(2)}_c=	-S(1,2) S(2,1) &=- \frac{1}{(2\pi)^2} \delta(0) \delta(\Delta \varphi_{\sigma}) \frac{1}{\Delta u_\sigma^2}.
			\end{split}
		\end{align}
	As for the connected $n$-point functions for $n \geq 3$, they vanish identically:
        \begin{align}
        \label{n>3van}
            G^{(n\geq 3)}_c = 0.
        \end{align}
   To prove the above claim, notice that the connected $n$-point function can be expressed explicitly as a sum over maximal length ``ring'' diagrams (see Appendix \ref{Wick})
		\begin{align}
		\label{Rnonsing}
			G_c^{(n)} &\sim \left(\prod_i \delta(\varphi_{ i ,i+1}) \right) (-1)^{n-1} R(u_1,\dots, u_n )\\
		R(u ) &= \sum_{\sigma \in C_n(I)} \prod_{i\in I } \frac{1}{u_i-u_{\sigma(i)}},~~~~ I = \{ 1,\dots, n\},
	\end{align}
where above $C_n(I)$ denotes the conjugacy class of $n$-cycles in the symmetric group $\Sigma_n(I)$ acting on the set $I = \{ 1,\dots ,n\}$. (Note that here and throughout this proof, for simplicity, we omit the subscripts $\sigma$ on the coordinates $u_\sigma, \varphi_{\sigma}$ and instead use $\sigma$ to denote an element of the symmetric group $\Sigma_n$.) Thus, we have reduced the proof to understanding the properties of the rational function $R(u)$. By construction, $R(u)$ is symmetric under permutation of the variables $u_i$ by elements of the permutation group $\Sigma_n$. To see this, select a permutation $\pi \in \Sigma_n$ and relabel the indices as $i \mapsto \pi(i)$. Then we have
	\begin{align}
		\prod_{i \in I} \frac{1}{u_{\pi(i)} - u_{\pi(\sigma(i))} }= 	\prod_{j \in I } \frac{1}{u_j - u_{\pi \sigma \pi^{-1}(j)}}.
	\end{align}
  Conjugation $\sigma \mapsto \pi \sigma \pi^{-1}$ is a bijection on $C_n(I)$. Next, consider a fixed term associated to some choice of $\sigma \in C_n(I)$:
  	\begin{align}
		R_\sigma(u) = \prod_{i \in I} \frac{1}{u_i - u_{\sigma(i)}}.
	\end{align}
We then multiply the Vandermonde determinant $\Delta(u) = \prod_{1 \leq i < j \leq n} (u_i - u_j)$ by the above expression.
For $n \geq 3$, the resulting product
	\begin{align}
		P_\sigma(u) =\Delta(u) R_\sigma(u)
	\end{align}
is a polynomial, as each pole appears once in the Vandermonde. It follows that
	\begin{align}
		P(u)= \Delta(u) R(u)
	\end{align}
is a polynomial, and moreover that $P(u)$ is antisymmetric in permutations of $u_i$ (since $\Delta(u)$ is antisymmetric and $R(u)$ is symmetric). Now, we use the fact that every antisymmetric polynomial is divisible by the Vandermonde determinant, yielding a symmetric polynomial as a result, hence
	\begin{align}
		\frac{P(u)}{\Delta(u)} =  R(u)
	\end{align}
is a symmetric polynomial. But we already know that $R(u)$ has degree $-n$, and since a nontrivial polynomial cannot have a negative degree, the only way that the above equation can be formally consistent is if all coefficients are equal to zero, hence $R(u) =0$, which implies that $G_c^{(n)} =0$ for $n\geq 3$ as claimed. Thus, we have found that the connected correlation functions of the operators $\rho(u_\sigma, \varphi_\sigma)$ vanish identically in the Fermi liquid description.

Consider as an example the connected $3$-point function. Using the determinant formula (\ref{FermiG}), the connected three-point function is given by
    \begin{align}
    \begin{split}
        G_c^{(3)} &= S(1,2)S(2,3) S(3,1) + S(1,3) S(3,2) S(2,1) \\
        &=\frac{i}{(2\pi)^3} \delta(\varphi_{\sigma 12})\delta(\varphi_{\sigma 23})\delta(\varphi_{\sigma 31}) \left( \frac{1}{u_{\sigma 12}u_{\sigma 23}u_{\sigma 31}}+ \frac{1}{u_{\sigma 13}u_{\sigma 32}u_{\sigma 21}}  \right) \\
        &= 0.
    \end{split}
    \end{align}
The same is true for the correlation functions of the operators $\rho(x, \tau_\sigma) = \rho(u_\sigma)$ in the Fermi liquid description of the $c=1$ string, as is straightforward to verify using the methods we have described above.

\section{Bosonization of Fermi Surfaces and Coadjoint Orbits}

\label{COM}

Bosonization provides a natural framework for describing the low-energy dynamics of fermionic systems in terms of collective bosonic degrees of freedom.
For a system with a Fermi surface, the low-energy excitations are particle-hole excitations localized near the Fermi surface.
Equivalently, these excitations may be viewed as shape fluctuations of the Fermi surface itself.
This observation suggests that the infrared dynamics of a Fermi liquid admits an effective description in terms of bosonic fields associated with density fluctuations or Fermi-surface deformations.

For the purposes of this paper, the relevant bosonic degrees of freedom are collective deformations of a Fermi surface in phase space.
The natural algebra generating such deformations is the algebra of fermion bilinears. At the full quantum level this algebra is the $W_\infty$ algebra, while its semiclassical limit is the $w_\infty$ algebra of functions on phase space equipped with the Poisson bracket.
The coadjoint orbit method provides a systematic way to construct the effective action for these Fermi-surface fluctuations.
This method is particularly well suited to noncritical M-theory, whose exact formulation is a nonrelativistic Fermi liquid in phase space.

\subsection{$W_{\infty}$ algebra and $w_{\infty}$ algebra}
\label{Walg}
In Fermi liquid theory, the fundamental low-energy degrees of freedom are deformations of the Fermi surface.
At the full quantum-mechanical level, such deformations are generated by moving fermions in phase space: one removes a fermion from one point and creates it at another.
This operation is naturally described by fermion bilinear operators, whose algebra is the $W_\infty$ algebra.

Let us first introduce the bilocal fermion bilinear
\begin{align}
B(\bm{\lambda},\bm{\lambda}') = \Psi^{\dagger}(\bm{\lambda})\Psi(\bm{\lambda}').
\end{align}
Using the canonical anticommutation relations of the fermions, these bilinears satisfy
\begin{align}
[B(\bm{\lambda},\bm{\lambda}'),B(\bm{\mu},\bm{\mu}')] &= \Psi^{\dagger}(\bm{\lambda})\Psi(\bm{\mu}') \delta^d(\bm{\lambda}'-\bm{\mu})- \Psi^{\dagger}(\bm{\mu})\Psi(\bm{\lambda}')\delta^d(\bm{\lambda}-\bm{\mu}') \notag \\
&=B(\bm{\lambda},\bm{\mu}')\delta^d(\bm{\lambda}'-\bm{\mu})  - B(\bm{\mu},\bm{\lambda}')\delta^d(\bm{\lambda}-\bm{\mu}').
\end{align}
It is useful to rewrite this algebra in a phase-space basis. We define the Fourier transform of the Wigner-transformed bilinear by
\begin{align}
\widetilde{B^W}(\bm{q}, \bm{\mu}) = \int  \mathrm d^d \lambda B(\bm{\lambda} + \frac{\bm{\mu}}{2},\bm{\lambda} - \frac{\bm{\mu}}{2}) e^{i \bm{q}\cdot \bm{\lambda}}.
\end{align}
Here $\bm q$ is conjugate to the center-of-mass coordinate, while $\bm\mu$ is the relative coordinate.
These operators obey
\begin{align}
[\widetilde{B^W}( \bm{q},\bm{\mu}),\widetilde{B^W}( \bm{q}',\bm{\mu}')] = -2i \sin  \frac{1 }{2}(\bm{q}' \cdot \bm{\mu}-\bm{q} \cdot \bm{\mu}') \widetilde{B^W}( \bm{q}+\bm{q}',\bm{\mu}+\bm{\mu}') .
\end{align}
This is the $W_\infty$ algebra written in a Fourier-Wigner basis.
The semiclassical limit is obtained by expanding the sine function at small phase-space momentum,
\begin{align}\bm q\cdot\bm\mu'\ll1, \qquad \bm q'\cdot\bm\mu\ll1.
\end{align}
In this limit,
\begin{align}
[\widetilde{B^W}( \bm{q},\bm{\mu}),\widetilde{B^W}( \bm{q}',\bm{\mu}')] \approx  -i (\bm{q}' \cdot \bm{\mu}-\bm{q} \cdot \bm{\mu}')\widetilde{B^W}( \bm{q}+\bm{q}',\bm{\mu}+\bm{\mu}'). \label{eq:WFapprox}
\end{align}
The algebra and its approximation are clearly visible in the basis $\widetilde{B^W}$.
To see the relation to the Poisson bracket more directly, we introduce the Wigner-space bilinear
\begin{align}
B^W(\bm{\lambda}, \bm{p}) =  -\frac{i}{(2\pi)^d}\int  \mathrm d^d \mu \mathrm d^dq\,\widetilde{B^W}( \bm{q},\bm{\mu}) e^{-i \bm{q}\cdot \bm{\lambda}} e^{i \bm{p}\cdot \bm{\mu}}.
\end{align}
For any phase-space function $F(\lambda,p)$, define the corresponding operator
\begin{align}
 F = \int \mathrm d^d\lambda \mathrm d^d p\, F (\bm{\lambda},\bm{p})B^W(\bm{\lambda}, \bm{p}).
\end{align}
Then the algebra takes the form
\begin{align}
[F,G] = \int \mathrm d^d\lambda \mathrm d^d p\, \{ F,G\}_{\text{Moyal}}(\bm{\lambda}, \bm{p}) B^W(\bm{\lambda}, \bm{p}),
\end{align}
where the Moyal bracket is
\begin{align}
\{ F,G\}_{\text{Moyal}} (\bm{\lambda}, \bm{p}) = 2 F(\bm{\lambda},\bm{p}) \sin \frac{1}{2}\left( \overleftarrow{\partial}_{\bm{\lambda}} \overrightarrow{\partial} _{\bm{p}}- \overleftarrow{\partial}_{\bm{p}} \overrightarrow{\partial} _{\bm{\lambda}} \right) G(\bm{\lambda},\bm{p}).
\end{align}
The limit \eqref{eq:WFapprox} corresponds to expanding the derivatives in the Moyal bracket to first order.
This is nothing but the Poisson bracket
\begin{align}
\{ F,G\}_{\text{Moyal}} (\bm{\lambda}, \bm{p}) \approx \sum_{i=1}^d \bigg( \frac{\partial F}{\partial \lambda^i}\frac{\partial G}{\partial p_i} -\frac{\partial F}{\partial p_i} \frac{\partial G}{\partial \lambda^i} \bigg) = \{ F,G\}.
\end{align}
Thus, in the semiclassical limit, the algebra of fermion bilinears becomes
\begin{align}
[F,G] \approx \int \mathrm d^d\lambda \mathrm d^d p\, \{ F,G\}(\bm{\lambda}, \bm{p}) B^W(\bm{\lambda}, \bm{p}).
\end{align}
We will refer to this semiclassical algebra as $w_\infty$.
Equivalently, $w_\infty$ is the Lie algebra of functions on phase space equipped with the Poisson bracket.
This is the algebra that will be used in the coadjoint-orbit construction of the bosonized Fermi surface theory below.

\subsection{Coadjoint orbit method}
\label{sec:COMgeneral}

In this subsection, we review the coadjoint orbit construction that will be used to bosonize Fermi surfaces. The basic idea is that the low-energy modes of a Fermi liquid can be viewed as collective deformations of the distribution function in phase space. Since these deformations are generated by canonical transformations, they are naturally organized by the coadjoint orbits of the semiclassical $w_\infty$ algebra.

In the usual theory of Nambu--Goldstone modes, a spontaneous symmetry breaking pattern $\mathcal{G}\to \mathcal{H}$ leads to a nonlinear sigma model whose target space is the coset $\mathcal{G}/\mathcal{H}$ \cite{Coleman:1969sm,Callan:1969sn}.
In Fermi liquid phases the low energy degrees of freedom can be viewed as Nambu-Goldstone modes associated with the breaking of the $w_{\infty}$ algebra by the presence of Fermi surfaces in phase space \cite{2005cond.mat..5529H}.\footnote{The $W_{1+\infty}$ algebra, its breaking and associated Nambu-Goldstone modes also arise in the context of the topological string \cite{Aganagic:2003qj}.}

In the context of Nambu-Goldstone modes, we first think about the action of a continuous symmetry $\mathcal G$ on the space of the order parameter $\Delta$.
Then we consider the orbit of the order parameter by $\mathcal G$ to identify the configuration of the order parameter in the ground states.
A subgroup $\mathcal H$ stabilizes the order parameter $\Delta$.
The orbit of the order parameter is identified with the coset space $\mathcal G/\mathcal H$, where $\mathcal H$ is the stabilizer group.
We follow this approach to write the effective theory based on symmetry breaking for the Fermi liquid.
A natural framework is the coadjoint orbit of the $w_{\infty}$ algebra.
This is because the distribution function $f$ is viewed as an element of the dual of the algebra as we shall see below.

\subsubsection{Distribution functions as elements of  coadjoint orbits}
Let $\mathfrak g$ denote the semiclassical $w_\infty$ algebra, namely the algebra of functions on phase space equipped with the Poisson bracket.
In the case of the Fermi liquid, the order parameter is the distribution function $f(\bm{\lambda},\bm{p})$.
A distribution $f(\bm{\lambda},\bm{p})$ on a phase space is understood as an element of the dual space $\mathfrak{g}^*$ of the algebra:\footnote{In quantum mechanics, the counterpart of a distribution is the density matrix $\rho$.
The expectation value of an observable $A$ is of course
\begin{align}
\braket{A} = \text{Tr} (\rho A).
\end{align}
In algebraic language, density matrices $\rho$ are viewed as linear functionals from the algebra of observables $\mathcal{A}$ to the complex numbers:
\begin{align}
\omega_{\rho} : \mathcal{A} \to \mathbb{C}.
\end{align}
From this perspective, it is natural to view distributions as elements of the dual space $\mathfrak{g}^*$.
}
\begin{align}
f \in \mathfrak{g}^*, \qquad f: \text{distribution on a phase space}
\end{align}
An element $f$ of the dual space $\mathfrak{g}^*$ maps $\mathfrak{g}$ to $\mathbb{R}$
\begin{align}
f : F \in \mathfrak{g} \mapsto \braket{f,F} \in \mathbb{R}
\end{align}
Indeed, it defines a linear functional on $\mathfrak g$ through the pairing
\begin{align}
\braket{f,F}  = \int \frac{\mathrm d^d \bm{\lambda} \mathrm d^d \bm{p}}{(2\pi)^d} f(\bm{\lambda},\bm{p}) F(\bm{\lambda},\bm{p}). \label{eq:sclarprod}
\end{align}
The adjoint action is given by the Lie bracket itself.
In the case of the $w_{\infty}$ algebra the Lie bracket is given by the Poisson bracket.
Therefore, the adjoint action of $G \in \mathfrak{g}$ on $F \in \mathfrak{g}$ is
\begin{align}
\text{ad}_G F = \{ G, F\} = \sum_{i = 1}^d \left(\frac{\partial G}{\partial \lambda^i}\frac{ \partial F}{\partial p_i} - \frac{\partial F}{\partial \lambda^i}\frac{ \partial G}{\partial p_i} \right) \label{eq:adaction}
\end{align}
An element of the identity component of the Lie group, $U \in \mathcal{G}$, may be written as an exponential of the element of the algebra $G \in \mathfrak{g}$ as $U = \exp G $.
The adjoint action of a group element is then given by
\begin{align}
\text{Ad}_U F = e^{\text{ad}_G} F = F + \{ G,F\} + \frac{1}{2} \{ G, \{ G, F\}\}  + \cdots \label{eq:Adaction}
\end{align}

The coadjoint action is deduced from the adjoint action \eqref{eq:adaction} and the scalar product  \eqref{eq:sclarprod} between $\mathfrak{g}^*$ and $\mathfrak{g}$.
The coadjoint action is determined through
\begin{align}
\braket{\text{ad}^*_{-G} f, F} = \braket{f, \text{ad}_G F}
\end{align}
Using partial integration, we can easily see
\begin{align}
\text{ad}^*_G f = \{G, f \}.
\end{align}
Similarly, the action of an element of the Lie group $U = \exp G$ is again defined by
\begin{align}
\braket{\text{Ad}_{U^{-1}}^* f, F } = \braket{ f, \text{Ad}_U F }
\end{align}
Explicitly, the coadjoint action is
\begin{align}
\text{Ad}_U^* f =  f + \{ G,f\} + \frac{1}{2} \{ G, \{ G, f\}\}  + \cdots
\end{align}

\subsubsection{Coadjoint orbits and  parametrizations }
\label{COMparam}

We now specialize to the zero-temperature ground-state distribution
\begin{align}
f_0(\bm{\lambda},\bm{p}) = \theta(- \mu - h(\bm{\lambda},\bm{p})),
\end{align}
where all states below the Fermi energy $-\mu$ are filled.
The coadjoint orbit $\mathcal{O}_{f_0}$ through $f_0$ is
\begin{align}
\mathcal{O}_{f_0} = \{ f | \exists U \in \mathcal{G},  f = \text{Ad}_U^* f_0\}
\end{align}
This space is equivalent to $\mathcal{G}/\mathcal{G}_0$ where $\mathcal{G}_0$ is the stabilizer of $f_0$:
\begin{align}
\mathcal{G}_0 = \{ V |    \text{Ad}_V^* f_0 = f_0\}.
\end{align}
This is the space of canonical transformations that do not deform the distribution $f_0$.

To parameterize the coadjoint orbits, we first write the canonical transformation in terms of the algebra element $\phi$:
\begin{align}
U = \exp(-\phi),
\end{align}
where $\phi(\bm{\lambda},\bm{p})$ is a function on the phase space.
The deformed distribution function by $\phi$ is then
\begin{align}
f &= \exp (-\phi) f_0 \exp(\phi) \notag \\
&= f_0 - \{ \phi,f_0\} + \frac{1}{2!} \{ \phi , \{ \phi, f_0\} \} - \frac{1}{3!}\{ \phi, \{ \phi , \{ \phi, f_0\} \} \}  + \cdots
\end{align}
An element of the stabilizer group $V = \exp(\alpha) \in \mathcal{G}_0$ should satisfy
\begin{align}
 0 = \exp (\alpha) f_0 \exp(-\alpha) -f_0 = \{ \alpha,f_0\}  + \cdots  .
\end{align}
At leading order this gives
\begin{align}
\{ \alpha,f_0\} = -\left( \partial_{\boldsymbol \lambda} \alpha \cdot \partial_{\boldsymbol p} h - \partial_{\boldsymbol \lambda} h \cdot \partial_{\boldsymbol p} \alpha\right)
 \delta ( - \mu - h) = 0.
 \end{align}
Thus, on the Fermi surface, stabilizer transformations are generated by functions that are constant along the Hamiltonian flow. The physical bosonic field is therefore not the full phase-space function $\phi$ itself, but rather its equivalence class under right multiplication by stabilizer transformations:
 \begin{align}
 \exp(-\phi)\sim \exp(-\phi)\exp(\alpha)
 \end{align}
 This gauge redundancy will play an important role below when choosing convenient representatives for the bosonic field.

\subsubsection{Bosonic action for free fermions in an external potential}
We now describe the bosonic action obtained from the coadjoint orbit method for free fermions in an external potential.
Let $h(\lambda,p)$ be the single-particle Hamiltonian and let
$ f_0(\lambda,p)=\theta(-\mu-h(\lambda,p)) $
denote the ground-state distribution function.

The resulting effective action takes the universal coadjoint-orbit form
\begin{align}
\label{bosonaction}
S = S_{\rm WZW} + S_{H}
\end{align}
The first term is the Wess--Zumino--Witten term on the coadjoint orbit,
\begin{align}
S_{\rm WZW} = \int \mathrm dt \, \braket{f_0 , U^{-1}\partial_t U}.
\end{align}
This term determines the symplectic structure of the space of Fermi-surface fluctuations.

The second term is the Hamiltonian evaluated on the deformed distribution function:
\begin{align}
S_H = - \int \mathrm d t \,\braket{f,h} = - \int \mathrm d t\, \braket{f_0, U^{-1}h U}.
\end{align}
Here $U^{-1}hU$ denotes the adjoint action of the canonical transformation $U$ on the phase-space function $h$. Equivalently, $S_H$ is the energy of the deformed Fermi sea.

It is sometimes convenient to use the following representation of the weak field expansion of the bosonic action:
	\begin{align}
	\begin{split}
    \label{Skexp}
		S &= S^{(0)} + \sum_{k=2}^{\infty} S^{(k)}\\
		S^{(k)} &= \frac{1}{k!} \int \mathrm dt \, \langle \mathrm{ad}_\phi^{k-2} D_t \phi, \{f_0,  \phi \} \rangle,~~~~k \geq 2.
	\end{split}
	\end{align}
Above, $D_t \equiv- \partial_t + \{ h, \cdot \}$ and $\mathrm{ad}_\phi \equiv \{ \phi, \cdot\}$. In action-angle coordinates $\{ \tau, h\}=1$, the   derivative can be expressed as $D_t = -\partial_t -\partial_{\tau}$ where $\partial_{\tau}$ generates the Hamiltonian flow---see Appendix \ref{app:weakfield} for a derivation of the above expression.

\subsubsection{Density operators}
The coadjoint orbit construction also gives a natural bosonized representation of density operators. Recall that the bosonic field $\phi$ parametrizes a deformation of the ground-state distribution function,
\begin{align}
f_0 \to f_{\phi} &= \exp (-\phi) f_0 \exp(\phi) \notag \\
&= f_0 - \{ \phi,f_0\} + \frac{1}{2!} \{ \phi , \{ \phi, f_0\} \} - \frac{1}{3!}\{ \phi, \{ \phi , \{ \phi, f_0\} \} \}  + \cdots
\end{align}
Since $f_\phi$ is an element of the dual space $\mathfrak g^*$, any element of the algebra $\mathfrak g$ defines an observable by pairing it with $f_\phi$.
In particular, a density operator localized at a point on the Fermi surface can be represented as a phase-space function. For example, in a coordinate chart $(\bm{\lambda},\xi)$ on the Fermi surface, we write\footnote{We can also consider density operators defined in other coordinate charts.
See, e.g., the density operators defined in  light-cone coordinates discussed  in Section~\ref{sec:LCdensitycorr}.}
\begin{align}
\rho(\bar{\bm{\lambda}},\bar{\xi}) = \delta^{d}(\bm{\lambda} - \bar{\bm{\lambda}}) \delta^{d-1}(\xi - \bar{\xi}) \in \mathfrak{g}.
\end{align}
Here $\xi$ denotes the residual coordinates on the Fermi surface at fixed $\lambda$.
The corresponding bosonized density operator is obtained by evaluating this algebra element on the deformed distribution function $f_{\phi}\in \mathfrak{g}^*$ by the boson field $\phi$:
 \begin{align}
 \rho[\phi](\bar{\bm{\lambda}},\bar{\xi})  &\equiv \braket{f_{\phi}, \rho(\bar{\bm{\lambda}},\bar{\xi})} \notag \\
 &= \int \frac{\mathrm d^d\bm{\lambda} \mathrm d^d\bm{p}}{(2\pi)^d} f_{\phi}(\bm{\lambda},\bm{p})\delta^{d}(\bm{\lambda} - \bar{\bm{\lambda}}) \delta^{d-1}(\xi - \bar{\xi}). \label{eq:Dgeneral}
 \end{align}
 This formula is the bosonized counterpart of the fermion bilinear density operator.

     Similar in spirit to (\ref{Skexp}), it is sometimes convenient to express the density operators as
	\begin{align}
		\begin{split}
        \label{rhokexp}
			\rho[\phi](\bar{\bm{\lambda}},\bar{\xi}) &=\rho^{(0)}[\phi](\bar{\bm{\lambda}},\bar{\xi})+\sum_{k=1}^\infty \rho^{(k)}[\phi](\bar{\bm{\lambda}},\bar{\xi})\\
			\rho^{(k)}[\phi](\bar{\bm{\lambda}},\bar{\xi}) &=-\frac{1}{k!} \langle \text{ad}_\phi^{k-1}\rho(\bar{\bm{\lambda}},\bar{\xi}) , \{ \phi, f_0\} \rangle,~~~~ k \geq 1.
		\end{split}
	\end{align}

More generally, one may integrate over some of the residual Fermi-surface coordinates in order to define projected density operators.
For example, the eigenvalue density operator is obtained schematically by integrating over $\bar\xi$, \begin{align} \rho[\phi](\bar{\bm{\lambda}}) = \int \mathrm d^{d-1}\bar\xi\, \rho[\phi](\bar{\bm{\lambda}},\bar\xi). \end{align}
Different choices of phase-space density operators correspond to different physical observables.
This distinction will become apparent below when we study the correlation functions of both the eigenvalue density operators and light-cone density operators in noncritical M-theory.

\subsection{Application of coadjoint orbit method to $c=1$ string theory }
\label{c=1COM}
In this section, we demonstrate the utility of the coadjoint orbit method systematized in \cite{Delacretaz:2022ocm}, by using it to rederive the Das-Jevicki collective field theory \cite{Das:1990kaa,Jevicki:1979mb,Sakita:1979gs} for the $c=1$ string. Note that the coadjoint orbits of the $W_{\infty}$ algebra were used to derive, by very similar means, an effective description of the double scaling limit of the $c=1$ matrix model in \cite{Dhar:1992hr}.

 The collective field method was first introduced by Jevicki and Sakita in \cite{Jevicki:1979mb} as a means to rewrite quantum mechanical systems with large numbers of (matrix) degrees of freedom in terms of gauge-invariant collective variables, with a central role being played by the collective eigenvalue density\footnote{The eigenvalue density can be regarded as the Fourier transform of the ``Wilson loop'' operator, i.e., $\phi(t,\bar \lambda ) = \int (\mathrm d k/2\pi) \, e^{-i k \bar \lambda} \Tr \exp(i k M(t) )$---see \cite{Jevicki:1993qn}.} $
        \phi(t,\bar \lambda) = \Tr \delta(\bar \lambda I -M(t))$.
As discussed briefly at the beginning of Section \ref{EFTFermi}, in \cite{Das:1990kaa} Das and Jevicki used the collective field method to derive an effective description of the $c=1$ matrix model, by rewriting the theory in terms of the eigenvalue density operator, leading to the effective action \eqref{eq:DJaction}. It is straightforward to show that with a suitable choice of gauge-fixing condition, the effective action that follows from the coadjoint orbit method is equivalent to the effective action derived via the Das-Jevicki-Sakita collective field method, as we now demonstrate. Recall that the Gross-Klebanov Hamiltonian (which is equivalent to the Das-Jevicki EFT) takes the form
    \begin{align}
        H= \frac{1}{2} \int _0 ^{\frac{T}{2}} \mathrm d\tau \left[  \Pi_S^2 + (S')^2 - \frac{\sqrt{\pi}}{N \bar v_{ F}^2} \left( \Pi_S S' \Pi_S + \frac{1}{3} (S')^3 \right) - \frac{1}{2N \bar v_{ F}^2 \sqrt{\pi} }S' \Sch ( \bar \lambda , \tau) \right].
    \end{align}
Expressed in terms of the dual field $\partial \tilde{S} / \partial \tau = \Pi_S$, the action associated to the above Hamiltonian can be written as
\begin{align}
\label{eq:Slag}
S &= \int \mathrm{d}t \int_0^{T/2} \mathrm{d}\tau \, \Biggl[
\frac{\partial \tilde{S}}{\partial \tau} \frac{\partial S}{\partial t}
- \frac{1}{2}\left(\frac{\partial \tilde{S}}{\partial \tau}\right)^2
- \frac{1}{2}\left(\frac{\partial S}{\partial \tau}\right)^2
+ \frac{\sqrt{\pi}}{2N \bar v_F^2}\left(
\frac{\partial S}{\partial \tau} \left(\frac{\partial \tilde S}{\partial \tau}\right)^2
+ \frac{1}{3}\left(\frac{\partial S}{\partial \tau}\right)^3
\right) \notag \\[4pt]
&\qquad + \frac{1}{4N \bar v_F^2 \sqrt{\pi}} \frac{\partial S}{\partial \tau} \, \mathrm{Sch}(\bar\lambda, \tau)
\Biggr].
\end{align}
Next, introduce a pair of chiral bosons, $\tilde{S} = \phi_L + \phi_R$, $S = -\phi_L + \phi_R$, so that up to total derivative terms in $t,\tau$, the above action becomes
\begin{align}
\label{eq:unfoldS}
S &= \int \mathrm{d}t \int_0^{T/2} \mathrm{d}\tau \, \Biggl[
- \frac{\partial \phi_L}{\partial \tau} \frac{\partial \phi_L}{\partial t}
+ \frac{\partial \phi_R}{\partial \tau} \frac{\partial \phi_R}{\partial t}
- \left(\frac{\partial \phi_L}{\partial \tau}\right)^2
- \left(\frac{\partial \phi_R}{\partial \tau}\right)^2 \notag \\[4pt]
&\qquad - \frac{2\sqrt{\pi}}{3N \bar v_F^2}\left(
\left(\frac{\partial \phi_L}{\partial \tau}\right)^3
- \left(\frac{\partial \phi_R}{\partial \tau}\right)^3
\right)
- \frac{1}{4N \bar v_F^2 \sqrt{\pi}}\left(\frac{\partial \phi_L}{\partial \tau} - \frac{\partial \phi_R}{\partial \tau}\right) \mathrm{Sch}(\bar\lambda, \tau)
\Biggr].
\end{align}
One can then unfold the theory by introducing the following chiral field:
\begin{equation}
\frac{1}{2\sqrt{\pi}} \phi(t,\tau) =
\begin{cases}
\phi_L(t,\tau) \qquad  &\tau > 0 \\
\phi_R(t,-\tau) \qquad  &\tau < 0.
\end{cases}
\end{equation}
In terms of the new field $\phi$, the action is now given by\footnote{At large Fermi velocity $\bar v_F \to \infty$, the above action reduces to the Floreanini-Jackiw action for a two-dimensional chiral boson.}
\begin{equation}
S = \frac{1}{4\pi} \int \mathrm d t\int _{-\frac{T}{2}}^{\frac{T}{2}} \mathrm d\tau  \biggl[ - \frac{\partial \phi}{\partial \tau} \frac{\partial \phi}{\partial t} - \left(\frac{\partial  \phi}{\partial \tau}\right)^2 - \frac{1}{3 N \bar v_{ F}^2} \left(\frac{\partial \phi}{\partial \tau}\right)^3 -  \frac{1}{2N \bar v_{ F}^2 }\frac{\partial\phi}{\partial \tau} \,\Sch ( \bar \lambda , \tau) \biggr]. \label{eq:GKboson}
\end{equation}
We will show below that up to coordinate redefinitions (and ignoring a tadpole term, which is the result of the normal-ordering prescription used in \cite{Gross:1990st}), the above action matches the action that follows from using the coadjoint orbit method to bosonize the Fermi surface.

 Our starting point for applying the coadjoint orbit method to the bosonization of the $c=1$ string Fermi liquid is to consider the classical Fermi sea, for which the single-particle dispersion is given by
\begin{align}
h(\lambda,p)  = \frac{1}{2} p^2 -  \frac{1}{2}\lambda^2.
\end{align}
The Fermi surface in the phase space is given by the surface where the single particle Hamiltonian takes a constant value
\begin{align}
h(\lambda,p) = - \mu.
\end{align}
The distribution function at zero temperature is
\begin{equation}
f_0(\lambda,p) = \theta(-\mu - h(\lambda,p) )
\end{equation}
where $\theta(x)$ is the Heaviside step function.
It is convenient to describe this system in terms of the action-angle variables $\tau,h$ defined by
\begin{align}
\lambda = \s{-2h} \cosh \tau, \quad p= \s{-2h} \sinh \tau. \label{eq:AAc1}
\end{align}
In  this coordinate system the Fermi sea droplet is very simple to describe, namely
\begin{align}
f_0 = \theta(-\mu - h).
\end{align}
Since the coordinates $\tau, h$ are canonical coordinates, the Poisson bracket takes exactly the same form in action-angle coordinates.
The action of a canonical transformation on the distribution function $f_0$ can therefore be expressed as follows in terms of Poisson brackets,
\begin{align}
\begin{split}
f &= f_0  - \{\phi, f_0 \} + \f{1}{2} \{\phi,\{ \phi, f_0\}\} + \cdots \\
&=  \theta (-\mu -h) + \delta( h + \mu) \partial_{\tau} \phi + \cdots
\end{split}
\end{align}
where $\phi(t,\tau,h)$ is a function in the phase space, which will play the role of the collective field on the Fermi surface.
The stabilizer of the Fermi surface is
\begin{equation}
\text{ad}_{\alpha}^* f_0 = \{\alpha, f_0\}=0 ~~\implies ~~ \left. \frac{\partial\alpha}{\partial\tau}(t,\tau,h)\right|_{h = -\mu} = 0.
\end{equation}
Then, on a coadjoint orbit, we have an identification of fields $\phi$, namely
\begin{equation}
\phi \sim \phi - \alpha + \cdots.
\end{equation}
This is a gauge redundancy, and before proceeding we first need to choose a gauge-fixing condition for the bosonic field $\phi$.
\paragraph{$h$-gauge}
One convenient gauge-fixing condition is the condition
\begin{equation}
\alpha(t,\tau,h) = \phi(t,\tau, h) - \phi(t,\tau, h = -\mu) \sim 0.
\end{equation}
At quadratic order, the equivalence class of fields is labeled by the convenient representative
\begin{equation}
 \phi(t,\tau, h= -\mu) \equiv \phi(t, \tau).
\end{equation}
\paragraph{$p$-gauge}
Another gauge-fixing condition is
\begin{align}
\alpha (t, \lambda, p ) = \phi(t,\lambda,p) -\phi(t,\lambda,\sgn(p) p_F(\lambda)) \sim 0,
\end{align}
where $p_F(\lambda)$ is the Fermi momentum as a function of $\lambda$ obtained by solving the condition $h(\lambda, p) = - \mu$ as
\begin{align}
p_F(\lambda) = \sqrt{\lambda^2 - 2\mu}.
\end{align}
At quadratic order, this leads to the representative
\begin{equation}
\phi(t,\lambda,\sgn(p) p_F(\lambda)) \equiv \phi(t, \lambda).
\end{equation}

\subsubsection{Quadratic action}
In this section, we expand the action \eqref{bosonaction} up to quadratic order in $\phi$.
At this order in the field expansion, the choice of gauge does not matter as we obtain the same action in any gauge.
\paragraph{WZW term}
The WZW term is given by
\begin{align}
S_{\rm WZW} &=  \int \mathrm dt \braket{f_0, U^{-1} \partial_t U}  \notag \\
&= \int \mathrm dt \braket{f_0,  - \dot{\phi} + \frac{1}{2} \{ \dot{\phi} , \phi \}+ \cdots }
\end{align}
The first term is
\begin{align}
\int \mathrm dt \braket{f_0,  - \dot{\phi} }
&= - \int \mathrm dt \int \frac{ \mathrm d \lambda  \mathrm dp}{2\pi} \theta(-\mu - h(\lambda,p)) \frac{\partial \phi}{\partial t} \notag \\
&= - \int \mathrm dt \frac{\partial}{\partial t} \int \frac{ \mathrm d \lambda  \mathrm dp}{2\pi} \theta(-\mu - h(\lambda,p))  \phi(t,\lambda, p) ,
\end{align}
and gives a total derivative term.
The next leading term is
\begin{align}
\frac{1}{2}\int \mathrm dt \braket{f_0,   \{ \dot{\phi} , \phi \} }  &=\frac{1}{2}\int \mathrm dt   \int \frac{ \mathrm d \lambda  \mathrm dp}{2\pi}\theta(-\mu - h(\lambda,p))  \left( \frac{\partial \dot{\phi}}{\partial \lambda} \frac{\partial \phi}{\partial p}-\frac{\partial \dot{\phi}}{\partial p} \frac{\partial \phi}{\partial \lambda}\right) \notag \\
&= \frac{1}{2}\int \mathrm dt   \int \frac{ \mathrm d \tau  \mathrm d h}{2\pi} \theta(-\mu - h) \left( \frac{\partial \dot{\phi}}{\partial \tau} \frac{\partial \phi}{\partial h}-\frac{\partial \dot{\phi}}{\partial h} \frac{\partial \phi}{\partial \tau}\right) \notag \\
&= -\frac{1}{2}\int \mathrm dt   \int \frac{ \mathrm d \tau  \mathrm d h}{2\pi} \delta(-\mu - h)  \dot{\phi} \frac{\partial \phi}{\partial \tau} \notag \\
&= -\frac{1}{4\pi} \int \mathrm dt \mathrm d\tau \, \frac{\partial \phi}{\partial t} \frac{\partial \phi}{\partial \tau}.
\end{align}
Here we do not use  the gauge fixing condition.
Therefore in any gauge at the quadratic order we obtain this expression.

\paragraph{Hamiltonian term}
The Hamiltonian term is expanded in the boson field $\phi$ as
\begin{equation}
S_H = -\int \mathrm dt\, \braket{f_0 , U^{-1} h U} = - \int \mathrm dt \, \braket{f_0 , h + \{\phi, h \} +  \frac{1}{2} \{ \phi, \{\phi, h \}\} + \cdots }
\end{equation}
The $O(\phi^0)$ term $-\int \mathrm d t \braket{f_0, h} $ gives (minus) the ground state energy, which is constant and can be ignored.
The $O(\phi^1)$ term $- \int \mathrm dt \braket{f_0, \{\phi,h \}}$ is a total derivative in phase space since
\begin{equation}
\{ \phi ,h \} = \f{\partial \phi}{\partial \tau}.
\end{equation}
Therefore the leading term is the quadratic term given by 
\begin{equation}
S_ H = -\frac{1}{2} \int \mathrm dt\,   \braket{ f_0, \{ \phi, \{\phi, h \}\}  } + \cdots  = \frac{1}{2} \int \mathrm dt \,   \braket{ \{  \phi, f_0\},  \{\phi, h \}  }+ \cdots
\end{equation}
The Poisson bracket $\{  \phi, f_0\}$ becomes
\begin{equation}
\{  \phi, f_0\} = \frac{\partial \phi}{\partial \tau}\frac{\partial f_0}{\partial h}- \frac{\partial \phi}{\partial h}\frac{\partial f_0}{\partial \tau} = -\delta(\mu +h)  \frac{\partial \phi}{\partial \tau}.
\end{equation}
It follows that the quadratic Hamiltonian term is given by
\begin{align}
 \frac{1}{2} \int \mathrm dt  \, \braket{ \{  \phi, f_0\},  \{\phi, h \}  } &= -\frac{1}{2}\int \mathrm dt \frac{\mathrm d\tau \mathrm d h}{2\pi}  \delta(\mu +h) \frac{\partial \phi}{\partial  \tau}\frac{\partial \phi}{\partial  \tau}  \notag \\
 &= - \frac{1}{4\pi}\int \mathrm dt \mathrm d\tau \,\left(\frac{\partial \phi}{\partial  \tau} \right)^2.
\end{align}

\paragraph{Total quadratic action}
Summarizing, the quadratic action takes the form
\begin{align}
S_{\text{WZW}} + S_H = -\frac{1}{4\pi} \int \mathrm dt \mathrm d\tau  \left[\frac{\partial \phi}{\partial  t} \frac{\partial \phi}{\partial  \tau} + \left( \frac{\partial \phi}{\partial  \tau} \right)^2 \right]. \label{eq:c1quad}
\end{align}
This reproduces the quadratic term in the bosonized action for the $c=1$ matrix model \cite{Das:1990kaa,Gross:1990st}.
The quadratic action is the same as that for a chiral boson propagating on a two-dimensional space with coordinates
$t,\tau$ \cite{Floreanini:1987as}.\footnote{We obtain a chiral boson at the edge of the integer quantum Hall effect after bosonization \cite{Wen:2007joe}.
Indeed the Fermi sea can be viewed as an integer quantum Hall effect on a phase space and the Fermi liquid theory is interpreted as a chiral edge mode of the phase space integer quantum Hall effect \cite{Lu:2023emm}. \label{qHe}}

\subsubsection{Full action}
In one-dimensional Fermi surface bosonization, there are no cubic or higher order contributions to the WZW term \cite{Delacretaz:2022ocm}.
For example, the cubic contribution to the WZW term is
\begin{align}
S^{(3)}_{\rm WZW} &= -\frac{1}{3!} \int \mathrm dt \,\braket{f_0,\{ \{\dot{\phi},\phi\}, \phi\}}= -\frac{1}{3!} \int \mathrm dt \,\braket{\{\phi, f_0\}, \{\dot{\phi},\phi\}}
\end{align}
However, using the  $p$-gauge condition $\partial \phi/\partial p =0$, we find that the Poisson bracket $\{ \dot{\phi}, \phi\}$ vanishes and therefore $S^{(3)}_{\text{WZW}}=0$. A straightforward application of this argument also implies that $S_{\text{WZW}}^{(k>3)}=0$. Therefore, we restrict our attention to the cubic and higher order contributions to the Hamiltonian term in $p$-gauge below.

\paragraph{Hamiltonian term}
The Hamiltonian term is expanded as
\begin{align}
S_H &= - \int \mathrm dt\, \braket{f_0, U^{-1} h U} = - \int \mathrm dt\, \braket{f_0 , h + \frac{1}{2} \{ \phi , \{ \phi, h\}\} +\frac{1}{3!} \{ \phi,\{ \phi , \{ \phi, h\}\} \} + \cdots} .
\end{align}
The cubic term is given by
\begin{align}
S_H ^{(3)} = \frac{1}{3!} \int  \mathrm dt \braket{\{ \phi, f_0\},\{ \phi , \{ \phi, h\}\}}.
\end{align}
Integrating by parts, the above term can be expressed as
\begin{align}
 \braket{\{ \phi, f_0\},\{ \phi , \{ \phi, h \}\}} &= -\int\frac{ \mathrm d\tau  \mathrm dh}{2\pi} \delta (h+ \mu) \frac{\partial \phi}{\partial \tau}
\left[ \frac{\partial \phi}{\partial \tau}\frac{\partial^2 \phi}{\partial \tau \partial h}  - \frac{\partial \phi}{\partial h} \frac{\partial^2\phi }{\partial \tau^2}\right] \notag \\
&=  \frac{1}{2}\int\frac{ \mathrm d\tau  \mathrm dh }{2\pi} \delta '(h+ \mu)
 \left(\frac{\partial \phi}{\partial \tau} \right)^3. \label{eqref:CubicGIndep}
\end{align}
\paragraph{$p$-gauge}
Using the relation
\begin{align}
\frac{\partial \phi}{\partial \tau}  = \{\phi, h\} =  \frac{\partial h}{\partial p} \frac{\partial \phi}{\partial \lambda} -  \frac{\partial h}{\partial \lambda} \frac{\partial \phi}{\partial p},
\end{align}
and the $p$-gauge condition $\partial \phi/\partial p = 0$, the cubic term is
\begin{align}
\frac{1}{2}\int\frac{\mathrm d\tau \mathrm dh }{2\pi} \delta '(h+ \mu)
 \left(\frac{\partial \phi}{\partial \tau} \right)^3 \notag
&=\frac{1}{2}\int\frac{\mathrm d\lambda \mathrm dp }{2\pi}  \frac{1}{\frac{\partial h}{\partial p}} \frac{\partial }{\partial p} \left( \frac{1}{\left|\frac{\partial h}{\partial p}\right|}  \left[\delta(p - p_F(\lambda)) + \delta(p + p_F(\lambda)) \right] \right)
 \left(p\frac{\partial \phi }{\partial \lambda} (\lambda) \right)^3 \notag \\
&=- \int _{\sqrt{2\mu}}^{\infty} \frac{\mathrm d\lambda}{2\pi} \left[ \left(\frac{\partial \phi_+}{\partial \lambda} (\lambda)\right)^3 - \left(\frac{\partial \phi_-}{\partial \lambda} (\lambda)\right)^3 \right].
\end{align}
Here $\phi_+(\lambda)$ and $\phi_-(\lambda)$ denote the restrictions of the bosonic field to the two branches of the Fermi surface, \[ p=+p_F(\lambda), \qquad p=-p_F(\lambda), \] respectively (compare this to the discussion around \eqref{eq:unfoldS}).
Equivalently, these two branches may be combined into a single unfolded coordinate $\tau\in\mathbb R$ on the Fermi surface. Expressed in terms of the parametrization
\begin{align} \lambda(\tau)=\sqrt{2\mu}\cosh\tau, \qquad p_F(\tau)=\sqrt{2\mu}\sinh\tau,
\end{align}
the branch $p=+p_F$ corresponds to $\tau>0$, while the branch $p=-p_F$ corresponds to $\tau<0$.
As functions of the coordinate $\tau$, $\phi_+(\lambda)$ and $\phi_-(\lambda)$ are represented by the positive- and negative-$\tau$ parts of a single chiral boson field $\phi(\tau)$. Thus we obtain
\begin{align} S_H^{(3)} = -\frac{1}{3!} \int \mathrm dt \frac{ \mathrm d\tau}{2\pi}\, \frac{1}{p_F(\tau)^2} \left(\frac{\partial \phi}{\partial \tau}\right)^3, \end{align}
where $p_F(\tau) = \sqrt{2\mu}\sinh \tau$.
This reproduces the cubic interaction terms for the $c=1$ theory \cite{Das:1990kaa,Gross:1990st,Sengupta:1990bt}.

We next show that all higher order terms $S^{(k>3)}_H$ vanish in $p$-gauge. A $k$th order contribution to the Hamiltonian is proportional to the following integral:
    \begin{align}
        \int \mathrm d vol_{PS} \, \left(\text{ad}_\phi^{k-2} \frac{\partial \phi}{\partial \tau} \right) \frac{\partial \phi}{\partial \tau} \delta(h+\mu).
    \end{align}
Using the fact that 
    \begin{align}
        \frac{\partial}{\partial \tau} &= p \frac{\partial}{\partial \lambda} + \lambda \frac{\partial}{\partial p},~~~~ \frac{\partial}{\partial h } = \frac{\lambda}{2h} \frac{\partial }{\partial \lambda}+ \frac{p}{2h} \frac{\partial}{\partial p},
    \end{align}
along with the $p$-gauge condition, it is straightforward to show that  
    \begin{align}
        \text{ad}_\phi^{k-2} \frac{\partial \phi}{\partial \tau} &= \text{ad}_\phi^{k-3} \left(\frac{\partial \phi}{\partial \lambda}\right)^2.
    \end{align}
The $p$-gauge condition implies that any derivatives of the term $(\partial \phi / \partial \lambda)^2$ with respect to $p$ must vanish on the Fermi surface. In particular, this means that
    \begin{align}
    \begin{split}
        \text{ad}_\phi \left(\frac{\partial \phi}{\partial \lambda}\right)^2 &= 0,
    \end{split}
    \end{align} 
hence $S^{(k>3)}_H=0$. Therefore, the total $p$-gauge action is cubic-exact:
\begin{align}
S_{\text{$p$-gauge}} = S_{\text{WZW}} + S_H
&=  -\frac{1}{4\pi} \int \mathrm dt \mathrm d\tau  \left[\frac{\partial \phi}{\partial  t} \frac{\partial \phi}{\partial  \tau} + \left( \frac{\partial \phi}{\partial  \tau} \right)^2  + \frac{2}{3!} \frac{1}{p_F(\tau)^2} \left(\frac{\partial \phi}{\partial \tau} \right)^3 \right].
\label{eq:c1qc}
\end{align}
Comparing the above action with \eqref{eq:GKboson}, we find that up to trivial redefinitions of the coordinates (and ignoring the Schwarzian term), the coadjoint orbit method correctly reproduces the double-scaling limit of the Das-Jevicki collective field theory.

\paragraph{$h$-gauge}
We next consider the $c=1$ bosonized action in $h$-gauge. Using the gauge-independent expression \eqref{eqref:CubicGIndep}, we have
\begin{align}
\begin{split}
\braket{\{ \phi,f_0\},\{ \phi , \{ \phi, h \}\}} &=  \frac{1}{2}\int\frac{ \mathrm d\tau \mathrm dh }{2\pi} \delta '(h+ \mu)
 \left(\frac{\partial \phi}{\partial \tau} \right)^3  \\
 &= - \frac{1}{2}\int\frac{ \mathrm d\tau  \mathrm dh }{2\pi} \delta (h+ \mu)
 \frac{\partial}{\partial h}\left(\frac{\partial \phi}{\partial \tau} \right)^3\\
 &= 0.
\end{split}
\end{align}
In the final line, we used the $h$-gauge condition $\partial \phi/\partial h = 0$, and thus we find that the  cubic interaction term vanishes. Therefore up to cubic order, the total action is that of a free chiral boson. In fact, it is not difficult to show that in $h$-gauge, all interaction terms $S^{(n>2)}$ vanish (see \eqref{eq:Shgauge}), hence
\begin{align}
\label{eq:c=1hgaugeaction}
S_{\text{$h$-gauge}} = S_{\text{WZW}} + S_H = -\frac{1}{4\pi} \int \mathrm dt \mathrm d\tau  \left[\frac{\partial \phi}{\partial  t} \frac{\partial \phi}{\partial  \tau} + \left( \frac{\partial \phi}{\partial  \tau} \right)^2 \right].
\end{align}
As we will see below in Section \ref{sec:quadcorr}, despite the fact that the effective description of the $c=1$ string appears free in $h$-gauge, the effective interactions of the eigenvalue density operator appear to be repackaged into the higher order corrections to the eigenvalue density operator. It would be interesting to compute the eigenvalue density correlation functions in $h$-gauge and carefully compare them to the eigenvalue density correlators in $p$-gauge order-by-order in an appropriate perturbative expansion; we leave this exercise for future work.

\subsubsection{Linearized density operator}
Here we apply the general expression for the density operator \eqref{eq:Dgeneral},  \eqref{rhokexp} to the special case of the bosonized $c=1$ matrix model.
First, we examine the linear term.  If we consider the density 
\begin{equation}
\rho(\bar{\lambda}) = \delta(\lambda - \bar{\lambda}) \in \mathfrak{g},
\end{equation}
then to first order in $\phi$ we obtain the following contribution to the density operator
\begin{align}
\begin{split}
\rho[\phi] (\bar{\lambda}) & \equiv \braket{f_\phi, \rho(\bar{\lambda})}  \\
&= \int \frac{ \mathrm d \lambda  \mathrm dp}{2\pi} f_{\phi}(\lambda,p) \delta (\lambda-\bar{\lambda})  \\
&= \int \frac{ \mathrm d \lambda  \mathrm dp}{2\pi} (\text{Ad}_{e^{-\phi}}^*\theta)\left( - \mu -  \left(\frac{1}{2}p^2 - \frac{1}{2} \lambda^2 \right) \right) \delta (\lambda-\bar{\lambda})  \\
   &=\int \frac{ \mathrm d \lambda  \mathrm dp}{2\pi} \left[ \theta (p - p_F^-(\lambda))\theta ( p_F^+(\lambda) - p) - \left(\frac{ \partial}{ \partial h }
  \theta (-\mu - h)\right) \frac{\partial \phi}{\partial \tau} + \cdots \right]\delta (\lambda-\bar{\lambda} )   \\
   &=\int \frac{ \mathrm d \lambda  \mathrm dp}{2\pi} \left[ \theta (p - p_F^-(\lambda))\theta ( p_F^+(\lambda) - p) +
  \delta (\mu + h)\frac{\partial \phi}{\partial \tau} + \cdots \right]\delta (\lambda-\bar{\lambda} )   \\
    &=\int \frac{ \mathrm d \lambda  \mathrm dp}{2\pi} \left[ \theta (p - p_F^-(\lambda))\theta ( p_F^+(\lambda) - p) \right]\delta (\lambda-\bar{\lambda} )   \\
     & \quad \quad + \int \frac{ \mathrm d \tau  \mathrm dh}{2\pi}\left[
  \delta (\mu + h)\partial_{\tau} \phi  + \cdots \right] \sum_{s = \pm} \left|\frac{ \mathrm d\tau}{ \mathrm d\lambda}\right|\delta (\tau-\tau(\bar{\lambda}, p_{F}^s(\bar{\lambda})) )  \\
 &= \frac{1}{2\pi} (p_F^+(\bar{\lambda})  - p_F^-(\bar{\lambda}) ) + \frac{1}{2\pi}\,
     \left|\frac{\mathrm{d}\tau}{\mathrm{d}\lambda}\right|
     \left(
       \frac{\mathrm{d}\phi}{\mathrm{d}\tau}\Bigl(\;\tau(\bar{\lambda},p_F(\bar{\lambda}))\Bigr)
       + \frac{\mathrm{d}\phi}{\mathrm{d}\tau}\Bigl(-\tau(\bar{\lambda},p_F(\bar{\lambda}))\Bigr)
     \right) + \cdots.
 \end{split}
\end{align}
Above,  $p_F^{\pm} =  \pm \sqrt{\lambda^2 - 2\mu}$ denotes the upper or the lower branch of the Fermi surface corresponding to one of the two solutions of the polynomial equation $p^2 - \lambda^2 = -2 \mu$, and
$\tau(\lambda,p)= \text{arctanh}(p/\lambda)$ as follows from the definition \eqref{eq:AAc1}.
In the language of symplectic geometry, $\delta (\mu+ h)$ can be regarded as a delta function supported on the Fermi surface, whereas $\delta(\lambda-\bar{\lambda})$ is a delta function supported on a Lagrangian submanifold in phase space.
Summarizing the above calculation, the expansion \eqref{rhokexp} leads to the following contributions to the density operator
\begin{align}
\rho^{(0)}[\phi](\bar{\lambda}) &= \frac{1}{2\pi} (p_F^+(\bar{\lambda})  - p_F^-(\bar{\lambda}) ) = \frac{1}{\pi} \sqrt{\bar\lambda^2 - 2\mu}, \notag \\
\rho^{(1)}[\phi](\bar{\lambda})
  &= \frac{1}{2\pi}\,
     \frac{\mathrm{d}\tau}{\mathrm{d}\lambda}
     \left[
       \frac{\mathrm{d}\phi}{\mathrm{d}\tau}\Bigl(\;\tau(\bar{\lambda},p_F(\bar{\lambda}))\Bigr)
       + \frac{\mathrm{d}\phi}{\mathrm{d}\tau}\Bigl(-\tau(\bar{\lambda},p_F(\bar{\lambda}))\Bigr)
     \right].
\end{align}
So far, we have not made use of the gauge condition. We next consider quadratic and higher order corrections to the eigenvalue density operator, which will differ depending on the choice of gauge. 

\subsubsection{Quadratic correction to the density operator }
\label{sec:quadcorr}
\paragraph{$p$-gauge}

The $O(\phi^2)$ contribution to the eigenvalue density in $p$-gauge vanishes, as we now show. Using $\rho = \delta(\lambda - \bar \lambda)$, we see that
    \begin{align}
    \begin{split}
      \rho^{(2)}[\phi](\bar \lambda)
      &=\frac{1}{2} \int \frac{\mathrm d \lambda \mathrm d p}{2\pi}\, \{ \phi, \{\phi,f_0\} \} \rho\\
         &=\frac{1}{2}\int \frac{\mathrm d \lambda \mathrm d p}{2\pi}\, \left[\frac{\partial^2 \phi}{\partial p \partial\lambda} \rho - \frac{\partial}{\partial \lambda}\left(\frac{\partial \phi}{\partial p} \rho\right ) \right] \frac{\partial \phi}{\partial \tau} \delta(h+\mu)\\
         &=0.
        \end{split}
    \end{align}
In going to the last line, we have used the $p$-gauge condition that  $\partial \phi / \partial p =0$. By similar calculations, it is straightforward to show that all higher order contributions to $\rho[\phi](\bar \lambda)$ also vanish in $p$-gauge:
    \begin{align}
        \rho^{(n)}[\phi](\bar \lambda) =0,~~~~ n>2.
    \end{align}

\paragraph{$h$-gauge}

The $O (\phi^2)$ contribution to the eigenvalue density in $h$-gauge is as follows:
\begin{align}
\begin{split}
\rho^{(2)}[\phi](\bar\lambda)
&=-\frac{1}{2} \int \frac{\mathrm d\tau\,\mathrm dh}{2\pi}\, \frac{\partial \phi}{\partial \tau}\,\frac{\partial}{\partial h}\left[\frac{\partial \phi}{\partial \tau} \delta(\mu+h)\right]\rho(\bar \lambda)\\
&= -\frac{1}{2} \int \frac{\mathrm d\tau}{2\pi}\,\left( \frac{\partial \phi}{\partial \tau}\right)^2\,\frac{\partial \lambda}{\partial \mu}\,\delta'(\lambda(\tau,-\mu)-\bar\lambda)\\
&=\frac{1}{8\pi \mu}\,\frac{\mathrm d}{\mathrm d\bar\lambda}\left\{\frac{\bar\lambda}{p_F(\bar \lambda)} \left[\left(\frac{\mathrm d\phi_+}{\mathrm d\tau}\right)^2+\left(\frac{\mathrm d\phi_-}{\mathrm d\tau}\right)^2\right]\right\},
\end{split}
\end{align}
where in the last line above $\phi_{\pm{}} = \phi(\pm{} \tau(\bar \lambda, p_F(\bar \lambda)))$. In principle, one can also consider higher-order corrections to the density operator in $h$-gauge. These, and higher order corrections to the density operators, appear to be responsible for encoding the non-trivial dynamics of the theory, despite the fact that the $h$-gauge action \eqref{eq:c=1hgaugeaction} is free.

\subsubsection{Deformed matrix models}
In the context of matrix models, we can have different types of potentials $V(\lambda)$.
For example, the deformed matrix model \cite{Jevicki:1993zg} has the potential
\begin{align}
    V(\lambda) = -\frac{1}{2}\lambda^2 + \frac{\alpha}{\lambda^2}.
\end{align}
The coadjoint orbit method based on the $w_{\infty}$ algebra can be applied to generic potentials. To apply the method we introduced above, we introduce action-angle coordinates.
The single particle dispersion is given by
\begin{align}
h = \frac{1}{2}p^2+ V(\lambda).
\end{align}
The time-of-flight variable is given by
\begin{align}
\tau(\lambda,h) = \pm \int _{\lambda_0}^{\lambda} \frac{\mathrm d\lambda}{\sqrt{-2V(\lambda)+ 2h}}.
\end{align}
Inverting the above function, we obtain a coordinate expression for $\lambda(\tau,h)$. We can then find the momentum as a function of $(\tau,h)$:
\begin{align}
p(\tau,h) = \sqrt{-2V(\lambda(\tau,h)) +2 h}.
\end{align}
In this way, we can rewrite the problem in terms of action-angle coordinates.
\begin{align}
(\lambda,p) \to (\tau,h).
\end{align}
For example, the action in $p$-gauge up to the cubic order in $\phi$ is
\begin{align}
S_{\text{$p$-gauge}} &= S_{\rm WZW}^{(2)} + S_H^{(2)} +S_H^{(3)}  \notag \\
 & = -\frac{1}{4\pi} \int \mathrm dt  \mathrm d\tau  \left[\frac{\partial \phi}{\partial  t} \frac{\partial \phi}{\partial  \tau} + \left( \frac{\partial \phi}{\partial  \tau} \right)^2 +\frac{1}{3} \frac{1}{p_F(\tau)^2} \left(\frac{\partial \phi}{\partial \tau} \right)^3 \right],\label{eq:SCubicGen}
\end{align}
where above
\begin{align}
p_F(\lambda) = \sqrt{-2V(\lambda)-2\mu}
\end{align}
is the Fermi momentum for a generic potential.
For the deformed matrix model, the action \eqref{eq:SCubicGen} is the same as the collective field  action in \cite{Jevicki:1993zg}.

\section{Bosonization of  Noncritical M-theory}

\label{sec:bosonizationM}
\label{Mtheoryboson} 

Having introduced the coadjoint orbit method in the previous section, we now turn our attention to its main application in this paper, namely the bosonization of the noncritical M-theory Fermi surface. We find that the elementary degrees of freedom of the bosonized theory consist of a family of (1+1)-dimensional chiral bosons parametrized by a pair of classically conserved quantities corresponding to canonical coordinates. In Section \ref{gfcond}, we describe the two main gauge-fixing conditions that we use to study the bosonized action, namely $p$-gauge and $h$-gauge. Then in Sections \ref{pgaugeact} and \ref{hgaugeact}, we present the leading contributions to the bosonic effective action in these two gauges. Various useful facts about the geometry of phase spaces and Fermi surfaces that we make use of throughout this section are collected in  Appendices \ref{classPS} and \ref{contact}.

\subsection{Gauge fixing conditions and physical observables}
\label{gfcond}

As described in Section \ref{COMparam}, the bosonized action produced by the coadjoint orbit method can be interpreted as a sigma model with gauge group corresponding to the stabilizer of the distribution function $f_0$. The gauge redundancy implies the existence of an equivalence class of bosonic fields $\phi$ that define the same EFT on the Fermi surface. Therefore, we must gauge fix the theory to isolate the true physical degrees of freedom. In principle, there are many possible  gauge-fixing conditions that all lead to physically-equivalent theories. However, some gauge-fixing conditions are more convenient than others, depending on the type of physical observables being studied, as we now describe in more detail.

When studying the correlation functions of the eigenvalue density operators $\rho(x,\boldsymbol \lambda)$ in the Fermi liquid description, it is most convenient to impose the $p$-gauge condition, for which the equivalence class of bosonic fields is labeled by the representative
    \begin{align}
     \phi(t, \boldsymbol \lambda,p = p_F(\boldsymbol \lambda,\xi),\xi) \equiv\phi(t,\boldsymbol \lambda, \xi).
    \end{align}
The bosonic field $\phi$ can then be regarded as an infinite number of three-dimensional fields indexed by the continuous variable $\xi$. Then, by summing $\rho[\phi](t,\boldsymbol{\lambda}, \xi)$ over all values of $\xi$, it is possible to define an effective density operator $\rho[\phi](t, \boldsymbol \lambda)$ that has the same coordinate dependence as its Fermi liquid counterpart---see (\ref{bosonrho}).

Similarly, when studying the correlation functions of the light-cone density operators $\rho(u_\sigma, \varphi_\sigma,\ell)$ described in Section \ref{exactLCFermi}, it is most convenient to impose the $h$-gauge condition, for which a representative of the equivalence class of bosonic fields is given by
    \begin{align}
       \phi(t,\tau_\sigma, h_\sigma = -\sigma \mu, \varphi_\sigma,\ell)\equiv \phi(t,\tau_\sigma, \varphi_\sigma, \ell) ,
    \end{align}
so that the bosonic field $\phi$ can be regarded as an infinite number of two-dimensional fields indexed by the conserved quantities $\varphi_\sigma, \ell$. Analogous to the above case, one can then define an effective density operator that exhibits the same coordinate dependence as its Fermi liquid counterpart by summing $\rho[\phi](u_\sigma, \varphi_\sigma, \ell)$ over all values of $\ell$---see (\ref{bosonrho}).

In the following sections, we use the above gauge-fixing conditions to explicitly calculate the bosonized effective action order-by-order in a weak field expansion.

\subsection{Bosonic action in $p$-gauge}
\label{pgaugeact}

In this subsection, we derive the bosonized effective action in $p$-gauge up to cubic order in the bosonic field.
This gauge is particularly convenient for studying the eigenvalue density operators, since the bosonic field is represented as a function on the Fermi surface parametrized by the position $\bm\lambda$ and the momentum angle $\xi$.
Equivalently, after passing to action-angle variables, the field may be viewed as
\begin{align}
\phi=\phi(t,\tau,m,\varphi),
\end{align}
where $m$ and $\varphi$ label classical trajectories on the Fermi surface. 
The result of this subsection can be summarized as (see \eqref{eq:cubicact})
\begin{align}
S_{p\text{-gauge}} = S^{(2)} + S^{(3)}_{\rm WZW} + S^{(3)}_{H} + O(\phi^4).
\end{align}
The quadratic term describes an infinite family of decoupled chiral bosons labeled by the conserved quantities $(m,\varphi)$.
At cubic order, there are two distinct contributions: one from the WZW term and one from the Hamiltonian term. The WZW contribution is absent in the ordinary one-dimensional Fermi surface bosonization of the $c=1$ matrix model, but it is nonzero for the higher-dimensional Fermi surface of noncritical M-theory.

\subsubsection{Quadratic terms}

We first derive the quadratic action. The canonical transformation generated by the bosonic field acts on the ground-state distribution as
\begin{align}
f &= f_0  - \{\phi, f_0 \} + \f{1}{2} \{\phi,\{ \phi, f_0\}\} + \cdots \notag \\
&=  \theta (-\mu -h) + \delta( h + \mu)\frac{ \partial \phi}{\partial \tau}(t,\tau,h, m,\varphi) + \cdots
\end{align}
where $\phi(t,\tau,h,m,\varphi)$ is a function on phase space.
The stabilizer of the Fermi surface is determined by
\begin{equation}
\text{ad}_{\alpha}^* f_0 = \{\alpha, f_0\}=0 ~~ \implies ~~\left. \frac{\partial \alpha}{\partial \tau}(t,\tau,h,m,\varphi)\right|_{FS} = 0.
\end{equation}
Thus, on each coadjoint orbit, the bosonic field is defined only up to the gauge equivalence
\begin{align}
\phi \sim \phi - \alpha + \cdots
\end{align}
In $p$-gauge, a convenient representative of this equivalence class is
\begin{equation}
 \phi(t,\bm{\lambda},\xi, p= p_F(\bm{\lambda},\xi)) \equiv \phi(t,\tau, m ,\varphi).
\end{equation}
The WZW term is
\begin{align}
S_{\text{WZW}} =  \int  \mathrm dt \braket{f_0, U^{-1} \partial_t U}
= \int \mathrm dt \braket{f_0,  - \dot{\phi} + \frac{1}{2} \{ \dot{\phi} , \phi \}+ \cdots }.
\end{align}
The first term is a total derivative,
\begin{align}
\int\mathrm dt \braket{f_0,  - \dot{\phi} }
&= - \int  \mathrm dt \int \frac{ \mathrm d^2 \bm{\lambda}  \mathrm d^2\bm{p}}{(2\pi)^2} \theta(-\mu - h(\bm{\lambda},\bm{p})) \frac{\partial \phi}{\partial t} \notag \\
&= - \int  \mathrm dt \frac{\partial}{\partial t} \int  \frac{ \mathrm d^2 \bm{\lambda}  \mathrm d^2\bm{p}}{(2\pi)^2}\theta(-\mu - h(\bm{\lambda},\bm{p}))  \phi(t,\bm{\lambda},\bm{p})
\end{align}
and will be dropped.
The leading nontrivial contribution from the WZW term is therefore
\begin{align}
\frac{1}{2}\int  \mathrm dt \braket{f_0,   \{ \dot{\phi} , \phi \} }
&= \frac{1}{2}\int  \mathrm dt   \int \frac{ \mathrm d \tau  \mathrm d h  \mathrm d m  \mathrm d \varphi}{(2\pi)^2} \theta(-\mu - h) \left( \frac{\partial \dot{\phi}}{\partial \tau} \frac{\partial \phi}{\partial h}-\frac{\partial \dot{\phi}}{\partial h} \frac{\partial \phi}{\partial \tau} + \frac{\partial \dot{\phi}}{\partial \varphi} \frac{\partial \phi}{\partial m}- \frac{\partial \dot{\phi}}{\partial m} \frac{\partial \phi}{\partial \varphi}  \right) \notag \\
&=- \frac{1}{2}\int  \mathrm dt   \int \frac{ \mathrm d \tau  \mathrm d h  \mathrm d m  \mathrm d \varphi}{(2\pi)^2}  \delta(-\mu - h)  \dot{\phi} \frac{\partial \phi}{\partial \tau} \notag \\
&= -\frac{1}{2(2\pi)^2} \int  \mathrm dt  \mathrm d\tau  \mathrm d m  \mathrm d \varphi \,  \frac{\partial \phi}{\partial t} \frac{\partial \phi}{\partial \tau}.
\end{align}
Thus, at quadratic order, the WZW term supplies the symplectic form for the chiral boson.

We next expand the Hamiltonian term,
\begin{align}
S_H &= - \int  \mathrm dt \braket{f_0, U^{-1} h U} = - \int  \mathrm dt \braket{f_0 , h + \frac{1}{2} \{ \phi , \{ \phi, h\}\} +\frac{1}{3!} \{ \phi,\{ \phi , \{ \phi, h\}\} \} + \cdots} .
\end{align}
The zeroth-order term is the ground-state energy, while the linear term is a total derivative on phase space.
The leading dynamical contribution is therefore quadratic:
\begin{align}
-\frac{1}{2} \int  \mathrm dt   \braket{ f_0, \{ \phi, \{\phi, h\}\}  } &= \frac{1}{2} \int  \mathrm dt   \braket{ \{  \phi, f_0\},  \{\phi, h \} }\notag \\
&=- \frac{1}{2}\int  \mathrm dt \frac{ \mathrm d\tau  \mathrm d h  \mathrm d m  \mathrm d\varphi}{(2\pi)^2}   \delta(\mu +h) \frac{\partial \phi}{\partial  \tau}\frac{\partial \phi}{\partial  \tau}  \notag \\
 &= - \frac{1}{2(2\pi)^2}\int  \mathrm dt  \mathrm d\tau  \mathrm dm \mathrm  d\varphi\left(\frac{\partial \phi}{\partial  \tau} \right)^2.
\end{align}
Combining the WZW and Hamiltonian contributions, the quadratic action is
\begin{align}
\label{eq:quadact}
S^{(2)} =S_{\text{WZW}}^{(2)} + S_{H}^{(2)} 
&= -\frac{1}{2(2\pi)^2} \int  \mathrm dt  \mathrm d\tau  \mathrm d m  \mathrm d \varphi \left( \frac{\partial \phi}{\partial t}  + \frac{\partial \phi}{\partial \tau}   \right)  \frac{\partial\phi}{\partial \tau}.
\end{align}
This is the action of decoupled chiral bosons labeled by the conserved quantities $(m,\varphi)$. After analytic continuation to Euclidean time $t = - ix  $, the corresponding propagator takes the form (cf. Appendix \ref{app:bosonprop})
\begin{align}
\braket{\phi(x_1,\tau_1,\varphi_1,m_1) \phi(x_2,\tau_2,\varphi_2,m_2)} =  F(\Delta x,\Delta \tau) \delta(\Delta m)\delta(\Delta \varphi), \label{eq:MBtwoPt}
\end{align}
where
\begin{align}
F(\Delta x,\Delta \tau)  =- (2\pi )\log (\Delta \tau +i \Delta x ).
\end{align}

\subsubsection{Cubic interaction terms for boson fields}
We now turn to the cubic interactions. There are two sources of cubic terms in $p$-gauge:
\begin{align}
 S^{(3)}_{\rm WZW} \qquad\text{and}\qquad S^{(3)}_H .\end{align}
The first comes from the WZW term on the coadjoint orbit, while the second comes from expanding the Hamiltonian on the deformed Fermi surface. Both terms contribute to density three-point functions in the bosonized theory.

\paragraph{Cubic WZW term}

The WZW term expands as
\begin{align}
S_{\rm WZW} = \int \mathrm dt \braket{f_0, U^{-1}\partial_tU} = \int \mathrm dt \braket{f_0,\frac{1}{2}\{ \dot{\phi},\phi\} - \frac{1}{3!}\{\{\dot{\phi},\phi\},\phi \} + \cdots}.
\end{align}
The quadratic contribution is
\begin{align}
\begin{split}
S^{(2)}_{\rm WZW}& = \frac{1}{2} \int \mathrm dt \braket{f_0,\{ \dot{\phi},\phi\}}\\ &= \frac{1}{2} \int\mathrm  dt \braket{\{\phi,f_0\}, \dot{\phi}}  \\
&= \frac{1}{2} \int \mathrm dt \braket {- \delta(h+\mu)\frac{ \partial \phi}{\partial  \tau} , \dot{\phi}}.
\end{split}
\end{align}
Above, we have used the Lie algebra relation $\braket{f,\{g,h\}} = -\braket{\{g,f\},h}$.
To evaluate the cubic term, it is useful to use
\begin{align}
\{\phi,f_0\} &= \frac{\partial \phi}{\partial \tau} \frac{\partial }{\partial h} \theta(-h - \mu) = - \delta(h+\mu) \frac{\partial \phi}{\partial \tau}.
\end{align}
In $p$-gauge, the Poisson bracket $\{\dot\phi,\phi\}$ can be written as
\begin{align}
\{\dot{\phi},\phi\} &= \omega^{\mu\nu}\frac{\partial \dot{\phi}}{\partial z^{\mu}} \frac{\partial \phi}{\partial z^{\nu}} = \frac{1}{p} \bm{t}_{\xi} \cdot \partial_{\bm{\lambda}} \dot{\phi}\frac{\partial \phi}{\partial \xi} - \frac{1}{p} \bm{t}_{\xi} \cdot \partial _{\bm\lambda}\phi \frac{\partial \dot {\phi}}{\partial \xi},
\end{align}
where $\omega^{\mu \nu}$ is the symplectic form. This identity depends only on the $p$-gauge condition
\begin{align}\phi(t,\bm\lambda,\bm p)\sim \phi(t,\bm\lambda,\xi),
\end{align}
and not on the specific form of the Hamiltonian.
We also use the identity for the delta function
\begin{align}
\delta(h + \mu) = \frac{1}{\frac{\partial h}{\partial p}} \delta(p- p_F(\bm{\lambda},\xi)).
\end{align}
Using these identities, the cubic contribution from the WZW term becomes
\begin{align}
\begin{split}
S^{(3)}_{\rm WZW} &= -\frac{1}{3!} \int \mathrm dt \braket{f_0,\{ \{\dot{\phi},\phi\}, \phi\}}\\
&= -\frac{1}{3!} \int \mathrm dt \braket{\{\phi, f_0\}, \{\dot{\phi},\phi\}}  \\
&= \frac{1}{3!} \int \mathrm dt\frac{\mathrm d^2 \boldsymbol \lambda \mathrm d^2 \boldsymbol p}{(2\pi)^2} \delta(h+\mu) \frac{\partial \phi} {\partial \tau}  \frac{1}{p} \bm{t}_{\xi} \cdot \left (\partial_{\bm{\lambda}} \dot \phi \frac{\partial \phi}{\partial \xi}-  \partial_{\bm{\lambda}}  \phi \frac{\partial \dot{\phi}}{\partial \xi} \right) \\
&=  \frac{1}{3!} \int \mathrm dt\frac{\mathrm d ^2 \boldsymbol \lambda   p \mathrm dp \mathrm d\xi}{(2\pi)^2}  \frac{1}{\frac{\partial h}{\partial p}}  \delta(p - p_F(\bm{\lambda},\xi))\frac{\partial \phi} {\partial \tau}  \frac{1}{p} \bm{t}_{\xi} \cdot \left(\partial_{\bm{\lambda}} \dot \phi  \frac{\partial \phi}{\partial \xi} -  \partial_{\bm{\lambda}}  \phi  \frac{\partial \dot{\phi}}{\partial \xi} \right)  \\
&= \frac{1}{3!} \int \mathrm dt\frac{\mathrm d^2 \boldsymbol \lambda  \mathrm  d\xi}{(2\pi)^2} \frac{1}{p_F(\bm{\lambda},\xi)} \frac{\partial \phi} {\partial \tau}   \bm{t}_{\xi} \cdot \left(\partial_{\bm{\lambda}} \dot \phi  \frac{\partial \phi}{\partial \xi} -  \partial_{\bm{\lambda}}  \phi  \frac{\partial \dot{\phi}}{\partial \xi} \right).
\end{split}
\end{align}
It is convenient to convert this expression to action-angle coordinates restricted to the Fermi surface $h = -\mu$, namely \begin{align}
w^\mu=(\tau,m,\varphi). 
\end{align}
Some useful relations for changing coordinates are
\begin{align}
\begin{split}
\frac{\partial \tau}{\partial \lambda_1} \frac{\partial m}{\partial \xi}-\frac{\partial m}{\partial \lambda_1} \frac{\partial \tau}{\partial \xi} &= - \sqrt{\mu + m} \sinh (\tau- \varphi)  \\
\frac{\partial \tau}{\partial \lambda_1} \frac{\partial \varphi}{\partial \xi}-\frac{\partial \varphi}{\partial \lambda_1} \frac{\partial \tau}{\partial \xi} &= - \frac{1}{2 \sqrt{\mu + m}} \cosh (\tau- \varphi)  \\
\frac{\partial m}{\partial \lambda_1} \frac{\partial \varphi}{\partial \xi}-\frac{\partial \varphi}{\partial \lambda_1} \frac{\partial m}{\partial \xi} &=  \sqrt{\mu + m} \sinh (\tau- \varphi)
\end{split}
\end{align}
and
\begin{align}
\begin{split}
\frac{\partial \tau}{\partial \lambda_2} \frac{\partial m}{\partial \xi}-\frac{\partial m}{\partial \lambda_2} \frac{\partial \tau}{\partial \xi} &= - \sqrt{\mu - m} \sinh (\tau+ \varphi)  \\
\frac{\partial \tau}{\partial \lambda_2} \frac{\partial \varphi}{\partial \xi}-\frac{\partial \varphi}{\partial \lambda_2} \frac{\partial \tau}{\partial \xi} &= - \frac{1}{2 \sqrt{\mu - m}} \cosh (\tau+ \varphi)  \\
\frac{\partial m}{\partial \lambda_2} \frac{\partial \varphi}{\partial \xi}-\frac{\partial \varphi}{\partial \lambda_2} \frac{\partial m}{\partial \xi} &=  -\sqrt{\mu - m} \sinh (\tau+ \varphi).
\end{split}
\end{align}
Compactly we can write
\begin{align}
\begin{split}
\frac{1}{2}\sum_{\nu,\rho}\epsilon _{\mu\nu\rho} \left(\frac{\partial w^{\nu}}{\partial \lambda_1} \frac{\partial w^{\rho}}{\partial \xi} -\frac{\partial w^{\rho}}{\partial \lambda_1} \frac{\partial w^{\nu}}{\partial \xi} \right)  &=  \partial_\mu \lambda_2(\tau,m,\varphi)  \\
\frac{1}{2}\sum_{\nu,\rho}\epsilon _{\mu\nu\rho} \left(\frac{\partial w^{\nu}}{\partial \lambda_2} \frac{\partial w^{\rho}}{\partial \xi} -\frac{\partial w^{\rho}}{\partial \lambda_2} \frac{\partial w^{\nu}}{\partial \xi} \right)  &=  -\partial_\mu \lambda_1(\tau,m,\varphi)
\end{split}
\end{align}
where $\partial_\mu \equiv \partial/\partial w^\mu$.
The most compact presentation is
\begin{align}
\frac{1}{2}\epsilon _{\mu\nu\rho} \left(\frac{\partial w^{\nu}}{\partial \lambda_i} \frac{\partial w^{\rho}}{\partial \xi} -\frac{\partial w^{\rho}}{\partial \lambda_i} \frac{\partial w^{\nu}}{\partial \xi} \right)=  \epsilon^{ij} \partial_\mu  \lambda_{j}(\tau,m,\varphi) .
\end{align}
This identity allows us to express the WZW cubic term in a coordinate-independent form on the Fermi surface:
\begin{align}
\begin{split}
S^{(3)}_{\rm WZW} &= \frac{1}{3!} \int \mathrm dt\frac{\mathrm d^2 \boldsymbol \lambda   \mathrm d\xi}{(2\pi)^2} \frac{1}{p_F} \frac{\partial \phi} {\partial \tau}   \bm{t}_{\xi} \cdot \left ( \partial_{\boldsymbol \lambda}  w^{\mu}\frac{\partial w^{\nu}}{\partial \xi} -  \partial_{\boldsymbol \lambda}  w^{\nu} \frac{\partial w^{\mu}}{\partial \xi} \right) \partial_{\mu} \dot{\phi} \partial_\nu \phi \\
&= -\frac{1}{3!} \int \mathrm dt\frac{\mathrm d \tau \mathrm dm   \mathrm d\varphi}{(2\pi)^2} \frac{1}{p_F} \frac{\partial \phi} {\partial \tau} \frac{\epsilon^{ik} p_{Fk}}{p_F}\tensor{\epsilon}{_i^j}\epsilon ^{\mu\nu\rho} \partial_\rho  \lambda_j \partial_\mu \dot{\phi} \partial_\nu \phi \\
&= -\frac{1}{ 3!} \int \mathrm dt\frac{\mathrm d \tau \mathrm dm   \mathrm  d\varphi}{(2\pi)^2}\frac{1}{p_F^2}\frac{\partial \phi} {\partial \tau}  p_{F}^j \partial_{\rho} \lambda_j   \epsilon ^{\mu\nu\rho}  \partial_\mu  \dot{\phi}\partial_\nu \phi\\
&= -\frac{1}{  3!} \int \mathrm dt\frac{\mathrm d vol_{FS}}{(2\pi)^2} R \phi \epsilon ^{\mu\nu\rho} p_{F}^j \partial_{\rho} \lambda_j   \partial_\mu \dot{\phi}\partial_\nu \phi  ,
\end{split}
\end{align}
where $ R = (1/p_F^{2}) \partial/\partial \tau$ is the Reeb vector field on the Fermi surface and we emphasize that we may either express the Fermi momentum as a function of rectilinear coordinates $p_F(\boldsymbol \lambda, \xi)$ or action-angle coordinates $p_F(\tau, \varphi,m)$, as the Fermi surface can always be parametrized in terms of either coordinate system. 
$p_{F1}(\tau,m,\varphi) = \sqrt{\mu - m} \sinh(\tau + \varphi)$ and $p_{F2}(\tau,m,\varphi) = \sqrt{\mu + m} \sinh(\tau - \varphi)$ are the components of the Fermi momentum vector $\bm{p}_F = (p_{F1} , p_{F2})^T$.
The last expression is written in a way that is manifestly coordinate-independent on the Fermi surface.
Written explicitly in terms of action-angle coordinates, the above cubic term takes the form
\begin{align}
\begin{split}
S^{(3)}_{\rm WZW} &= - \frac{1}{   3!} \int \mathrm dt \frac{\mathrm d\tau \mathrm dm \mathrm d\varphi}{(2\pi)^2} \frac{1}{p_F^2}\frac{\partial \phi} {\partial \tau}\Bigg[ p_F^j \frac{\partial \lambda_j}{\partial \tau}  \left( \frac{\partial \dot{\phi}}{\partial m}\frac{\partial \phi}{\partial \varphi} - \frac{\partial \dot{\phi}}{\partial \varphi}\frac{\partial \phi}{\partial m} \right)   \\
& \quad +  p_F^j \frac{\partial \lambda_j}{\partial m} \left( \frac{\partial \dot{\phi}}{\partial \varphi}\frac{\partial \phi}{\partial \tau} - \frac{\partial \dot{\phi}}{\partial \tau }\frac{\partial \phi}{\partial \varphi} \right)   +  p_F^j\frac{\partial \lambda_j}{\partial \varphi} \left( \frac{\partial \dot{\phi}}{\partial \tau}\frac{\partial \phi}{\partial m} - \frac{\partial \dot{\phi}}{\partial m}\frac{\partial \phi}{\partial \tau} \right) \Bigg] .\label{eq:WZW3}
\end{split}
\end{align}
This is a new cubic interaction that has no direct counterpart in the collective field theory action that results from bosonizing the $c=1$ string Fermi surface.

\paragraph{Cubic term from the Hamiltonian term}

The Hamiltonian term is expanded as
\begin{align}
\begin{split}
S_H &= - \int \mathrm dt \braket{f_0, U^{-1}h U}  \\
&= - \int \mathrm dt \braket{f_0 , h+ \{\phi, h\} + \frac{1}{2} \{ \phi , \{ \phi, h\}\} +\frac{1}{3!} \{ \phi,\{ \phi , \{ \phi, h\}\} \} + \cdots} .
\end{split}
\end{align}
The cubic term is
\begin{align}
S_H ^{(3)} = \frac{1}{3!} \int \mathrm dt \braket{\{ \phi,f_0\},\{ \phi , \{ \phi, h\}\}}.
\end{align}
After repeated integration by parts, the integrand reduces to
\begin{align}
 &\braket{\{ \phi,f_0\},\{ \phi , \{ \phi, h \}\}} \notag \\
&= -\int\frac{\mathrm d\tau \mathrm dh\mathrm dm \mathrm d\varphi}{(2\pi)^2} \delta (h+ \mu) \frac{\partial \phi}{\partial \tau}
\left[ \left(\frac{\partial \phi}{\partial \tau}\frac{\partial^2 \phi}{\partial \tau \partial h}  - \frac{\partial \phi}{\partial h} \frac{\partial^2\phi }{\partial \tau^2}\right)+\left(  \frac{\partial \phi}{\partial \varphi} \frac{\partial^2 \phi}{\partial m \partial  \tau}-\frac{\partial \phi}{\partial m}\frac{\partial^2 \phi}{\partial \varphi \partial \tau} \right)\right] \notag \\
&=  \frac{1}{2}\int\frac{\mathrm d\tau \mathrm dh \mathrm dm \mathrm d\varphi}{(2\pi)^2} \delta '(h+ \mu)
 \left(\frac{\partial \phi}{\partial \tau} \right)^3.
\end{align}
Furthermore, after performing additional integrations by parts and imposing the $p$-gauge condition $\partial\phi/\partial p=0$, the cubic Hamiltonian term can be written as
\begin{align}
S_H ^{(3)}
&=  -\frac{3}{ 2 \cdot 3!} \int \mathrm dt \frac{\mathrm d^2 \boldsymbol \lambda \mathrm d\xi}{(2\pi)^2} \frac{1}{p_F^2} \left( \frac{\partial \phi}{\partial \tau} \right)^2 \left ( \frac{\partial \phi}{\partial \tau}- 2\frac{ \bm{t}_{\xi} \cdot \partial_{\bm{\lambda}} p_F}{p_F}\frac{\partial \phi}{\partial \xi} \right).\label{eq:H3}
\end{align}
The details of these manipulations are given in Appendix~\ref{sec:DerBA}, in particular in \eqref{eq:h3detail} and \eqref{eq:H3derivation}.
The first term is the direct higher-dimensional generalization of the familiar cubic interaction of the $c=1$ matrix model. The second term is new:
it is proportional to the variation of the Fermi momentum along the tangential direction of the Fermi surface $ \frac{1}{p_F^2} ( \frac{\partial \phi}{\partial \tau} )^2 \frac{ \bm{t}_{\xi} \cdot \partial_{\bm{\lambda}} p_F}{p_F}\frac{\partial \phi}{ \partial \xi}$ and therefore has no analogue in the one-dimensional case, where there is no residual angular coordinate $\xi$.
This term is proportional to $\bm t_{\xi}\cdot\partial_{\bm\lambda} p_F$ and therefore vanishes when $p_F$ has no tangential dependence along the Fermi surface.
In such a case the term $(\partial \phi/\partial \tau)^3$ is consistent with the cubic term in \cite{Delacretaz:2022ocm}.

The total action up to the cubic order is
\begin{align}
\label{eq:cubicact}
S &= S_{\rm WZW} ^{(2)} + S_{\rm WZW} ^{(3)} + S_{H} ^{(2)} + S_{H} ^{(3)} \notag\\
&= -\frac{1}{2} \int \mathrm dt \frac{\mathrm d\tau \mathrm d m \mathrm d \varphi}{(2\pi)^2} \left( \frac{\partial \phi}{\partial t}  +\frac{\partial \phi}{\partial \tau}  \right)  \frac{\partial \phi}{\partial \tau} \notag \\
& \quad -\frac{1}{3!}
\int \mathrm dt\,
\frac{\mathrm d\tau\,\mathrm dm\,\mathrm d\varphi}{(2\pi)^2}
\frac{1}{p_F^2}
\frac{\partial\phi}{\partial\tau}
\epsilon^{\mu\nu\rho}
p_F^j\partial_\rho\lambda_j\,
\partial_\mu\dot\phi\, \notag \\
& \quad  -\frac{3}{ 2 \cdot 3!} \int \mathrm dt \frac{\mathrm d\tau \mathrm d m \mathrm d\varphi}{(2\pi)^2} \frac{1}{p_F^2} \left( \frac{\partial \phi}{\partial \tau} \right)^2 \left ( \frac{\partial \phi}{\partial \tau}- 2\frac{ \bm{t}_{\xi} \cdot \partial_{\bm{\lambda}} p_F}{p_F}\frac{\partial \phi}{\partial \xi} \right).
\end{align}

\subsection{Bosonic action in $h$-gauge }
\label{hgaugeact}

In this section, we describe bosonized noncritical M-theory in $h$-gauge, where we impose the gauge-fixing condition
    \begin{align}
            \left. \frac{\partial \phi}{\partial h}\right|_{FS} = 0.
    \end{align}
This condition can be interpreted as a gauge in which the field $\phi$ living on the classical Fermi surface $h = -\mu$ is independent of normal fluctuations. In this gauge, it is convenient to work in a set of generic action-angle coordinates $\tau,h, \varphi, m$---see Section \ref{actionangle}.\footnote{We stress that this coordinate system is not uniquely fixed by the defining properties described in Section \ref{actionangle}, and that there are multiple sets of coordinates with these properties.} In this set of coordinates, the stabilizer of the Fermi surface $h = -\mu$ is given by
    \begin{align}
    \begin{split}
        \text{ad}_\alpha^* f_0 = \{\alpha, f_0\} = 0~~   \implies ~~ \left.\frac{\partial \alpha}{\partial \tau}\right|_{h=-\mu} = 0.
    \end{split}
    \end{align}
It follows that the above gauge-fixing condition $\alpha \sim 0$ implies that the gauge orbits of the scalar field $\phi$ can be represented by
    \begin{align}
         \phi(t,\tau,h=-\mu,m,\varphi) \equiv \phi(t, \tau, \varphi, m).
    \end{align}
Using $\{ f_0 , \phi\} = \delta(h+\mu) \partial \phi / \partial \tau$ along with the fact that  the volume form on the Fermi surface is $\mathrm{d}{vol}_{{FS}} =  \mathrm d\tau \wedge  \mathrm d\varphi \wedge \mathrm dm$ (see Section \ref{aavol}), the weak field expansion of the bosonic action takes the form (\ref{Skexp})
    \begin{align}
        \begin{split}
        \label{eq:Shgauge}
        S_{\text{$h$-gauge}} &= S^{(0)} -\sum_{n=2}^{\infty} \frac{1}{n!} \int \mathrm dt \frac{\mathrm  d \tau \mathrm d \varphi  \mathrm dm}{(2\pi)^2} \,\frac{\partial \phi}{\partial \tau} \, \text{ad}_\phi^{n-2} \left(\frac{\partial \phi}{\partial t} +\frac{\partial \phi}{\partial \tau}\right),
        \end{split}
    \end{align}
where the adjoint action on derivatives of $\phi$ is restricted to the $\varphi,m$ sector by the $h$-gauge fixing condition.
The quadratic action is
    \begin{align}
        \begin{split}
            S^{(2)}&= -\int \mathrm d t \frac{\mathrm  d \tau \mathrm d \varphi  \mathrm dm}{(2\pi)^2}  \, \frac{1}{2} \frac{\partial \phi}{\partial \tau}  \left(\frac{\partial \phi}{\partial t} + \frac{\partial \phi}{\partial \tau}\right).
        \end{split}
    \end{align}
The Euclidean continuation of the  time-ordered propagator is given by (\ref{aaprop}),
    \begin{align}
    \label{hgaugeprop}
       \langle T ( \phi(x_1,  \tau_1, \varphi_1 , m_1) \phi(x_2,  \tau_2, \varphi_2 , m_2)) \rangle =- 2\pi \delta(\Delta \varphi) \delta(\Delta m) \log \Delta u,
    \end{align}
where $u = i x + \tau$, with $x$ denoting Euclidean time. Although it is straightforward to work out the higher order contributions to the effective action, we do not bother computing them explicitly because it turns out that the connected $n$-point correlation functions of the light-cone density operators vanish for $n\geq 3$. We explain why this is the case in Section \ref{sec:LCdensitycorr}.

\section{Bosonized Density Correlation Functions}

\label{bosonicdensitycorr}

\begin{figure}
\begin{center}
$
\begin{tikzpicture}
\definecolor{vtxcol}{RGB}{38,50,99}    
\definecolor{linkcol}{RGB}{198,76,42}  
\definecolor{grpcol}{RGB}{23,128,128}  
\tikzset{
    vtx/.style={draw=vtxcol!70!black, circle, fill=vtxcol, scale=.5},
    grp/.style={draw=grpcol, line width=1pt, fill=grpcol, fill opacity=.12},
    link/.style={draw=linkcol, line width=1.2pt},
    hdr/.style={text=vtxcol},
}
\node(L0)[hdr] at (-7.5,0) {\underline{$N_\phi = 4$}};
\node(1)[scale=.8,label=below:{$\substack{ \sim p_F^{-4}
}$}] at (3,0) {$
            \begin{tikzpicture}
			\node(A)[vtx] at (-1,1) {};
			\node(B)[vtx] at (-1,-1) {};
			\node(C)[vtx] at (1,1) {};
			\node(D)[vtx] at (1,-1) {};
			\node[grp, ellipse, fit=(A)(B)] {};
			\node[grp, ellipse, fit=(C)(D)] {};
			\draw[link] (A) -- (C);
			\draw[link] (B) -- (D);
		\end{tikzpicture}
        $};
\node(2)[scale=.8,label=below:{$\substack{\sim p_F^{-2}}$}] at (-3,0) {$        \begin{tikzpicture}
			\node(A)[vtx] at (-1,1) {};
			\node(B)[vtx] at (-1,-1) {};
			\node(C)[vtx] at (1,1) {};
			\node(D)[vtx] at (1,-1) {};
			\node[grp, ellipse, fit=(A)] {};
			\node[grp, ellipse, fit=(B)] {};
			\node[grp, ellipse, fit=(C)(D)] {};
			\draw[link] (A) -- (C);
			\draw[link] (B) -- (D);
		\end{tikzpicture}
        $};
    \node(3)[scale=.8,label=below:{$\substack{ \sim p_F^{-6}}$}] at (3,-12) {$
         		\begin{tikzpicture}
			\node(A1)[vtx] at (-0.2887*2.5, 0.3333*3) {};
			\node(A2)[vtx] at (-0.5774*2.5, -0.1667*2.5) {};
			\node(B1)[vtx] at (-0.5774*2.5, -0.6667*2.5) {};
			\node(B2)[vtx] at (0.5774*2.5, -0.6667*2.5) {};
			\node(C1)[vtx] at (0.5774*2.5, -0.1667*2.5) {};
			\node(C2)[vtx] at (0.2887*2.5, 0.3333*3){};
			\node[grp, ellipse, fit=(A2)(B1)] {};
			\node[grp, ellipse, fit=(B2)(C1)] {};
			\node[grp, ellipse, fit=(C2)(A1)] {};
			\draw[link] (A1) -- (A2);
			\draw[link] (B1) -- (B2);
			\draw[link] (C1) -- (C2);
		\end{tikzpicture}
    $};
    \node(L-6)[hdr] at (-7.5,-6) {\underline{$N_\phi = 6$}};
    \node(4)[scale=.8,label=below:{$\substack{ \sim p_F^{-4}}$}] at (3,-6) {$
         		\begin{tikzpicture}
			\node(A1)[vtx] at (-0.2887*2.5, 0.3333*3) {};
			\node(A2)[vtx] at (-0.5774*2.5, -0.1667*2.5) {};
			\node(B1)[vtx] at (-0.5774*2.5, -0.6667*2.5) {};
			\node(B2)[vtx] at (0.5774*2.5, -0.6667*2.5) {};
			\node(C1)[vtx] at (0.5774*2.5, -0.1667*2.5) {};
			\node(C2)[vtx] at (0.2887*2.5, 0.3333*3){};
			\node[grp, ellipse, fit=(A2)] {};
            \node[grp, ellipse, fit=(B1)] {};
			\node[grp, ellipse, fit=(B2)(C1)] {};
			\node[grp, ellipse, fit=(C2)(A1)] {};
			\draw[link] (A1) -- (A2);
			\draw[link] (B1) -- (B2);
			\draw[link] (C1) -- (C2);
		\end{tikzpicture}
    $};
    \node(5)[scale=.8,label=below:{$\substack{ \sim p_F^{-2} }$}] at (-3,-6) {$
         			\begin{tikzpicture}
			\node(A)[vtx] at (0, 1*2) {};
			\node(B)[vtx] at (-0.8660*2, -0.5*2) {};
			\node(C)[vtx] at (0.8660*2, -0.5*2) {};
			\node[grp, ellipse, fit=(A)] {};
			\node[grp, ellipse, fit=(B)] {};
    		  \node[grp, ellipse, fit=(C)] {};
			\draw[link] (0,0) -- (A);
			\draw[link] (0,0) -- (B);
			\draw[link] (0,0) -- (C);
		\end{tikzpicture}
    $};
    \node(6)[scale=.8,label=below:{$\substack{\sim p_F^{-4}}$}] at (-3,-12) {$
         	\begin{tikzpicture}
			\node(A1)[vtx] at (0, 1*2) {};
			\node(B1)[vtx] at (-0.8660*2, -0.5*2) {};
			\node(C1)[vtx] at (0.8660*2, -0.5*2) {};
			\node(A2)[vtx] at (0, 1*.5) {};
			\node(B2)[vtx] at (-0.8660*.5, -0.5*.5) {};
			\node(C2)[vtx] at (0.8660*.5, -0.5*.5) {};
			\draw[link] (A1) -- (A2);
			\draw[link] (B1) -- (B2);
			\draw[link] (C1) -- (C2);
			\node[grp, circle, fit=(A2)(B2)(C2)] {};
            	\node[grp, ellipse, fit=(A1)] {};
			\node[grp, ellipse, fit=(B1)] {};
    		  \node[grp, ellipse, fit=(C1)] {};
		\end{tikzpicture}
    $};
    \end{tikzpicture}
    $
\end{center}
\caption{Diagrams representing various tree level contributions to the connected correlators of the eigenvalue     density operators $\rho[\phi](x,\boldsymbol \lambda) $ in bosonized noncritical M-theory---see Section \ref{3ptfn}. A circle containing $k$ dots represents an operator $\rho^{(k)}[\phi]$, a $k$-valent vertex represents an interaction $S^{(k)}$, and a line connecting a pair of dots corresponds to a propagator $\langle T \phi \phi \rangle$. We organize the diagrams according to the number of fields, $N_\phi$, and the corresponding order in the inverse Fermi momentum $1/p_F$, which at large $\lambda_r$ goes like $1/\lambda_r$. For $N_{\phi} = 4$, the leading diagram at large $\lambda_r$ is $\langle \rho^{(1)}[\phi] \rho^{(1)}[\phi] \rho^{(2)}[\phi]\rangle$, whereas for $N_\phi = 6$, the leading diagram is $\langle \rho^{(1)}[\phi] \rho^{(1)}[\phi] \rho^{(1)}[\phi] \rangle$; these leading contributions correspond to the diagrams contributing to the connected three-point function in Figure 3 of \cite{Delacretaz:2022ocm}.}
\label{hgaugegraph}
\end{figure}

In this section, we study the correlation functions of density operators in bosonized M-theory. We study two different kinds of density operators, namely the eigenvalue density operators and the light-cone density operators, defined by the respective bases
    \begin{alignat}{2}
        \label{bosonrho}
            \rho(\bar{\boldsymbol{\lambda}}) &= \int \mathrm d\bar \xi \, \rho(\bar{\boldsymbol{\lambda}}, \bar \xi)& \qquad  \rho(\bar{\boldsymbol{\lambda}}, \bar \xi)& = \delta^2(\bar{\boldsymbol{\lambda}} - \boldsymbol{\lambda})\delta(\bar \xi - \xi)\\
            \rho(\bar \tau_\sigma, \bar \varphi_\sigma) &=\int \mathrm d\bar \ell\, \rho(\bar \tau_\sigma, \bar \varphi_\sigma,\bar \ell) & \qquad \rho(\bar \tau_\sigma, \bar \varphi_\sigma,\bar \ell)&= \delta(\bar\tau_\sigma - \tau_\sigma) \delta(\bar \varphi_\sigma - \varphi_\sigma) \delta(\bar \ell -\ell).
    \end{alignat}
These density operators correspond to distinct physical observables when expanded in a basis of the dynamical boson field $\phi$. The eigenvalue density operators $\rho(\bar{\boldsymbol{\lambda}})$ are most easily studied in $p$-gauge, whereas the light-cone density operators $\rho(\bar\tau_\sigma,\bar\varphi_\sigma)$ are most easily studied in $h$-gauge. Throughout this section, the correlation functions in the bosonized theory are evaluated at tree level in the effective bosonic description.

Below, we provide evidence that the $n$-point correlation functions of $\rho[\phi](x,\bar{\boldsymbol{\lambda}})$ agree with the corresponding density correlation functions in the Fermi liquid description to leading order in the perturbative expansion of the bosonized theory, for $n=1,2,3$.
Furthermore, we show that the correlation functions of $\rho[\phi](x,\bar \tau_\sigma, \bar \varphi_\sigma)$ match those in the Fermi liquid description to all orders in perturbation theory; in particular, in both the Fermi liquid description and the bosonized theory, the connected density $n$-point correlation functions vanish identically for $n \geq 3$.

\subsection{Eigenvalue density one-point function}

We begin by computing the density one-point function in the bosonized description.
As explained in Section~\ref{sec:COMgeneral}, density operators are naturally represented as elements of the algebra acting on the coadjoint orbit.
For the eigenvalue density, we take
\begin{align}
\rho(\bar{\bm\lambda}) = \delta^2(\bm\lambda-\bar{\bm\lambda})\in \mathfrak g .
\end{align}
The corresponding bosonized density operator is obtained by pairing this algebra element with the deformed distribution function,
\begin{align}
\rho[\phi](\bar{\bm{\lambda}}) = \int \frac{\mathrm d \bm{\lambda} \mathrm d \bm{p}}{(2\pi)^2} f_{\phi}(\bm{\lambda},\bm{p })\delta^2(\bm{\lambda} - \bar{\bm{\lambda}}). \label{eq:bosonrho1pt}
\end{align}
The one-point function is obtained by evaluating this expression on the undeformed Fermi sea.
Equivalently, at this order we set
\begin{align}
\phi=0, \qquad f_\phi=f_0 .
\end{align}
It is also useful to introduce a more refined phase-space density localized not only at $\bar{\bm\lambda}$ but also at a fixed momentum angle $\bar{\xi}$.
In the semiclassical bosonized description, we may represent this operator as
\begin{align} \rho(\bar{\bm\lambda},\bar{\xi}) = \delta^{(2)}(\bm\lambda-\bar{\bm\lambda}) \delta(\xi-\bar{\xi})\in \mathfrak g ,
\end{align} and define
\begin{align}
\rho[\phi](\bar{\bm{\lambda}},\bar{\xi}) = \int \frac{\mathrm d \bm{\lambda} \mathrm d \bm{p}}{(2\pi)^2} f_{\phi}(\bm{\lambda},\bm{p })\delta^2(\bm{\lambda} - \bar{\bm{\lambda}}) \delta(\xi- \bar{\xi}). \label{eq:bosonrho1ptxi}
\end{align}
The eigenvalue density \eqref{eq:bosonrho1pt} is obtained by integrating \eqref{eq:bosonrho1ptxi} over $\bar{\xi}$.

\subsubsection{One-point function from bosonization}
Let us now evaluate the semiclassical one-point function. For the undeformed distribution,
\begin{align} f_0(\bm\lambda,\bm p) = \theta\!\left( -\mu-\frac{1}{2}(\bm{p}^2-\bm{\lambda}^2) \right), \end{align}
we find
\begin{align}
\begin{split}
\rho[\phi = 0](\bar{\bm{\lambda}}) &= \int \frac{\mathrm d \bm{\lambda} \mathrm d \bm{p}}{(2\pi)^2} \theta \left(
 - \mu - \left( \frac{1}{2} \bm{p}^2 - \frac{1}{2} \bm{\lambda}^2 \right)\right)\delta^2(\bm{\lambda} - \bar{\bm{\lambda}})  \\
 &= \int_0^{\infty} \frac{ \mathrm dp}{2\pi} \, p \int _0^{2\pi} \frac{\mathrm d \xi}{2\pi}  \theta  \left(
 - \mu + \left(  \frac{1}{2} \bar{\bm{\lambda}}^2 - \frac{1}{2} \bm{p}^2\right)\right) \\
 &= \frac{1}{2\pi} \int_0^{ \sqrt{\bar{\bm{\lambda}}^2 -2 \mu}}  \mathrm dp \, p   \\
 &=
 \begin{cases}\displaystyle
 \frac{1}{4\pi} (\bar{\bm{\lambda}}^2 -2 \mu),  \qquad &|\bar{\bm{\lambda}}| > \sqrt{2\mu} \\[0.5em]
 0, \qquad &|\bar{\bm{\lambda}}| < \sqrt{2\mu}
 \end{cases}.
 \end{split}
 \label{eq:boson1pt}
\end{align}
This is the semiclassical density obtained from the volume of the filled Fermi sea at fixed $\bar{\bm\lambda}$.
\begin{figure}[t]
\begin{center}
\includegraphics[width=7cm]{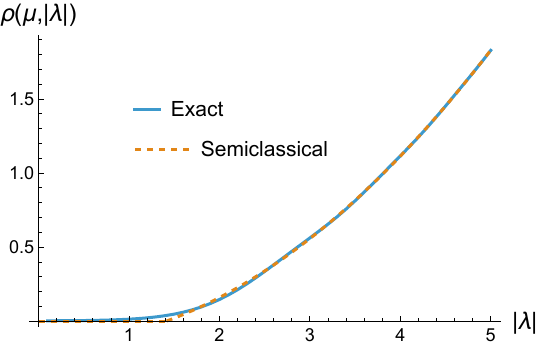}
\includegraphics[width=7cm]{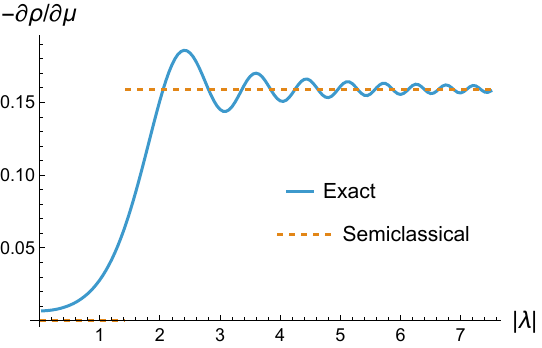}
\caption{The plot of the exact one-point function compared to the semiclassical one-point function.
We set $\mu =1$.
The left panel shows the density $\rho$, while the right panel shows its derivative with respect to the Fermi energy, $-\partial \rho / \partial \mu$.
}
\label{fig:MonePtCompare}
\end{center}
\end{figure}
Similarly, for the density operator localized also at fixed $\bar{\xi}$, we obtain
\begin{align}
\begin{split}
\rho[\phi = 0](\bar{\bm{\lambda}},\bar{\xi}) &= \int \frac{\mathrm d \bm{\lambda} \mathrm d \bm{p}}{(2\pi)^2} \theta \left (
 - \mu - \left( \frac{1}{2} \bm{p}^2 - \frac{1}{2} \bm{\lambda}^2 \right)\right)\delta^2(\bm{\lambda} - \bar{\bm{\lambda}}) \delta(\xi - \bar{\xi})   \\
 &=
\begin{cases}\displaystyle
 \frac{1}{2(2\pi)^2} (\bar{\bm{\lambda}}^2 -2 \mu),  \qquad &|\bar{\bm{\lambda}}| > \sqrt{2\mu} \\[0.5em]
 0, \qquad &|\bar{\bm{\lambda}}| < \sqrt{2\mu}
 \end{cases}.
  \end{split}
\end{align}
Integrating the above expression over $\bar{\xi}$ reproduces \eqref{eq:boson1pt}, as expected.

In Figure~\ref{fig:MonePtCompare}, we compare the semiclassical result \eqref{eq:boson1pt} with the exact one-point function computed from the Fermi liquid description in \eqref{0A1pt}. The two results agree well away from the classical turning point $ |\bar{\bm\lambda}|=\sqrt{2\mu}$.  Near the turning point, a discrepancy is expected: the semiclassical approximation treats the Fermi surface as sharp, whereas the exact expression includes the quantum smoothing of the edge of the Fermi sea.

\subsection{Eigenvalue density two-point function}

We next compute the two-point function of the eigenvalue density operator in the bosonized description and compare it with the exact Fermi liquid result discussed in Section~\ref{Mrectdensity}. The starting point is the weak-field expansion of the density operator
\begin{align}
\begin{split}
\rho[\phi](\bar{\bm{\lambda}}, \bar{\xi}) &= \braket{ f_0 - \{\phi, f_0\} + \frac{1}{2}\{ \phi, \{ \phi,f_0\}\}+ \cdots,\rho(\bar{\bm{\lambda}},\bar{\xi})}   \\
&= \braket{f_0,\rho(\bar{\bm{\lambda}},\bar{\xi})} - \braket{\{\phi, f_0\},\rho(\bar{\bm{\lambda}},\bar{\xi})} - \frac{1}{2} \braket{\{\phi, f_0\}, \{\phi , \rho(\bar{\bm{\lambda}},\bar{\xi})\}} + \cdots  .
\end{split}
\end{align}
Using $\{\phi, f_0\} = -\delta(h+ \mu) (\partial \phi / \partial \tau)$ and working to linear order in $\phi$, we obtain
\begin{align}
\begin{split}
\rho^{(1)}[\phi](\bar{\bm{\lambda}}, \bar{\xi}) 
&= \int\frac{\mathrm dh \mathrm d\tau \mathrm dm \mathrm d\varphi }{(2\pi)^2} \delta(h+\mu) \frac{\partial\phi}{\partial \tau} \delta^2(\bm{\lambda}- \bar{\bm{\lambda}}) \delta(\xi - \bar{\xi})  \\
&= \int\frac{\mathrm d\tau \mathrm dm \mathrm d\varphi }{(2\pi)^2} \frac{\partial \phi}{\partial \tau} \delta^2(\bm{\lambda}- \bar{\bm{\lambda}}) \delta(\xi - \bar{\xi})  \\
&=\frac{1}{(2\pi)^2} \frac{\partial \phi}{\partial \tau}(\bar{\bm{\lambda}},\bar{\xi}) \\
&=\frac{1}{(2\pi)^2} p_F(\bar{\bm{\lambda}},\bar{\xi})^2 R \phi(\bar{\bm{\lambda}},\bar{\xi}).
\end{split}
\end{align}
In the last line above, $R = (1/p_F^2)\partial/\partial \tau$ is the Reeb vector field  \eqref{eq:ReebV}.
Thus, at leading order in $\phi$, the density two-point function reduces to the two-point function of a chiral boson. Using the propagator \eqref{eq:MBtwoPt}, the leading contribution is, schematically,
\begin{align}
     \int \mathrm d\xi_i \mathrm d\xi_f  \left\langle\frac{\partial \phi}{\partial \tau_i} (\tau_i,m_i,\varphi_i)\frac{\partial \phi}{\partial \tau_f}(\tau_f,m_f,\varphi_f) \right\rangle
     &= \int \mathrm d\xi_i \mathrm d\xi_f \frac{\partial}{\partial \tau_i}\frac{\partial}{\partial \tau_f}F(\tau_i,\tau_f) \delta(\Delta m)\delta(\Delta \varphi ).
 \end{align}
To evaluate this expression as a correlation function of the eigenvalue densities, we insert delta functions fixing the two coordinate-space positions:
\begin{equation}
\int \mathrm d\xi_i \mathrm d^2\bm{\lambda}_i \mathrm d\xi_f  \mathrm d^2\bm{\lambda}_f \delta^2(\bm{\lambda}_i - {\bar{\bm{\lambda}}}_i)\delta^2(\bm{\lambda}_f - \bar{\bm{\lambda}}_f)  \frac{\partial}{\partial \tau_i}\frac{\partial}{\partial \tau_f}F(\tau_i,\tau_f) \delta(\Delta m)\delta(\Delta \varphi ). \label{eq:IntInTwo}
\end{equation}
We then change variables from the Fermi-surface coordinates
$(\bm\lambda_i,\xi_i), (\bm\lambda_f,\xi_f)$
to the corresponding action-angle variables $(\tau_i,m_i,\varphi_i), (\tau_f,m_f,\varphi_f)$.
Using the unit Jacobian relation for the Fermi-surface volume form, \eqref{eq:FSmeasure}, this gives
\begin{equation}
\int \mathrm d\tau_i \mathrm dm _i \mathrm d\varphi_i  \mathrm d\tau_f\mathrm dm _f \mathrm d\varphi_f \delta^2(\bm{\lambda}_i(\tau_i,m_i,\varphi_i) - \bar{\bm{\lambda}}_i)\delta^2(\bm{\lambda}_f(\tau_f,m_f,\varphi_f) - \bar{\bm{\lambda}}_f) \frac{\partial}{\partial \tau_i}\frac{\partial}{\partial \tau_f} F(\tau_i,\tau_f) \delta(\Delta m)\delta(\Delta \varphi ).
\end{equation}
The delta functions $\delta(m_i-m_f)\delta(\varphi_i-\varphi_f)$ impose the condition that the two insertions lie on a common classical trajectory labeled by the conserved quantities $(m,\varphi)$.
After carrying out the integrals over $m_i,\varphi_i$, and writing $m_f\to m$, $\varphi_f\to\varphi$, we find that the above integral becomes
\begin{equation}
\int \mathrm d\tau_i  \mathrm d\tau_f \mathrm dm \mathrm d\varphi \,\delta^2(\bm{\lambda}_i(\tau_i,m ,\varphi) - \bar{\bm{\lambda}}_i)\delta^2(\bm{\lambda}_f(\tau_f,m,\varphi) - \bar{\bm{\lambda}}_f)  \frac{\partial}{\partial \tau_i}\frac{\partial}{\partial \tau_f}F(\tau_i,\tau_f).
\end{equation}
The remaining delta functions produce the Jacobian associated with the change of variables \begin{align}
(\tau_i,\tau_f,m,\varphi) \longleftrightarrow (\bm\lambda_i,\bm\lambda_f).
\end{align}
Geometrically, both sets of variables parametrize two points on the same classical trajectory. Equivalently, the inverse map may be constructed using the Hamilton--Jacobi momenta
\begin{align}
\bm{p}_i(\boldsymbol\lambda_i,\boldsymbol\lambda_f;-\mu), \qquad \bm{p}_f(\boldsymbol\lambda_i,\boldsymbol\lambda_f;-\mu).
\end{align}
The inverse map should be constructed using $\bm{\lambda}_i, \bm{p}_i(\bm{\lambda}_i,\bm{\lambda}_f; -\mu)$ and  $\bm{\lambda}_f, \bm{p}_f(\bm{\lambda}_i,\bm{\lambda}_f; -\mu)$.
Since the functions
\begin{align}
\bm{\lambda}_i(\tau_i,m,\varphi), \qquad \bm{\lambda}_f(\tau_f,m,\varphi)
\end{align} are known explicitly in terms of action-angle coordinates, the Jacobian can be computed directly. Thus the bosonized two-point function becomes
\begin{align}
\begin{split}
&\braket{\rho^{(1)}[\phi] (\bar{\bm{\lambda}}_f) \rho^{(1)}[\phi] (\bar{\bm{\lambda}}_i) }   \\
& = \frac{1}{(2\pi)^4}\int \mathrm d\tau_i  \mathrm d\tau_f \mathrm dm \mathrm d\varphi \,\delta^2(\bm{\lambda}_i(\tau_i,m ,\varphi) - \bar{\bm{\lambda}}_i)\delta^2(\bm{\lambda}_f(\tau_f,m,\varphi) - \bar{\bm{\lambda}}_f) \frac{\partial}{\partial \tau_i}\frac{\partial}{\partial \tau_f}F(\tau_i,\tau_f)   \\
&  =\frac{1}{(2\pi)^4} \sum_{s = \pm}\left[\left.\frac{\partial (\bm{\lambda}_i, \bm{\lambda}_f)}{ \partial (\tau_i,\tau_f, m ,\varphi)}  \right|_{(\tau_i,\tau_f,m,\varphi) = (s\bar{\tau}_i, s\bar{\tau}_f, \bar{m},s\bar{\varphi})} \right]^{-1}
\frac{\partial}{\partial \tau_i}\frac{\partial}{\partial \tau_f}F(\bar{\tau}_i^s,\bar{\tau}_f^s)  \\
&= - \bigg[\frac{\partial (\bm{\lambda}_i, \bm{\lambda}_f)}{ \partial (\tau_i,\tau_f, m ,\varphi)}  \bigg]^{-1}  \frac{2\pi}{(2\pi)^4} \frac{\partial}{\partial \tau_i} \frac{\partial}{\partial \tau_f}\Big[\log (\Delta \tau - i \Delta x )+ \log (\Delta \tau + i \Delta x)\Big]  \\
&= - \frac{1}{8\pi^3} \bigg[\frac{\partial (\bm{\lambda}_i, \bm{\lambda}_f)}{ \partial (\tau_i,\tau_f, m ,\varphi)}  \bigg]^{-1} \bigg[ \frac{1}{(\Delta \tau - i \Delta x)^2}  + \frac{1}{(\Delta \tau + i \Delta x)^2}\bigg]  \\
&= - \frac{1}{4\pi^3} \bigg[\frac{\partial (\bm{\lambda}_i, \bm{\lambda}_f)}{ \partial (\tau_i,\tau_f, m ,\varphi)}  \bigg]^{-1}  \frac{(\Delta \tau)^2 - (\Delta x)^2}{((\Delta \tau)^2  + (\Delta x)^2)^2},
\end{split}
\label{eq:boson2ptfull}
\end{align}
with $\Delta x = x_f-x_i$, etc. The two terms in the sum over $s=\pm$ correspond to the two possible orientations of the classical trajectory connecting the two insertion points. The above Jacobian determinant is
\begin{align}
&\frac{\partial (\bm{\lambda}_i, \bm{\lambda}_f)}{ \partial (\tau_i,\tau_f, m ,\varphi)}  \notag \\
&=  \sinh (\tau_f - \tau_i) \Big[ -\mu \cosh(\tau_f - \tau_i)  + \mu\cosh(\tau_f + \tau_i) \cosh(2\varphi) - m \sinh(\tau_f + \tau_i) \sinh(2\varphi)\Big],
\end{align}
where in the second line we set $h=-\mu$ on the Fermi surface. For small $\tau_f-\tau_i$, this reduces to
\begin{align}
\frac{\partial (\bm{\lambda}_i, \bm{\lambda}_f)}{ \partial (\tau_i,\tau_f, m ,\varphi)} &\approx (\tau_f - \tau_i) \Big[ h  - h\cosh(2 \tau_i) \cosh(2\varphi) - m \sinh(2\tau_i) \sinh(2\varphi)\Big] \notag \\
&\approx (\tau_f - \tau_i) p_F(\tau_i,m,\varphi)p_F(\tau_f,m,\varphi) .
\end{align}
\begin{figure}[t]
\begin{center}
\includegraphics[width=8.0cm]{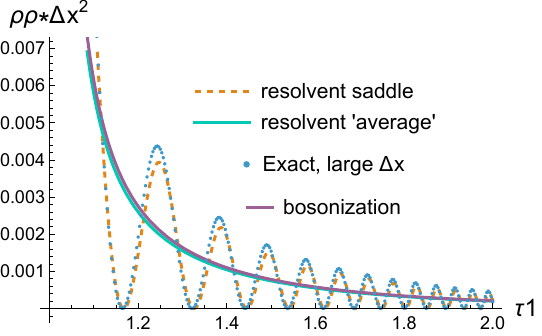}
\end{center}
\caption{
Comparison of the density two-point function obtained from the exact fermionic description, the saddle-point approximation, and bosonization in the $\Delta x\to\infty$ limit.
The correlators are shown as functions of $\tau_1$ for
$\mu=1$, $\tau_2=0$, $m=0.1$, and $\varphi=1$, where $(\tau_1,\tau_2)=(\tau_f,\tau_i)$.
\label{fig:ExactVsBosonization2pt}}
\end{figure}
We can now compare the bosonized two-point function with the exact Fermi liquid result. For this purpose, we parametrize the two insertion points in coordinate space as
\begin{align}
\bar{\bm\lambda}_i=\bm{\lambda}(\tau_i,m,\varphi), \qquad \bar{\bm\lambda}_f= \bm{\lambda}(\tau_f,m,\varphi).
\end{align}
This allows both the exact and bosonized two-point functions to be expressed as functions of the same variables $(\tau_i,\tau_f,m,\varphi)$.  The comparison is shown in Figure~\ref{fig:ExactVsBosonization2pt}. At large Euclidean time separation, \eqref{eq:boson2ptfull} becomes
\begin{align}
&\braket{\rho^{(1)}[\phi] (\bm{\lambda}_f) \rho^{(1)}[\phi] (\bm{\lambda}_i) } \notag \\
 &\approx \frac{1}{4\pi^3} \frac{1}{(\Delta x)^2}  \bigg[\frac{\partial (\bm{\lambda}_i, \bm{\lambda}_f)}{ \partial (\tau_i,\tau_f, m ,\varphi)} \bigg]^{-1} \notag \\
 &= \frac{1}{4\pi^3} \frac{1}{(\Delta x)^2} \frac{1}{\sinh (\tau_f - \tau_i) \big[ -\mu \cosh(\tau_f - \tau_i)  + \mu\cosh(\tau_f + \tau_i) \cosh(2\varphi) - m \sinh(\tau_f + \tau_i) \sinh(2\varphi) \big] }.
\end{align}
This expression should be compared with the exact Fermi liquid result in \eqref{eq:twoptSemiAve}, obtained in the $\Delta x\to\infty$ limit using the stationary-phase approximation for the resolvent and after dropping the rapidly oscillating term.
The ratio between the bosonized result and the averaged exact result is
\begin{align}
\begin{split}
 &\lim_{\Delta x \to \infty}\frac{\braket{\rho^{(1)}[\phi] (\bm{\lambda}_f) \rho^{(1)}[\phi] (\bm{\lambda}_i) }}{\braket{\rho (x_f,\bm{\lambda}_f) \rho (x_i,\bm{\lambda}_i) }}   \\
 & \approx \frac{\sqrt{ -\mu^2  + \big[ \mu \cosh (\tau_i + \tau_f) \cosh (2\varphi) - m \sinh (\tau_i + \tau_f) \sinh (2\varphi)\big]^2}}{ -\mu \cosh(\tau_f - \tau_i)  + \mu\cosh(\tau_f + \tau_i) \cosh(2\varphi) - m \sinh(\tau_f + \tau_i) \sinh(2\varphi)  }.
\end{split}
\end{align}
In the asymptotic regime $\tau_i+\tau_f\to\infty$, this ratio approaches one.
This is precisely the regime in which the saddle-point approximation to the resolvent \eqref{eq:RNMsaddle1} is reliable.
Thus, in the common regime of validity, the bosonized two-point function agrees with the exact Fermi liquid result.

\subsection{Eigenvalue density three-point function}
\label{3ptfn}
In this section, we compute the density three-point function using bosonization.
The calculation is performed at tree level within the bosonized effective theory. Compared to the Das--Jevicki--Sakita collective field theory for the $c=1$ noncritical string \cite{Das:1990kaa,Jevicki:1979mb,Sakita:1979gs}, the present theory exhibits two additional features.
First, the WZW term receives a cubic-order correction.
Second, the density operator acquires nonlinear corrections.
Both contributions enter the connected three-point function and are essential for obtaining the correct result.

In the following sections, we will mainly focus on the causal structure of the three-point function, in particular showing it agrees with the causal structure of the corresponding Fermi liquid three-point function, and we do not explicitly perform the integrals needed to evaluate the three-point function exactly. Nevertheless, we stress that it is possible (albeit cumbersome) to do an exact computation of the three-point function thanks to the underlying integrability of noncritical M-theory.

\subsubsection{Quadratic correction to the density operator }
In this section, we derive the $O(\phi^2)$ correction to the eigenvalue density operator of the bosonized theory.
We first recall that the density operator is expanded as
\begin{align}
\rho[\phi](\bar{\bm{\lambda}}, \bar{\xi}) &= \braket{ f_0 - \{\phi, f_0\} + \frac{1}{2}\{ \phi, \{ \phi,f_0\}\},\rho(\bar{\bm{\lambda}},\bar{\xi})} + \cdots \notag \\
&= \braket{f_0,\rho(\bar{\bm{\lambda}},\bar{\xi})} - \braket{\{\phi, f_0\},\rho(\bar{\bm{\lambda}},\bar{\xi})} - \frac{1}{2} \braket{\{\phi, f_0\}, \{\phi , \rho(\bar{\bm{\lambda}},\bar{\xi})\}} + \cdots.
\end{align}
The quadratic term in $\phi$ is
\begin{align}
\begin{split}
\rho^{(2)}[\phi](\bar{\bm\lambda}, \bar{\xi})
&= - \frac{1}{2} \braket{\{\phi, f_0\}, \{\phi , \rho(\bar{\bm{\lambda}},\bar{\xi})\}}  \\
&= \frac{1}{2}\int \frac{\mathrm d^2 \boldsymbol \lambda  p \mathrm dp \mathrm d\xi }{(2\pi)^2}
 \frac{1}{p} \delta(p - p_F(\bm\lambda,\xi))\frac{\partial \phi}{\partial \tau} \Bigg[ \bm{t}_{\xi} \cdot \partial_{\bm \lambda} \phi\frac{1}{p}\frac{\partial }{\partial \xi} \left(\delta^2(\bm{\lambda}- \bar{\bm{\lambda}})\delta(\xi - \bar{\xi}) \right)  \\
 & \qquad \qquad -
 \frac{1}{p} \frac{\partial \phi}{\partial \xi} \bm{t}_{\xi} \cdot  \partial_{\bm \lambda}\left(\delta^2(\bm\lambda- \bar{\bm{\lambda}})\delta(\xi - \bar{\xi}) \right) \Bigg]  \\
 &=  \frac{1}{2 (2\pi)^2} \partial_{\bar{\bm{\lambda}}} \cdot \left( \bm{t}_{\bar{\xi}} \frac{\partial \phi}{\partial \bar{\tau}}\frac{1}{p_F(\bm{\bar \lambda},\bar \xi)} \frac{\partial \phi}{\partial \bar{\xi}} \right)- \frac{1}{2(2\pi)^2} \frac{\partial}{\partial \bar{\xi}}\left( \frac{1}{p_F(\bm{\bar \lambda},\bar \xi)} \frac{\partial \phi}{\partial \bar{\tau}} \bm{t}_{\bar{\xi}} \cdot \partial_{\bar{\bm{\lambda}}} \phi   \right),
 \end{split}
\end{align}
where above $\boldsymbol t_\xi$ is defined in the doubly-polar coordinate system. We can present the above expression in a more covariant form as follows:\footnote{
The density is understood as a 3-form on the Fermi surface and the correction terms at second order are locally exact forms:
\begin{align*}
\rho^{(2)} [\phi] & \equiv \rho^{(2)} [\phi](\bm{\lambda},\xi) \mathrm d\lambda_1 \wedge \mathrm d\lambda_2 \wedge d\xi = \mathrm d \alpha ,~~~~
\alpha = \frac{1}{2(2\pi)^2}
p_F^j\partial_\nu\lambda_j\,
R^\sigma\partial_\sigma\phi\,
\partial_\rho\phi\,
\mathrm dw^\nu\wedge\mathrm dw^\rho
.
\end{align*}
}
\begin{align}
\begin{split}
\rho^{(2)}[\phi](\bar{\bm{\lambda}}, \bar{\xi})
&=  \left.
\frac{1}{2(2\pi)^2}
\partial_\mu
\left[
 \epsilon^{\mu\nu\rho}
 \frac{p_F^j\partial_\nu\lambda_j}{p_F^2}
 \frac{\partial\phi}{\partial\tau}
 \partial_\rho\phi
\right]
\right|_{w=\bar w}
\\
&= \left.
\frac{1}{2(2\pi)^2}
\partial_\mu
\left[
 \epsilon^{\mu\nu\rho}
 p_F^j\partial_\nu\lambda_j\,
 R\phi\,
 \partial_\rho\phi
\right]
\right|_{w=\bar w},
\end{split}
\end{align}
where $\partial_\mu = \partial / \partial w^\mu$ with $w^\mu$ denoting action angle coordinates, and where $R = R^{\sigma}\partial_{\sigma} = (1/p_F^2)\partial/\partial \tau$ is the Reeb vector field \eqref{eq:ReebV}.

\subsubsection{Structure of three-point functions}
In the bosonized theory, we find that there are two contributions to the eigenvalue density three-point functions; these two contributions are represented by the two $ O(1/p_F^2)$ diagrams in Figure \ref{hgaugegraph}.
The first contribution combines the linearized eigenvalue densities with the cubic interaction terms in the bosonic action:
\begin{align}
\left\langle \rho^{(1)}[\phi](\bm{\lambda}_1)\rho^{(1)}[\phi](\bm{\lambda}_2)\rho^{(1)}[\phi](\bm{\lambda}_3) i S^{(3)}[\phi]\right\rangle, ~~~~ S^{(3)} = S^{(3)}_{\rm WZW} + S^{(3)}_H.
\end{align}
Above, $S_{\rm WZW}^{(3)}$ is given in \eqref{eq:WZW3} and $S_H^{(3)}$  is given in \eqref{eq:H3}.
Both the WZW term $S_{\rm WZW}^{(3)}$ and the Hamiltonian term $S_{H}^{(3)}$ are given as the integral of local functions  on the Fermi surface, of the form
\begin{align}
\begin{split}
&\int \mathrm d\xi_1 \mathrm d \xi_2 \mathrm d\xi_3 \left\langle \left( \prod_{a=1}^3 \frac{\partial \phi}{\partial \tau_a}(\bm{\lambda}_a,\xi_a) \right) \int \frac{\mathrm d \tau \mathrm d m \mathrm d\varphi}{(2\pi)^2}  A_3^{\mu\nu\rho}(\tau,m,\varphi)\partial_{\mu}\phi\partial_{\nu}\phi\partial_{\rho}\phi\right\rangle  \\
&= -\int \mathrm d\xi_1 \mathrm d\xi_2 \mathrm d\xi_3 \frac{\mathrm d \tau \mathrm d m \mathrm d\varphi}{(2\pi)^2} \partial_{\mu}\partial_{\nu}\partial_{\rho}A_3^{\mu\nu\rho}(\tau,m,\varphi)
\prod_{a=1}^3 \left\langle \frac{\partial \phi}{\partial \tau_a} (\boldsymbol \lambda_a , \xi_a) \phi(\tau, \varphi,m ) \right\rangle. 
\end{split}
\end{align}
Above, $A_3^{\mu\nu\rho}$ are coefficients obtained from the cubic interaction terms in \eqref{eq:WZW3} and \eqref{eq:H3}. The above expression is schematic: the time coordinate $t$ and the vertex integral $\int \mathrm dt$ are suppressed throughout this section, and for $S^{(3)}_{\rm WZW}$ one leg is $\partial_\mu \dot\phi$, so an extra $\partial_t$ acts on the propagator $F$ in \eqref{eq:S3terms}.
Then using the two-point function for bosons \eqref{eq:MBtwoPt}, the three-point functions are given by
\begin{align}
\begin{split}
&
\left\langle \rho^{(1)}[\phi](\bm{\lambda}_1)\rho^{(1)}[\phi](\bm{\lambda}_2)\rho^{(1)}[\phi] (\bm{\lambda}_3)i S^{(3)}[\phi] \right\rangle\\
& = -\int \frac{\mathrm d\xi_1 \mathrm d\xi_2 \mathrm d\xi_3}{(2\pi)^6} \mathrm d m \mathrm d\varphi \int\frac{\mathrm d \tau }{(2\pi)^2} \partial_{\mu}\partial_{\nu}\partial_{\rho}A_3^{\mu\nu\rho}(\tau,m,\varphi) \prod_{a=1}^3 \frac{\partial F}{\partial \tau_a}(\tau_a,\tau) \delta(m_a-m)\delta(\varphi_a - \varphi) . \label{eq:S3terms}
\end{split}
\end{align}
Next, we consider the contribution from the second-order correction to the density operator.
Since each of the three density insertions can contribute a $\rho^{(2)}[\phi]$ term, the total contribution of this type is
\begin{align}
\begin{split}
&\braket{\rho^{(1)}[\phi](\bm{\lambda}_1)\rho^{(1)}[\phi](\bm{\lambda}_2)\rho^{(2)}[\phi] (\bm{\lambda}_3)} \\
+ &\braket{\rho^{(1)}[\phi](\bm{\lambda}_1)\rho^{(2)}[\phi](\bm{\lambda}_2)\rho^{(1)} [\phi](\bm{\lambda}_3)}\\
+ &\braket{\rho^{(2)}[\phi](\bm{\lambda}_1)\rho^{(1)}[\phi](\bm{\lambda}_2)\rho^{(1)}[\phi] (\bm{\lambda}_3)}.
\end{split}
\end{align}
For example, the first term is
\begin{align}
\begin{split}
    &\braket{\rho^{(1)}[\phi](\bm{\lambda}_1)\rho^{(1)}[\phi](\bm{\lambda}_2)\rho^{(2)}[\phi](\bm{\lambda}_3)}  \\
& = \frac{1}{2} \int \left( \prod_{a=1}^3 \frac{\mathrm d\xi_a}{(2\pi)^2}\right) \left\langle\frac{\partial\phi}{\partial \tau_1}(\bm{\lambda}_1,\xi_1)\frac{\partial\phi}{\partial \tau_2}(\bm{\lambda}_2,\xi_2) \partial_{\mu} \Bigl[\epsilon^{\mu \nu\rho} \partial_{\nu}p_F(\bm{\lambda}_3,\xi_3) R^{\sigma}\partial_{\sigma}^3\phi(\bm{\lambda}_3,\xi_3) \partial_{\rho}\phi(\bm{\lambda}_3,\xi_3)\Bigr]\right\rangle  \\
&= \frac{1}{2} \int  \left( \prod_{a=1}^3 \frac{\mathrm d\xi_a}{(2\pi)^2}\right)  \partial_{\mu}^3 \bigg(  \epsilon^{\mu \nu\rho} \partial_{\nu}^3p_F(\bm{\lambda}_3,\xi_3) R^{\sigma} \partial_{\sigma}^3\left\langle\frac{\partial \phi}{\partial \tau_1}(\bm{\lambda}_1,\xi_1) \phi(\bm{\lambda}_3,\xi_3)\right\rangle \\
&~~~~~~~~~~~~~~~~~~~~~~~~\times \partial_\rho^3\left\langle\frac{\partial \phi}{\partial \tau_2}(\bm{\lambda}_2,\xi_2) \phi(\bm{\lambda}_3,\xi_3)\right\rangle \bigg)  + (1\leftrightarrow 2)  \\
&=\frac{1}{2} \int  \left( \prod_{a=1}^3 \frac{\mathrm d\xi_a}{(2\pi)^2}\right)  \partial_{\mu}^3 \bigg(  \epsilon^{\mu \nu\rho} \partial_{\nu}^3p_F(\bm{\lambda}_3,\xi_3) R^{\sigma} \partial_{\sigma}^3 \Big[ \partial_{\tau_1}F(\tau_1,\tau_3) \delta(m_1-m_3) \delta(\varphi_1-\varphi_3) \Big]   \\
& \qquad \times \partial_\rho ^3\Big[ \partial_{\tau_2}F(\tau_2,\tau_3) \delta(m_2-m_3) \delta(\varphi_2-\varphi_3) \Big]  \bigg) + (1\leftrightarrow 2).
\end{split}
\end{align}
In the above expression, $\partial^3$ denotes a derivative with respect to the third insertion point $(\bm{\lambda}_3, \xi_3)$.
Using the delta function identity $f(x) \delta'(x-x_0) = - f'(x)\delta(x-x_0)$, we obtain the integral
\begin{align}
&\braket{\rho^{(1)}[\phi](\bm{\lambda}_1)\rho^{(1)}[\phi](\bm{\lambda}_2)\rho^{(2)}[\phi] (\bm{\lambda}_3)} \notag  \\
&=\int \mathrm d\xi_1 \mathrm d\xi_2 \mathrm d\xi_3 B^{112}(\bm{\lambda}_1,\xi_1,\bm{\lambda}_2,\xi_2,\bm{\lambda}_3,\xi_3) \delta(m_1-m_3) \delta(\varphi_1-\varphi_3)\delta(m_2-m_3) \delta(\varphi_2-\varphi_3).
\end{align}
Integrating in the coordinates $m,\varphi$, we have
\begin{align}
&\braket{\rho^{(1)}[\phi](\bm{\lambda}_1)\rho^{(1)}[\phi](\bm{\lambda}_2)\rho^{(2)}[\phi] (\bm{\lambda}_3)} \notag \\
&=
\int \mathrm  d\xi_1  \mathrm d\xi_2 \mathrm   d\xi_3 \mathrm dm \mathrm d\varphi B^{112}(\bm{\lambda}_1,\xi_1,\bm{\lambda}_2,\xi_2,\bm{\lambda}_3,\xi_3) \prod _{a=1}^3 \delta(m_a -m) \delta(\varphi_a -\varphi). \label{eq:rho2term}
\end{align}
In this form, we can see that the integrals over $\xi_1,\xi_2,\xi_3,m,\varphi$ and the delta functions $\prod _{a=1}^3 \delta(m_a -m) \delta(\varphi_a -\varphi) $ take the same form as the contribution from $S^{(3)}_H,S^{(3)}_{\rm WZW}$. The terms
        $\braket{\rho^{(1)}[\phi]\rho^{(2)}[\phi]\rho^{(1)}[\phi]}$ and $\braket{\rho^{(2)}[\phi]\rho^{(1)}[\phi]\rho^{(1)}[\phi]}$ are evaluated in the same manner and give the same type of contributions, namely
\begin{align}
\begin{split}
&\braket{\rho^{(1)}[\phi](\bm{\lambda}_1)\rho^{(2)}[\phi](\bm{\lambda}_2)\rho^{(1)}[\phi] (\bm{\lambda}_3)}  \\
&=
\int \mathrm d\xi_1  \mathrm d\xi_2 \mathrm  d\xi_3  \mathrm dm \mathrm d\varphi\, B^{121}(\bm{\lambda}_1,\xi_1,\bm{\lambda}_2,\xi_2,\bm{\lambda}_3,\xi_3) \prod _{a=1}^3 \delta(m_a -m) \delta(\varphi_a -\varphi)   \\
&\braket{\rho^{(2)}[\phi](\bm{\lambda}_1)\rho^{(1)}[\phi](\bm{\lambda}_2)\rho^{(1)}[\phi] (\bm{\lambda}_3)}  \\
&= \int\mathrm  d\xi_1 \mathrm d\xi_2 \mathrm d\xi_3 \mathrm dm \mathrm d\varphi \, B^{211}(\bm{\lambda}_1,\xi_1,\bm{\lambda}_2,\xi_2,\bm{\lambda}_3,\xi_3) \prod _{a=1}^3 \delta(m_a -m) \delta(\varphi_a -\varphi).
\end{split}
\end{align}
To evaluate the above integrals, one needs explicit expressions for
$A_3^{\mu\nu\rho}(\tau,m,\varphi)$,
$B^{112}$,
$B^{121}$,
and $B^{211}$.
Deriving these coefficients in closed form, although straightforward, is beyond the scope of the present work and will be left for future investigation. Nevertheless, the integrals over
$\xi_1,\xi_2,\xi_3,m$, and $\varphi$
can be carried out using the delta functions.
In the following, we focus on the remaining delta function constraints and their associated Jacobian factors, and compare them with the exact three-point function obtained from the Fermi liquid description.

\subsubsection{Integrals and causal structure}
To evaluate the integrals 
\begin{align}
\int \mathrm d\xi_1\mathrm d\xi_2\mathrm d\xi_3 \mathrm dm' \mathrm d\varphi' \prod_{a=1}^3 \delta(m_a(\bm{\lambda}_a,\xi_a)-m') \delta(\varphi_a(\bm{\lambda}_a,\xi_a)-\varphi'),
\end{align}
in \eqref{eq:S3terms} and \eqref{eq:rho2term},
it is useful to introduce a parametrization adapted to the classical trajectories.
The three insertion points $\bm{\lambda}_1,\bm{\lambda}_2,\bm{\lambda}_3$,  which together define six coordinates, can be parametrized by $(\tau_1,\tau_2,\tau_3,m,\varphi,s)$ as
\begin{align}
\begin{split}
\bm{\lambda}_1 &= \bm{\lambda} (\tau_1,m,\varphi)  \\
\bm{\lambda}_2 &= \bm{\lambda} (\tau_2,m,\varphi)  \\
\bm{\lambda}_3 &= \bm{\lambda} (\tau_3,m,\varphi) + s  \bm{n} (\tau_3,m,\varphi),
\end{split}
\end{align}
where
\begin{align}
\begin{split}
&\bm{\lambda} (\tau,m,\varphi) =
\begin{pmatrix}
\sqrt{\mu - m} \cosh (\tau + \varphi) \\
\sqrt{\mu + m} \cosh (\tau - \varphi)
\end{pmatrix},  \\
&\bm{n}(\tau,m,\varphi) =
\begin{pmatrix}
\lambda_2(\tau,m,\varphi) \\
-\lambda_1(\tau,m,\varphi)
\end{pmatrix} =
\begin{pmatrix}
\sqrt{\mu + m} \cosh (\tau - \varphi) \\
-\sqrt{\mu - m} \cosh (\tau + \varphi)
\end{pmatrix}.
\end{split}
\end{align}
Here $\bm{\lambda}(\tau,m,\varphi)$ gives the position at time $\tau$ on the classical trajectory labeled by the conserved quantities $m$ and $\varphi$. The parameter $s$ measures the deviation of the third insertion point $\bm{\lambda}_3$ from this classical trajectory. This is the same parametrization as the one used in \eqref{eq:sixparam3pt} to analyze the three-point function in the Fermi liquid description.
Using the delta function identity $\delta(x-y)f(x) = \delta(x-y)f(y)$, we have
\begin{align}
\begin{split}
&\prod_{a=1}^3 \delta(m_a-m') \delta(\varphi_a-\varphi')\\
&= \delta(m_1-m_2)\delta(\varphi_1-\varphi_2) \delta(m_1-m')\delta(\varphi_1-\varphi')  \delta(m_3-m_1)\delta(\varphi_3-\varphi_1).
\end{split}
\end{align}
The integrals over $m'$ and $\varphi'$ can then be performed straightforwardly, leaving the following four delta functions:
\begin{align}
 \delta(m_1-m_2)\delta(\varphi_1-\varphi_2)  \delta(m_3-m_1)\delta(\varphi_3-\varphi_1).
\end{align}
To perform the integral $\int \mathrm d\xi_1 \mathrm d\xi_2 \,\delta(m_1-m_2)\delta(\varphi_1-\varphi_2)$,
we apply the same integrating-in technique \eqref{eq:IntInTwo} used in the two-point function. This yields the same Jacobian factor as in the two-point-function computation:
\begin{align}
\begin{split}
&\int \mathrm d\xi_1 \mathrm d\xi_2 \delta(m_1-m_2)\delta(\varphi_1-\varphi_2)  \delta(m_3-m_1)\delta(\varphi_3-\varphi_1)\\
&=\int \mathrm d ^2 \bm{\lambda}_1'\mathrm d ^2 \bm{\lambda}_2' \mathrm d\xi_1 \mathrm d\xi_2 \delta^2(\bm{\lambda}_1' -\bm{\lambda}_1)\delta^2(\bm{\lambda}_2' -\bm{\lambda}_2)\delta(m_1-m_2)\delta(\varphi_1-\varphi_2)  \delta(m_3-m_1)\delta(\varphi_3-\varphi_1)\\
&= \left[ \frac{\partial (\bm{\lambda}_1, \bm{\lambda}_2)}{ \partial (\tau_1 ,\tau_2, m,\varphi)} \right]^{-1} \delta(m_3-m)\delta(\varphi_3-\varphi).
\end{split}
\end{align}
We are therefore left with two delta-function constraints for a single integration variable:
\begin{align}
\int \mathrm d\xi_3\, \delta(m_3-m)\delta(\varphi_3-\varphi).
\end{align}
Integrating in $\bm{\lambda}_3'$,
\begin{align}
\begin{split}
&\int \mathrm d\xi_3 \mathrm d^2\bm{\lambda}_3'\, \delta^2(\bm{\lambda}_3' -\bm{\lambda}_3) \delta(m_3(\bm{\lambda}_3',\xi_3)-m)\delta(\varphi_3(\bm{\lambda}_3',\xi_3)-\varphi) \\
&= \int \mathrm d\tau_3 ' \mathrm dm_3 \mathrm d\varphi_3 \delta^2(\bm{\lambda}_3'(\tau_3',m_3,\varphi_3) -\bm{\lambda}_3) \delta(m_3- m)\delta(\varphi_3-\varphi) \\
&= \int \mathrm d\tau_3 '  \delta^2(\bm{\lambda}_3'(\tau_3',m,\varphi) -\bm{\lambda}_3) \\
&= \int \mathrm d\tau_3 '  \delta^2(\bm{\lambda}_3'(\tau_3',m,\varphi) -\bm{\lambda}(\tau_3,m,\varphi) - s \bm{n}(\tau_3,m,\varphi)) \\
&= [\bm{p}_F(\tau_3,m,\varphi)\cdot \bm{\lambda}(\tau_3,m,\varphi)]^{-1} \delta(s).
\end{split}
\end{align}
Finally, we obtain
\begin{align}
&\int \mathrm d\xi_1\mathrm d\xi_2\mathrm d\xi_3 \mathrm dm' \mathrm d\varphi' \prod_{a=1}^3 \delta(m_a(\bm{\lambda}_a,\xi_a)-m') \delta(\varphi_a(\bm{\lambda}_a,\xi_a)-\varphi') \notag \\
&=\left[ \frac{\partial (\bm{\lambda}_1, \bm{\lambda}_2)}{ \partial (\tau_1 ,\tau_2, m,\varphi)} \big(\bm{p}_F(\tau_3,m,\varphi)\cdot \bm{\lambda}(\tau_3,m,\varphi)\big)\right]^{-1} \delta(s).
\end{align}
This delta function $\delta(s)$ implies that the three-point function vanishes unless all three points
$\bm{\lambda}_1$, $\bm{\lambda}_2$, and $\bm{\lambda}_3$ lie on the same classical trajectory.
Thus, the bosonized description reproduces the same delta-function support as the semiclassical Fermi liquid result in \eqref{eq:Caus3Fermi} in Section \ref{sec:3ptFermi}.
In the Fermi liquid description, when $\bm{\lambda}_1,\bm{\lambda}_2,\bm{\lambda}_3$ do not lie on the same classical trajectory, the oscillatory phases in
\begin{align}
 \sin(W(\bm{\lambda}_1,\bm{\lambda}_2; -\mu)) \sin(W(\bm{\lambda}_2,\bm{\lambda}_3;-\mu))\sin(W(\bm{\lambda}_1,\bm{\lambda}_3;-\mu))
\end{align}
do not cancel and average to zero.
The bosonized description incorporates this support condition from the outset.

\subsection{Light-cone density $n$-point functions}
\label{sec:LCdensitycorr}

In this section, we study the correlation functions of the density operators
    \begin{align}
        \rho(\bar \tau_\sigma, \bar \varphi_\sigma,\bar \ell) = \delta(\tau_\sigma - \bar \tau_\sigma) \delta(\varphi_\sigma - \bar \varphi_\sigma) \delta(\bar \ell -\ell),
    \end{align}
summed over the angular momentum $\ell$:
    \begin{align}
\rho(\bar\tau_\sigma, \bar\varphi_\sigma) = \delta(\tau_\sigma - \bar \tau_\sigma) \delta(\varphi_\sigma - \bar \varphi_\sigma) \equiv \int \mathrm d\bar \ell\,   \rho(\bar \tau_\sigma, \bar \varphi_\sigma,\bar \ell) .
    \end{align}
We show below that the correlation functions of the above density operators match with the correlation functions of the fermion bilinear operators (\ref{LCFermirho}) to all orders in the perturbative expansion of the bosonized theory.

In order to analyze the correlators of these density operators, it is most convenient to work in $h$-gauge, i.e., to impose the gauge-fixing condition
    \begin{align}
    \label{LChgauge}
        \phi(t, \tau_\sigma, h_\sigma = -\sigma \mu, \varphi_\sigma, \ell) \equiv \phi(t, \tau_\sigma, \varphi_\sigma, \ell)
    \end{align}
on the Fermi surface $h = \sigma h_\sigma = -\mu$. In $h$-gauge, it is convenient to represent the effective action as  (\ref{Skexp}):
	\begin{align}
	\begin{split}
    \label{eq:hgaugeaction}
		S &= S^{(0)} + \sum_{k=2}^{\infty} S^{(k)},~~~~~
		S^{(k)}[\phi] 
         =- \frac{\sigma}{k!} \int \mathrm dt \, \frac{\mathrm d \tau_\sigma \mathrm d\varphi_\sigma  \mathrm d\ell}{(2\pi)^2} \, \frac{\partial \phi}{\partial \tau_\sigma} \, \mathrm{ad}_\phi^{k-2} \left(\frac{\partial \phi}{\partial t} + \sigma  \frac{\partial \phi}{\partial \tau_\sigma}\right)
	\end{split}
	\end{align}
where $\mathrm{ad}_\phi \equiv \{ \phi, \cdot\}$. By writing the action in this way, we can see clearly that each term in $S^{(k)}$ contains $ k-2$ derivatives with respect to $\ell$ and $ k$ copies of the field $\phi$.
Similarly, using (\ref{rhokexp}), we can express the density operators in $h$-gauge as
	\begin{align}
		\begin{split}
        \label{rhoknew}
			\rho[\phi](\tau_\sigma, \varphi_\sigma)  &=\rho^{(0)}[\phi](\tau_\sigma, \varphi_\sigma) +\sum_{k=1}^\infty \rho^{(k)}[\phi](\tau_\sigma, \varphi_\sigma)\\
            \rho^{(k)}[\phi](\tau_\sigma, \varphi_\sigma)  &=- \frac{\sigma}{k!} \int \frac{\mathrm d\ell}{(2\pi)^2} \mathrm{ad}_\phi^{k-1} \frac{\partial \phi}{\partial \tau_\sigma} .
		\end{split}
	\end{align}
The operator $\rho^{(k)}[\phi]$ contains $k-1$ derivatives with respect to $\ell$ and  $k$ copies of $\phi$.
Notice that the action of $\text{ad}_\phi$ on the field $\phi$ only includes Poisson brackets computed with respect to the coordinates $\varphi_\sigma, \ell$ because any derivatives of $\phi$ with respect to $h_\sigma$ vanish on the Fermi surface due to the gauge-fixing condition (\ref{LChgauge}).

We now compare the correlation functions of the above density operators to the correlation functions of the corresponding density operators in the Fermi liquid description. We show that connected correlation functions of the light-cone density operators in the bosonized theory exactly match those of the Fermi liquid theory.

First, note that the exact one point function,
    \begin{align}
            G^{(1)} = \rho^{(0)}[\phi]=\frac{1}{(2\pi)^2} \int \mathrm d\ell \int_{-\infty}^{-\mu} \mathrm dh,
        \end{align}
   precisely matches the expression for the Fermi liquid one-point function (\ref{LCFermi1pt}) under the identification
        \begin{align*}
            \sum_q \to \int \mathrm d \ell.
        \end{align*}
    Next, consider the connected $n$-point density correlation functions for $n \geq 2$. The Fermi liquid connected two-point function is given by (\ref{LC2ptFermi}), and the connected $n$-point functions for $n\geq 3$ vanish identically---see (\ref{n>3van}) and below. We will prove that in the bosonized theory,
        \begin{align}
        \begin{split}
        \label{rhovanish}
           G_c^{(2)}(1,2) &= \langle T (\rho^{(1)}[\phi](u_{\sigma 1}, \varphi_{\sigma 1}) \rho^{(1)}[\phi](u_{\sigma 2}, \varphi_{\sigma 2})) \rangle_c\\
            G_c^{(n)}(1,\dots,n) &= \langle T (\rho[\phi](u_{\sigma 1}, \varphi_{\sigma 1} )\cdots \rho[\phi](u_{\sigma n}, \varphi_{\sigma n}))\rangle_c = 0,~~~~ n \geq 3,
        \end{split}
        \end{align}
    which exactly matches the light-cone density correlation functions in the Fermi liquid to all orders.

    To prove (\ref{rhovanish}), we make use of the fact that the Euclidean two-point function is given by (see (\ref{aaprop}))
        \begin{align}
            \langle T( \phi(x_1,\tau_{\sigma 1}, \varphi_{\sigma 1},\ell_1) \phi(x_2,\tau_{\sigma 2}, \varphi_{\sigma 2},\ell_2)) \rangle  =  -2\pi \delta(\Delta \varphi_\sigma) \delta(\Delta \ell) \log \Delta u_\sigma,
        \end{align}
where $u_\sigma = i x + \sigma \tau_\sigma $.
The key point is that the presence of angular momentum-conserving delta functions in $\langle T (\phi \phi) \rangle$  forces contributions to the density correlators that contain derivatives with respect to the angular momentum to vanish, as we now explain. First, assume that $n \geq 2$; then, each term contributing to a connected $n$-point function of the form
    \begin{align}
    \label{Gcterm}
        G_c^{(n)}(1,\dots,n) \supset \left\langle T \left(\prod_{j=1}^n \rho^{(k_j)}[\phi](j)\right)\right\rangle_c
    \end{align}
for which $k_j \geq 2$ for some $j$ can be made to take the following schematic form after integrating by parts and evaluating integrals over a maximal number of delta functions in $\ell$:
    \begin{align}
        \left\langle T \left(\prod_{j=1}^n \rho^{(k_j)}[\phi](j)\right) \right\rangle_c \sim \sum ( \cdots  ) \int \mathrm d\ell \,  \partial_{\ell}^{ \sum_j k_j - n} \delta(\ell -\ell')   =0,
    \end{align}
where the prefactor $(\cdots)$ in the above expression is independent of $\ell$. Since the above expression is proportional to the integral of a total derivative of a delta function, it vanishes.\footnote{We assume that we can always choose a regularization scheme for the delta functions that ensures that they and their derivatives decay rapidly at infinity.} Notice that the above statement holds true for any number of interaction vertices $S^{(k_j)}$ (including none) since every order-$k$ contribution to a density operator for which $k>1$ contains at least one derivative with respect to $\ell$---see \eqref{rhoknew}. Consider, for example, the following free contribution to the connected density three-point function:
    \begin{align}
    \begin{split}
    &  G_c^{(3)}(1,2,3) \\&\supset \int \mathrm d\ell_1 \mathrm d\ell_2 \mathrm d\ell_3 \, \left\langle \frac{\partial \phi}{\partial \tau_{\sigma 1}}  \text{ad}_{\phi} \left(\frac{\partial \phi}{\partial \tau_{\sigma 2}} \right)\frac{\partial \phi}{\partial \tau_{\sigma 3}}  \right \rangle_0\\
      &\sim \frac{\partial}{\partial \varphi_{\sigma 2}} \delta(\varphi_{\sigma 12}) \delta(\varphi_{\sigma 23}) \frac{1}{u_{\sigma 12}} \frac{1}{u_{\sigma 23}^2}  \int d\ell_1d\ell_3  \, \delta(\ell_{12}) \int d\ell_2 \,  \frac{\partial}{\partial \ell_2} \delta(\ell_{23}) + \cdots.
    \end{split}
    \end{align}
Above, the subscript `0' indicates that the correlation function is evaluated in the free limit. The term shown above is proportional to an integral of a total derivative of a delta function and thus vanishes; each of the remaining suppressed terms can also be written in a similar form and thus also vanish.

The above discussion implies that the only nonvanishing contribution to connected $n$-point functions must come from correlators of the form (\ref{Gcterm}) with all $k_j = 1$. But notice that any connected $n$-point function of this form that involves interactions $S^{(k_j)}$ with $k_j \geq 3$ for some $j$ also contains derivatives with respect to $\ell$, and thus it also vanishes by analogous reasoning. It follows that the only non-zero connected $n$-point function is the two-point function
    \begin{align}
    \begin{split}
         G_c(1,2) &=-\frac{1}{(2\pi)^2}\left( \frac{1}{2\pi} \int \mathrm d \ell \right) \delta(\Delta \varphi_\sigma) \frac{1}{\Delta u_\sigma^2}.
    \end{split}
    \end{align}
The above expression exactly matches the Fermi liquid light-cone density two-point function (\ref{LC2ptFermi}) under the  identification\footnote{The divergence $\delta(0)$ in both the Fermi liquid and bosonized descriptions arises due to a sum (integral) over angular momentum.}
    \begin{align}
        \delta(0) \to \frac{1}{2\pi} \int \mathrm d \ell.
    \end{align}
Thus, we see that the connected $n$-point functions of light-cone densities in the bosonized theory match their Fermi liquid counterparts to all orders in the perturbative expansion.

\section{Summary and Future Directions}
\label{sec:conclusion}

In this paper we have proposed an effective field theory description of the Fermi liquid formulation of noncritical M-theory introduced in \cite{Horava:2005tt}. We constructed our proposal using the coadjoint orbit method to bosonize the noncritical M-theory Fermi surface, largely following the approach systematized in \cite{Delacretaz:2022ocm}. We further provided evidence that the correlation functions of Fermi surface density perturbations in the bosonized theory match their counterparts in the exact Fermi liquid description in an appropriate semiclassical limit, an important check on the consistency of our proposal. In more detail, we considered two distinct sets of density operators in the bosonized theory whose elementary quanta we expect to correspond to particle-hole excitations of the Fermi surface in different single-particle bases: for the so-called eigenvalue density operators, we showed that the $n$-point correlation functions agree with their Fermi liquid counterparts for $n \leq 3$; for the light-cone density operators, we argued that their connected correlation functions match their Fermi liquid counterparts for all $n$, as they vanish identically for $n \geq 3$. This evidence is a clear signal that the bosonized theory correctly captures the physics of the (2+1)-dimensional Fermi liquid and deserves further study as a candidate effective description of the canonical vacuum of noncritical M-theory. This is an important step towards uncovering a unified effective gravity description of the noncritical string landscape, very much analogous to the role of critical M-theory in 11 spacetime dimensions.

A generic Fermi surface EFT resulting from the coadjoint orbit approach to bosonization could in principle be rather complicated, despite providing a systematic weak field and derivative expansion. Consequently, there is no good reason to expect that one can solve the EFT beyond leading order in perturbation theory. In the case of noncritical M-theory, however, the integrability of the underlying Fermi liquid description plays an essential role in making the EFT more tractable. Indeed, by choosing a basis of action-angle coordinates, we found that the dynamics of the EFT is captured by an infinite family of two-dimensional chiral bosons, each labeled by a pair of classical constants of motion (i.e., two of the four action-angle coordinates that commute with the single particle Hamiltonian).

For the light-cone density operators, the underlying integrability of the theory was sufficient to determine their $n$-point correlation functions for all $n$, making the comparison with the corresponding correlation functions in the Fermi liquid description straightforward. The $n$-point correlation functions of the eigenvalue density operators turned out to be much more involved to compute explicitly for $n \geq 3$, but they too can in principle be evaluated exactly thanks to the underlying symmetry of the theory. The real challenge for the eigenvalue density operators is finding a clever way to demonstrate that their correlation functions match those of their exact Fermi liquid counterparts---this is crucial for understanding precisely what physics of the underlying Fermi liquid description is captured by the bosonized theory. We defer a more complete analysis of the eigenvalue density operators and other observables to future work.

In addition to a more complete analysis of the eigenvalue density operators, there are numerous other future avenues we hope to explore, a few of which we describe below:

\begin{itemize}

\item One of the primary motivations for this work was identifying an effective three-dimensional spacetime gravity description of the canonical vacuum of noncritical M-theory. For noncritical string theories, the conceptual problem of recovering the spacetime effective description from the Fermi liquid description is essentially solved by bosonization, since the spacetime tachyon is related to the collective field up to a nonlocal field redefinition. It is natural to guess that a similar idea might work for noncritical M-theory, as mentioned in \cite{Horava:2005tt}. If so, then the EFT description of noncritical M-theory should be related to a sector of some three-dimensional gravitational theory, possibly up to a complicated field redefinition similar to the leg pole transform. Concretely relating bosonized noncritical M-theory to a more conventional effective description of three-dimensional gravity would be illuminating for many reasons---for instance, a more conventional geometric description of noncritical M-theory might permit a more intuitive and exhaustive exploration of the landscape of two-dimensional noncritical string theories, much as eleven-dimensional (critical) M-theory has played a central role in studies of the critical string landscape. A subtlety here is that our bosonized description of noncritical M-theory is not a conventional local field theory, due to the fact that the phase space dependence of the bosonic field essentially builds noncommutativity into the theory.\footnote{For a generic three-dimensional Fermi surface EFT in $p$-gauge, this is closely related to the fact that the bosonic field depends on a residual angular ``momentum'' coordinate $\xi$ that does not Poisson-commute with the position coordinates on which the bosonic field also depends. A similar subtlety is expected to be absent in two spacetime dimensions (i.e., 1d Fermi surfaces), where no continuous residual coordinate dependence remains after gauge-fixing the $p$ dependence. This issue presumably does not affect the bosonized description of two-dimensional noncritical string theories.
} Nevertheless, it may be possible to find a suitable limit of the bosonized theory that coincides with a local gravitational theory.

\item Related to the above, there has been partial progress in identifying a spacetime effective gravity description of noncritical M-theory in the conformal limit \cite{Horava:2007ds}. Analogous to the conformal limit of the type 0A noncritical string discussed in \cite{Strominger:2003tm,ho2004isometry,Aharony:2005hm}, this is the high energy regime of the theory that exhibits accidental conformal symmetry due to the disappearance of all dimensionful scales. The authors of \cite{Horava:2007ds} proposed that the conformal limit of noncritical M-theory corresponds to higher spin conformal gravity coupled to a Dirac fermion propagating on AdS$_2 \times S^1$, with the $S^1$ factor corresponding to the M-theory circle. It would be very interesting to understand whether the conformal limit can be probed directly in the bosonized description of noncritical M-theory, and moreover whether there is a systematic way to devise an effective action that validates the predictions of \cite{Horava:2007ds}. It would further be interesting to ask whether one can incorporate controlled corrections into this description order-by-order in a $1/\mu$ expansion, as these expressions might shed light on the spacetime geometry characterizing the noncritical M-theory vacuum.

\item Another special limit of noncritical M-theory, the high temperature limit, was studied in \cite{Horava:2005wm}. There, the authors derived an exact expression for the free energy of the vacuum solution and showed that its high temperature limit matches the weak coupling expansion of the partition function of the closed topological A model on the resolved conifold; this was interpreted as a hint that noncritical M-theory is dual to the closed topological A model on the conifold. It would again be illuminating to explore this regime in the context of the bosonized EFT description (for instance, by bosonizing the underlying fermionic many body system at finite temperature), especially in view of this possible duality. In particular, since there is compelling evidence that the large $N$ limit of U($N$) Chern-Simons on $S^3$ provides a non-perturbative definition of the closed topological A model via Gopakumar-Vafa duality (see the excellent review \cite{Marino:2004uf} and references therein), it would be especially interesting to explore the relationship between the Chern-Simons description and the bosonized EFT description of noncritical M-theory.

\item A common theme of the above directions is that they aim to uncover a more conventional spacetime gravity description of noncritical M-theory. Assuming such a description can be found, a longer term aspiration is to use the spacetime geometry to explore the landscape of noncritical two-dimensional string theories, much as geometric operations like Kaluza-Klein compactification and orbifolds of critical M-theory have been used to identify different critical string theory vacua. This would in particular complement the direct identification of hydrodynamic solutions in the Fermi liquid description discussed in \cite{Horava:2005tt,Horava:2007ds}.

\item Much also remains to be understood within the exact Fermi liquid description of noncritical M-theory itself. One loose end is an exact calculation of the noncritical M-theory S-matrix, for instance by using the $\mathcal W_{\infty}$ symmetry of the Fermi liquid, in analogy with Polchinski's calculation of the S-matrix for the $c=1$ string \cite{Polchinski:1994mb}. This calculation is particularly relevant for further study of the EFT description of noncritical M-theory, as it would be desirable to show that the S-matrix of the bosonized theory agrees with the Fermi liquid S-matrix in an appropriate limit.

\item
In condensed matter systems, spinful fermions provide an important generalization of ordinary Fermi surface bosonization, since the spin sector gives rise to non-Abelian collective modes.
For example, in one-dimensional systems, fermions with an internal $N$-component index are naturally bosonized in terms of a $U(N)_1$ WZW model \cite{Witten:1983ar, Fradkin:2013sab}.
 In the context of higher dimensional Fermi surface bosonization, the directions tangent to the Fermi surface may be treated as internal degrees of freedom \cite{Chen:2025jdv}.
 The large-$N$ limit provides a useful viewpoint on higher-dimensional Fermi surface bosonization.
This is closely related to regarding the angular-momentum direction as an internal label.
From this point of view, the bosonization of noncritical M-theory may be viewed as a generalization of spinful one-dimensional bosonization.
In the context of noncritical M-theory, this question is particularly suggestive because the relation between noncritical M-theory and the type 0A string is known exactly at the level of the underlying Fermi liquid.
It would therefore be valuable to understand how this exact relation is reflected in the relation between the type 0A boson and the bosonic collective field of noncritical M-theory. This may provide a solvable lower-dimensional analogue of the relation between type IIA supergravity and eleven-dimensional supergravity, and could offer useful lessons for formulating the bosonic effective theory of critical M-theory.

\item In this paper, we have studied  density operators other than those in the $\lambda$-representation, namely the light-cone density operators.
Similarly, density operators in addition to those in the $\lambda$-representation could also be studied in the context of the $c=1$ matrix model.
An interesting example is the momentum-space density
\begin{equation*}
\rho(p)=\Psi^\dagger(p)\Psi(p).
\end{equation*}
The momentum representation has several potential advantages over the $\lambda$-representation. First, the leg-pole factor, which appears nonlocally in the $\lambda$-representation, may become more local in the momentum representation. Second, since there is no classical turning point in momentum space, the semiclassical description can be simpler. Moreover, even in situations where the Fermi surface develops folds and becomes multivalued in the $\lambda$-representation, it may still admit a single-valued description in momentum space.

This issue is also important for understanding the regime of validity of the bosonized action as a candidate string field theory \cite{Polchinski:1991uq,Ginsparg:1993is}. However, even if the momentum-space description is simpler, the relation between the $\lambda$-space density and the momentum-space density is not straightforward. Although the fermion fields themselves are related by a linear transformation, a density operator is not mapped, in general, only to another density operator. Rather, one must include more general fermion bilinears that are not themselves density operators. Understanding the relation between density operators in different representations within the larger space of fermion bilinears is therefore an important problem for future work.
\end{itemize}
These are just a handful of potential research directions, mostly intended to illustrate noncritical M-theory's rich connections to various corners of the noncritical string landscape, and to highlight opportunities for exploring this landscape that might be afforded by an effective three-dimensional gravity description. A clear theme that hopefully emerges from these examples is that noncritical M-theory occupies a privileged position within the two-dimensional string landscape due to the large amount of symmetry encoded in the underlying phase space geometry, and that further study of this theory will benefit immensely from exploiting its integrability as much as possible. The prospect of a non-perturbatively well-defined, solvable unified model of strings is particularly appealing, and may eventually help clarify outstanding questions in critical string theory related to the emergence of spacetime geometry and causality---questions that remain to be understood due in part to our ignorance about the  fundamental degrees of freedom that define string theory on the interior of the moduli space, away from asymptotic limits. We hope to answer many of these questions in future work as we continue to explore noncritical M-theory and its implications for the landscape of noncritical strings.

\section*{Acknowledgements}
We have benefited from discussions with Benjamin Heidenreich, Tigran Sedrakyan, Eric Sharpe, Shu-Heng Shao, Victor Rodriguez,  Masaki Oshikawa, Satoshi Iso, Tomonori Ugajin, and Tadashi Takayanagi while preparing this manuscript. PJ was supported in part by the Simons Collaboration on Global Categorical Symmetries and also by
the NSF grant PHY-2412361 while completing some parts of this work. TN was supported by MEXT KAKENHI Grant Nos.~23K13094 and 24H00944 and by JST PRESTO Grant No.~JPMJPR2359.

\appendix

\section{Wick's Theorem for Fermions}
\label{Wick}

Consider a vacuum state $\ket{\mu}$ given by filling the Fermi sea up to the Fermi energy $\varepsilon_F = -\mu$ where $ \mu \geq 0$. Given a basis of creation and annihilation operators $ a^\dagger_{\boldsymbol n},  a_{\boldsymbol n}$, where $\boldsymbol n$ collectively refers to conserved quantum numbers, the vacuum $\ket{\mu}$ is conventionally defined by the condition
	\begin{align}
	\begin{split}
		 a_{\boldsymbol n} \ket{\mu } &= 0 ,~~~~\boldsymbol n \supset E  > -\mu\\
		 a^\dagger_{\boldsymbol n} \ket{\mu} &=0,~~~~ \boldsymbol n \supset E <-\mu.
	\end{split}
	\end{align}
Given the following mode expansion of a fermionic operator $\Psi$,
    \begin{align}
        \Psi = \sum_{\boldsymbol n} a_{\boldsymbol n} \psi_{\boldsymbol n},
    \end{align}
where $\psi_{\boldsymbol n}$ are a set of eigenfunctions with quantum numbers $\boldsymbol n$, we can split $\Psi$ into ``$+$'' and ``$-$'' components as follows:
	\begin{align}
		 \Psi &=  \Psi^+ +  \Psi^-,~~~~  \Psi^- \ket{\mu } =0, ~~~~ \bra{\mu}  \Psi^+ =0,
	\end{align}
where
    \begin{align}
        \begin{split}
            \Psi^+ = \sum_{\{\boldsymbol n | E < -\mu \}} a_{\boldsymbol n} \psi_{\boldsymbol n},~~~~ \Psi^- = \sum_{\{\boldsymbol n |  E > -\mu\}} a_{\boldsymbol n} \psi_{\boldsymbol n}.
        \end{split}
    \end{align}
We next recall some facts about normal ordering for fermions. We define normal order to place all factors $\Psi_i^-$ on the right:
		\begin{align}
			 : \Psi_1 \cdots \Psi_n : = (-1)^{P} \Psi_{i_1}^+ \cdots \Psi_{i_k}^+ \cdots \Psi_{i_n}^{-},
		\end{align}
	where $P$ is the parity of the permutation that brings the sequence $1\cdots n$ to the sequence $i_1 \dots i_n$. It follows that
		\begin{align}
			 :\Psi_1 \cdots \Psi_n : = (-1)^{P}  : \Psi_{i_1} \cdots \Psi_{i_n} :
		\end{align}
	Wick contraction is given by
		\begin{align}
			\wick{\c{\Psi} \Psi_1 \cdots \Psi_n \c{{\Psi'}} } = (-1)^n \wick{\c{\Psi} \c{\Psi'} } \Psi_1 \cdots \Psi_n,
		\end{align}
	where
		\begin{align}
			\wick{\c{ \Psi_1} \c{{ \Psi_2}} } &= \{ \Psi_1^- ,\Psi_2^+ \} = \Psi_1^- \Psi_2^+ + \Psi_2^+ \Psi_1^-.
		\end{align}
	Note that by construction $\braket{\mu| :\!\Psi_1\cdots \Psi_n\!: |\mu} =0$, hence
		\begin{align}
				\wick{\c{ \Psi_1} \c{{ \Psi_2}} } =\braket{\mu|	\wick{\c{ \Psi_1} \c{{ \Psi_2}} } |\mu} = \braket{\mu|\Psi_1 \Psi_2 |\mu}.
		\end{align}
	The static Wick's theorem is
		\begin{align}
		\begin{split}
			\Psi_1 \cdots \Psi_n &=  \, :\Psi_1 \cdots \Psi_n: \\
			&+\sum_{(ij)}  :\Psi_1\cdots \wick{\c{\Psi_i} \cdots \c{\Psi_j} } \cdots \Psi_n : \\
			&+\sum_{(ij) (kl)}  :\Psi_1 \cdots \wick{\c1{\Psi_i} \cdots \c2{\Psi_k} \cdots \c1{\Psi_j} \cdots \c2{\Psi_{l}}} \cdots \Psi_n : + \cdots.
		\end{split}
		\end{align}
	It follows that for products consisting of an even number of $+$ and $-$ operators,
		\begin{align}
		\begin{split}
        \label{Pf}
			\braket{\mu | \Psi_1 \cdots \Psi_{2n}| \mu} &= \sum (-1)^P \braket{\mu| \Psi_{i_1} \Psi_{j_1}| \mu} \cdots \braket{\mu | \Psi_{i_n} \Psi_{j_n} |\mu}=\Pf(M)
		\end{split}
		\end{align}
	where the matrix $M$ is defined by
		\begin{align}
		\label{Mdef}
			M_{ab} = \begin{cases}
				\braket{\mu | \Psi_a \Psi_b |\mu}~~~~ &a< b\\
				-\braket{\mu | \Psi_b \Psi_a |\mu}  ~~~~& a >b \\
				0 ~~~~ &a = b.
			\end{cases}
		\end{align}
We next introduce time dependence $\Psi_i \to \Psi_i(t_i)$. The time ordering operator acts as
	\begin{align}
		T  \Psi_1(t_1) \Psi_2(t_2)  &= : \Psi_1(t_1) \Psi_2 (t_2) : + \wick{\c{{\Psi_1(t_1)}} \c{\Psi_2({t_2)}}}
	\end{align}
where the contraction above is the time-ordered Wick contraction
	\begin{align}
	\begin{split}
		\wick{\c{{\Psi_1(t_1)}} \c{\Psi_2({t_2)}}}  &= \braket{\mu | T \Psi_1(t_1) \Psi_2(t_2)  |\mu}.
	\end{split}
	\end{align}
The time-ordered Wick contraction has the same ordering properties as the non-time-ordered Wick contraction, but there is also a new property (for fermions)
	\begin{align}
	\label{Tanti}
		\wick{\c{{\Psi_1(t_1)}} \c{\Psi_2({t_2)}}} = - \wick{\c{{\Psi_2(t_2)}} \c{\Psi_1({t_1)}}}.
	\end{align}
Wick's theorem for time-ordered products is the same as the static Wick's theorem, with the static Wick contractions replaced by the time-ordered Wick contraction.

We will apply the above rules to expectation values of products of the fermion bilinear operators
    \begin{align}
    \rho  = \Psi^\dagger  \Psi,~~~~
		 \Psi&= \sum_{\boldsymbol n}  a_{\boldsymbol n} \psi_{\boldsymbol n} = \Psi^+ +  \Psi^-.
	\end{align}
The time-ordered $n$-point function takes the same form as the expression in (\ref{Pf}), with static Wick contractions replaced by time-ordered Wick contractions. Consider the time-ordered $n$-point function
	\begin{align}
		\begin{split}
			G^{(n)}(1,\dots,n) &= \braket{\mu| T (\rho (1) \cdots \rho (n)) |\mu}.
		\end{split}
	\end{align}
That is, we take $\Psi_{2k-1} =  \Psi^\dagger(k), \Psi_{2k} =  \Psi(k)$. Since the ``anomalous'' contractions vanish,
	\begin{align}
		\braket{\mu|  \Psi(i)  \Psi(j) |\mu} = 0,~~~~\braket{\mu|  \Psi^\dagger(i)  \Psi^\dagger(j) |\mu} =0 ,
	\end{align}
 this means that we can reorganize the components of the matrix $M$ in (\ref{Mdef}) 
so that
    \begin{align}
		M' = P M P^{T} = \begin{pmatrix} 0 & S \\
		- S^{T} & 0 \end{pmatrix},~~~~
	\Pf(M) = \det(S).
	\end{align}
The components of the matrix $S$ are given by
	\begin{align}
		S(i,j) = \braket{\mu| T ( \Psi^\dagger(i)  \Psi(j) )|\mu},
	\end{align}
thanks to the fact that time-ordered Wick contractions are anti-symmetric in exchange (see (\ref{Tanti})). However, this is not true for non-time-ordered Wick contractions. For the non-time-ordered case, we must use the definition of  $M_{ab}$ given in (\ref{Mdef}).

A more explicit expression for $G^{(n)}$ follows from the properties of the determinant:
	\begin{align}
	\begin{split}
    \label{Gdet}
		G^{(n)}(1,\dots,n)
		&=\sum_{\sigma \in \Sigma_n} \sgn(\sigma) \prod_{i=1}^{n} S(i, \sigma(i))\\
        &=\sum_{ \mathcal P } \prod_{I \in  \mathcal P} \left[\sum_{\sigma \in C_m(I)} (-1)^{m-1} \prod_{i\in I} S(i, \sigma(i))\right],~~~~ n \geq m \equiv |I|.
	\end{split}
	\end{align}
Above,
 $C_m(I)$ denotes the conjugacy class of $m$-cycles in the symmetric group $\Sigma_m(I)$, and the sum runs over all set partitions $\mathcal P = \{ I_1, \cdots, I_k\}$ of $\{ 1,\dots, n\}$ into nonempty, pairwise disjoint subsets, i.e., $I_j \ne \emptyset, I_j \cap I_k = \emptyset$ if $j \ne k$, and $\cup I_j = \{ 1, \dots, n\}$.\footnote{Note that we have used the fact that every $m$-cycle $\sigma$ has the same $\sgn(\sigma) = (-1)^{m-1}$, as any $m$-cycle can be decomposed into $m-1$ transpositions.} Using (\ref{Gdet}), we can see that the connected correlation functions $G_c^{(n)}$ correspond to sums over maximal length ``ring'' diagrams,
	\begin{align}
		G^{(n)}_c(1,\dots, n) = \sum_{\sigma \in C_n} (-1)^{n-1} \prod_{i=1}^n S(i, \sigma(i) ).
	\end{align}
	For example, the connected density two-point function is
		\begin{align}
        \begin{split}
		      G^{(2)}_c  = - S(1,2) S(2,1).
         \end{split}
		\end{align}

For some purposes, it is also useful to define the hole and particle propagators, respectively:
	\begin{align}
	\label{holepart}
		S^h(i,j) &\equiv \braket{\mu|  \Psi^\dagger(i) \Psi(j) |\mu} = \sum_{\alpha < -\mu} \psi_\alpha^*(i)\psi_\alpha(j)\\
	-	S^p(i,j) & \equiv \braket{\mu|  \Psi(i)  \Psi^\dagger(j)  |\mu} = \sum_{ \alpha > -\mu} \psi_\alpha(i) \psi^*_\alpha(j).
	\end{align}
As an example, in terms of the particle and hole propagators, the non-time-ordered two-point function is given by
	\begin{align}
		\begin{split}
			 \braket{\mu| \rho(1)  \rho(2) |\mu}
			&=S^h(1,1) S^h(2,2) - S^h(1,2) S^p(1,2),
		\end{split}
	\end{align}
and the non-time-ordered three-point function is given by
	\begin{align}
		\begin{split}
			\braket{\mu| \rho(1)\rho(2)\rho(3)|\mu}&=S^h(1,1) S^h(2,2) S^h(3,3) - S^h(1,1) S^h(2,3) S^p(2,3) \\
            &- S^h(1,3) S^h(2,2) S^p(1,3)- S^h(1,2) S^p(1,2) S^h(3,3)
			 \\
            &+ S^h(1,2) S^h(2,3) S^p(1,3) + S^h(1,3) S^p(1,2) S^p(2,3).
		\end{split}
	\end{align}

\section{Phase Space Geometry of Noncritical M-theory}

\subsection{Classical phase space and canonical coordinates}

\label{classPS}

The classical phase space for noncritical M-theory is isomorphic to $\mathbb R^4$ parametrized by coordinates $(\boldsymbol \lambda, \boldsymbol p) $, equipped with the symplectic form
    \begin{align}
        \omega =  \mathrm d  \lambda_1 \wedge \mathrm d  p_1 +  \mathrm d  \lambda_2 \wedge \mathrm d  p_2.
    \end{align}
The single particle Hamiltonian is
    \begin{align}
        h = \frac{1}{2} \boldsymbol p^2 - \frac{1}{2} \boldsymbol \lambda^2,
    \end{align}
in terms of which the Fermi surface is defined by
    \begin{align}
        h = -\mu, ~~~~ \mu \geq 0.
    \end{align}
We formulate the zero temperature Fermi sea as a droplet, defined by the density
    \begin{align}
        f_0(\boldsymbol \lambda, \boldsymbol p) = \theta( - \mu - h(\boldsymbol \lambda,\boldsymbol p)).
    \end{align}
There are various coordinate systems that one can use to parametrize this Fermi surface. When discussing the connection to type 0A string theory, it is useful to work in polar coordinates, whereas for the $c=1$ and type 0B string theories, it is most natural to work in rectilinear coordinates. For some purposes, however, we find it useful to work in more exotic coordinate systems, which we describe in more detail below.

\subsubsection{Doubly-polar coordinates}

  Doubly-polar coordinates, which are not canonical, are defined by
    \begin{align}
    \label{app:dub}
        \boldsymbol \lambda = (\lambda_r \cos \theta, \lambda_r \sin \theta),~~~~\boldsymbol p = (p \cos \xi, p \sin \xi).
    \end{align}
We sometimes write  
    \begin{align}
        \boldsymbol p = p \boldsymbol n_\xi, ~~~~ \boldsymbol t_{\xi} \equiv \partial_\xi \boldsymbol n_\xi~~\implies ~~ \boldsymbol p \cdot \boldsymbol t_\xi = 0.
    \end{align}
    In terms of these coordinates, the symplectic form is given by
    \begin{align}
        \omega = \cos ( \theta - \xi) (\mathrm d\lambda_r \wedge \mathrm dp + \lambda_r \mathrm d \theta \wedge p \mathrm d\xi) + \sin ( \theta - \xi) ( \mathrm d\lambda_r \wedge p \mathrm d\xi + \mathrm dp \wedge \lambda_r \mathrm d\theta),
    \end{align}
and the single particle Hamiltonian is given by
    \begin{align}
        h = \frac{1}{2} p^2- \frac{1}{2} \lambda_r^2.
    \end{align}
In this case, we parametrize the Fermi surface by writing
    \begin{align}
        p_{\text F}(\lambda_r) = \sqrt{\lambda_r^2 - 2 \mu},
    \end{align}
where we impose the restriction $\lambda_r \geq \sqrt{2\mu}$.

\subsubsection{Action-angle coordinates}

\label{actionangle}

We denote by ``action-angle coordinates'' a class of canonical coordinate systems $\tau, h, \varphi, m$ satisfying
    \begin{align}
    \label{genaa2}
        \{ \tau, h\} =\{ \varphi, m\} =1
    \end{align}
with all other Poisson brackets vanishing, and where $h$ is the Hamiltonian. Note that this coordinate system is not unique, and there are many choices of canonical coordinates for which one of the canonical ``momenta'' is the Hamiltonian. It follows that the coordinates $\varphi,m$ are constants of motion:
    \begin{align}
       \{ h, \varphi\} = \{ h, m\} =0.
    \end{align}
The dynamical behavior of the bosonic field $\phi$ capturing the fluctuations of the noncritical M-theory Fermi surface can be parametrized entirely in terms of the ``angle''\footnote{Since $h$ is unbounded for the inverted harmonic oscillator, its conjugate variable $\tau$ is not compact.} $\tau$, while the coordinates $\varphi, m$ play the role of continuous labels. Below, we describe two specific sets of action-angle coordinates that we use in this paper.

One useful set of action-angle coordinates is given by the polar representation of the light-cone coordinates $z_\sigma = [ p_1 + \sigma \lambda_1 + i (p_2 + \sigma \lambda_2)]/2 = e^{\tau_\sigma + i \varphi_\sigma} / \sqrt{2}$, namely
\begin{alignat}{2}
    \tau_\sigma &= \log\left(\sqrt{2 z_\sigma z_\sigma^*}\right) &\qquad
    h_\sigma &= \sigma \left(z_\sigma z_{-\sigma}^* + z_\sigma^* z_{-\sigma}\right) = \sigma h \\
    \varphi_\sigma &= -i \arctanh\left(\frac{z_\sigma - z_\sigma^*}{z_\sigma + z_\sigma^*}\right) &\qquad
    \ell &= \sigma i \left(z_\sigma z_{-\sigma}^* - z_\sigma^* z_{-\sigma}\right),
\end{alignat}
where $\sigma =\pm{}1$ distinguishes ``left-moving'' from ``right-moving'' coordinates.\footnote{Although $\ell$ is the same angular momentum that appears in polar coordinates,  $\varphi_\sigma$ is not equal to the polar angle $\theta$. In particular, $\varphi_\sigma$ is a constant of motion satisfying $\{ h, \varphi_\sigma \} = 0 $, whereas $\{ h, \theta\} \ne 0$.
} The non-vanishing Poisson brackets are
    \begin{align}
        \{ \tau_\sigma, h_\sigma\} =  \{\varphi_\sigma, \ell\} = 1      .
    \end{align}
It follows that $\varphi_\sigma, \ell$ Poisson-commute with $h = \sigma h_\sigma$.

Another useful set of action-angle coordinates is
    \begin{alignat}{2}
        \tau &= \tau_1 +\tau_2 & \qquad h&= h_1 +h_2 \\
        \varphi &= \tau_1-\tau_2&\qquad m&= h_1-h_2,
    \end{alignat}
where
    \begin{align}
        \tau_i = \frac{1}{2} \arctanh\left( \frac{p_i}{\lambda_i} \right),~~~~h_i =\frac{1}{2} (p_i^2 - \lambda_i^2).
    \end{align}
The non-vanishing Poisson brackets in this case are
    \begin{align}
        \{ \tau, h \} = \{ \varphi , m \} =1.
    \end{align}

\subsection{Contact geometry of the Fermi surface}

\label{contact}

In this section, we collect some useful results concerning the geometry of Fermi surfaces. The starting point for understanding the geometry of a Fermi surface is the geometry of the phase space in which it is embedded as a constant energy hypersurface.
Given a classical phase space $\mathbb R^{2d}$ (where in our case $d=2$) with standard canonical coordinates $\boldsymbol \lambda, \boldsymbol p$, there exists a Liouville 1-form given by
    \begin{align}
    \label{L1}
       \eta =\boldsymbol p \cdot  \mathrm d \boldsymbol \lambda.
    \end{align}
The Liouville 1-form defines a canonical symplectic form $- \mathrm{d}  \eta$, which up to normalization determines the canonical volume form on phase space:
    \begin{align}
    \omega \equiv - \mathrm d \eta, ~~~~\text{d}vol_{{PS}} =   \frac{ \omega^d}{(2\pi)^dd!}.
    \end{align}
 The volume form on the Fermi surface can in turn be obtained from the pullback of the Liouville 1-form to the Fermi surface, namely
    \begin{align}
        \eta_{F}\equiv \left. \eta \right|_{{FS}}.
    \end{align}
The 1-form $\eta_{F}$ is a contact form in that it defines a non-degenerate top form on the Fermi surface,
    \begin{align}
      -  \eta_{F} \wedge \mathrm{d} \eta_{F},
    \end{align}
which up to proportionality can be taken to be the volume form on the Fermi surface. Note that one has some freedom in how to define $\eta_{F}$---if $\eta_{F}$ is a contact form, and $f$ is a nowhere vanishing function on the Fermi surface, then $\eta_{F}' = f \eta_{F}$ is also a contact form, as $- \eta_{F}' \wedge \mathrm{d} \eta_{F}' =- f^2 \eta_{F} \wedge \mathrm{d} \eta_{F}$.
The contact structure of the Fermi surface also uniquely defines a Reeb vector field $R$ by the conditions $\eta_{F}(R)=1, R \cdot \mathrm d\eta_{F} =0 $, which in this case generates the Hamiltonian flow up to reparametrization.\footnote{
There is an interesting analogy with three-dimensional Chern-Simons theory. The Liouville 1-form is similar to a classical gauge field in that any two Liouville 1-forms that differ by an exact term define the same canonical symplectic form. Thus, one can interpret $\eta_{F} \wedge \mathrm d \eta_{F}$ as a classical Chern-Simons 3-form.}

\subsubsection{Rectilinear coordinates}
In terms of rectilinear coordinates,
 the contact form is given by
    \begin{align}
      \eta_{F} = p_{ F} \boldsymbol n_{\xi} \cdot \mathrm{d} \boldsymbol \lambda,~~~~ \boldsymbol n_\xi = \boldsymbol p/p_{ F},
    \end{align}
and it is straightforward to show that the  top form defined by $\eta_{F}$ can be expressed as follows:
    \begin{align}
    -  \eta_{F} \wedge \mathrm{d} \eta_{F} = p_{ F}^2 \mathrm{d} \lambda_1 \wedge \mathrm{d} \lambda_2 \wedge \mathrm{d} \xi.
\end{align}
It is sometimes useful to introduce a rescaled contact form $\eta_{F}' = \eta_{F}/p_{ F}$ in terms of which we may define a  flat volume form on the Fermi surface:
    \begin{align}
    \label{cancon}
     \eta_{F}' = \boldsymbol n_\xi \cdot \mathrm{d} \boldsymbol \lambda,~~~~    \mathrm{d}vol_{{FS}} = -  \eta_{F}' \wedge \mathrm d \eta_{F}' =\mathrm{d} \lambda_1 \wedge \mathrm{d} \lambda_2 \wedge \mathrm{d} \xi.
    \end{align}
The above volume form is the standard volume form that appears when using the coadjoint orbit method to define various physical quantities living on the classical Fermi surface.  Using the fact that $\mathrm d \eta_{{F}}' =\mathrm d\xi \wedge (\boldsymbol t_\xi \cdot \mathrm{d} \boldsymbol \lambda)$ where $\boldsymbol t_{\xi} = \partial \boldsymbol n_\xi/\partial \xi$, a straightforward calculation shows  that the Reeb vector field associated to the contact form $\eta_{{F}}'$ in (\ref{cancon}) is given by
    \begin{align}
         R' = \boldsymbol n_\xi \cdot \partial_{\boldsymbol \lambda}.
    \end{align}

\subsubsection{Action-angle coordinates}
\label{aavol}

Another useful set of canonical coordinates is action-angle coordinates,  defined in Section \ref{actionangle}. Since $h$ is one of the canonical momenta in this coordinate system, it follows that the generator of the classical Hamiltonian flow is given by $\partial/\partial \tau$:
    \begin{align}
  \frac{ \partial f}{\partial \tau} = \{f,h\}.
    \end{align}
In terms of these coordinates, the Liouville 1-form (\ref{L1}) and its restriction to the Fermi surface are given by\footnote{This follows from the Poincar\'e lemma, since $\mathrm d(\eta-(h\,\mathrm d\tau+m\,\mathrm d\varphi)) = -\omega + \omega = 0$.}
    \begin{align}
    \begin{split}
     \eta &= h \mathrm d\tau + m \mathrm d\varphi + \mathrm d f\\
		\eta_{F} & =-\mu \mathrm d\tau + m\mathrm d\varphi + \mathrm d f.
    \end{split}
	\end{align}
Therefore, the standard top form on the Fermi surface can be expressed as
	\begin{align}
		 -  \eta_{F}  \wedge \mathrm{d} \eta_{F} =   (  -\mu + \partial_{\tau} f) \mathrm{d} \tau\wedge \mathrm{d} \varphi\wedge \mathrm{d} m.
	\end{align}
But notice that
    \begin{align}
    \begin{split}
    \label{pFproof}
        -\mu +  \frac{\partial f}{\partial \tau} &= \eta\left(\frac{\partial}{ \partial \tau}\right)\\
        &=\boldsymbol p \cdot \mathrm d\boldsymbol \lambda\left(\frac{\partial}{\partial\tau}\right)\\
        &=\boldsymbol p \cdot \frac{\partial \boldsymbol \lambda}{\partial \tau}\\
        &=p_{ F}^2.
    \end{split}
    \end{align}
Thus, as above, in action-angle coordinates we may also rescale $\eta_{F}$ in order to define a flat volume form on the Fermi surface:
    \begin{align}
    \label{eq:FSmeasure}
            \mathrm{d}vol_{{FS}} = \mathrm d\tau \wedge \mathrm d \varphi \wedge \mathrm d m = \mathrm d^2 \boldsymbol \lambda  \wedge \mathrm d \xi .
    \end{align}
In particular, this shows that there always exists a change of coordinates to action-angle coordinates on the Fermi surface with unit Jacobian.
Moreover,
the Reeb vector field associated to $\eta_F$ takes the following simple form:
    \begin{align}
        R =    \frac{1}{p_{ F}^2} \frac{\partial}{\partial \tau} . \label{eq:ReebV}
    \end{align}
We can prove that the above expression is the Reeb vector field by checking the defining conditions: The correct normalization is immediate---from (\ref{pFproof}), we see that
    \begin{align}
        \begin{split}
          R \cdot  \eta_{F} = \frac{1}{p_{ F}^2} \eta_{F}\left(\frac{\partial}{\partial \tau}\right) = 1,
        \end{split}
    \end{align}
where in the above equation $\cdot$ denotes the interior product. Furthermore, because $\mathrm d \eta_{F}$ is minus the pullback of the symplectic form $\omega$ to the Fermi surface and since $h$ is constant on the Fermi surface, it follows that
    \begin{align}
       R \cdot \mathrm d \eta_{F} = -\left.\frac{1}{p_F^2} \frac{\partial}{\partial \tau}\cdot \omega\right|_{{FS}} =0.
    \end{align}

\subsubsection{Hamilton's principal function}
\label{on-shell}
Hamilton's principal function is the on-shell action evaluated on the classical trajectory connecting the initial and final positions:
\begin{align}
\label{eq:Sprin}
S(\bm{\lambda}_f,\bm{\lambda}_i;t,t_0)
&= \frac{1}{2 \sinh  \Delta t} \left((\bm{\lambda}_f^2 + \bm{\lambda}_i^2)\cosh \Delta t  - 2 \bm{\lambda}_f \cdot \bm{\lambda}_i  \right).
\end{align}
This function determines the phase of the single-particle propagator
$\braket{\bm{\lambda}_f|e^{-ih(t-t_0)}|\bm{\lambda}_i}$.
In the semiclassical approximation, the propagator takes the form \cite{Dirac:1958qm, Gutzwiller:1990}
\begin{align}
\braket{\bm{\lambda}_f|e^{- i h (t-t_0) }|\bm{\lambda}_i} \approx  \exp \left(  i S(\bm{\lambda}_f,\bm{\lambda}_i;t,t_0)\right).
\end{align}
From Hamilton's principal function, the conjugate momenta are obtained as
\begin{align}
\begin{split}
\bm{p}_f(\bm{\lambda}_f,\bm{\lambda}_i;t,t_0)
&=
\partial_{\bm{\lambda}_f} S(\bm{\lambda}_f,\bm{\lambda}_i;t,t_0)
=
\frac{1}{\sinh\Delta t}
\left(
\bm{\lambda}_f\cosh\Delta t-\bm{\lambda}_i
\right),
\\
\bm{p}_i(\bm{\lambda}_f,\bm{\lambda}_i;t,t_0)
&=
-\partial_{\bm{\lambda}_i} S(\bm{\lambda}_f,\bm{\lambda}_i;t,t_0)
=
-\frac{1}{\sinh\Delta t}
\left(
\bm{\lambda}_i\cosh\Delta t-\bm{\lambda}_f
\right).
\end{split}
\end{align}
These satisfy the relation
\begin{align}
&|\bm{p}_i(\bm{\lambda}_f,\bm{\lambda}_i;t,t_0)|^2 |\bm{p}_f(\bm{\lambda}_f,\bm{\lambda}_i;t,t_0)|^2 - [\bm{p}_i(\bm{\lambda}_f,\bm{\lambda}_i;t,t_0) \cdot \bm{p}_f(\bm{\lambda}_f,\bm{\lambda}_i;t,t_0)]^2  \notag \\
&= |\bm{\lambda}_i|^2 |\bm{\lambda}_f|^2 -  (\bm{\lambda}_i \cdot \bm{\lambda}_f)^2.
\end{align}
Hamilton's characteristic function is obtained by the Legendre transform of Hamilton's principal function,
\begin{equation}
W(\bm{\lambda}_f,\bm{\lambda}_i; E) = S(\bm{\lambda}_f,\bm{\lambda}_i; t, t_0) + E (t-t_0),
\end{equation}
where $\Delta t=t-t_0$ is determined by the stationarity condition
\begin{equation}
\frac{\partial }{\partial t} \left(  S(\bm{\lambda}_f,\bm{\lambda}_i; t, t_0) + E (t-t_0) \right) = 0.
\end{equation}
In the noncritical M-theory case, this condition becomes
\begin{align}
& \frac{\partial }{\partial t} \left(S(\bm{\lambda}_f,\bm{\lambda}_i; t, t_0) + E (t-t_0) \right) \notag \\
&= -\frac{1}{2\sinh^2 \Delta t} \Big[(\bm{\lambda}_i^2 + \bm{\lambda}_f^2) - 2\bm{\lambda}_i \cdot \bm{\lambda}_f \cosh (\Delta t) \Big] + E = 0.
\end{align}
Solving this extremality condition gives
\begin{align}
\cosh \Delta t &= -\frac{1}{2E}\left( \pm{} \sqrt{(\bm{\lambda}_i^2 + 2E)(\bm{\lambda}_f^2 + 2E) -\bm{\lambda}_i^2\bm{\lambda}_f^2 +  (\bm{\lambda}_i \cdot \bm{\lambda}_f )^2} + \bm{\lambda}_i \cdot \bm{\lambda}_f \right) .
\end{align}
In principle, one could use the above equation to explicitly calculate $W(\boldsymbol \lambda_f, \boldsymbol \lambda_i;E)$ by eliminating the dependence on $\Delta t$  in (\ref{eq:Sprin}), but the resulting expression is rather complicated.  We note, however, that Hamilton's characteristic function can also be evaluated explicitly in action-angle coordinates as follows:
\begin{align}
W  &= \int _{\tau_i}^{\tau_f} \bm{p} \cdot \frac{d\bm{\lambda}}{d\tau}d\tau  \notag \\
&= E (\tau_f - \tau_i) + \sinh(\tau_f - \tau_i) [ - E \cosh(\tau_f + \tau_i ) \cosh(2 \varphi) - m \sinh(\tau_f + \tau_i) \sinh(2 \varphi) ]. \label{eq:HamiltonPFM}
\end{align}
Taking derivatives with respect to the endpoint coordinates gives
\begin{align}
&\partial_{\bm{\lambda}_f}W = \partial_{\bm{\lambda}_f} y^{\mu}\frac{\partial W}{\partial y^{\mu}} =  \bm{p}_f(\bm{\lambda}_f,\bm{\lambda}_i; -\mu)  =\begin{pmatrix}
    \sqrt{\mu - m} \sinh (\tau_f + \varphi) \\
     \sqrt{\mu + m} \sinh (\tau_f - \varphi)
\end{pmatrix} \notag \\
&\partial_{\bm{\lambda}_i}W =\partial_{\bm{\lambda}_i} y^{\mu}\frac{\partial W}{\partial y^{\mu}}=- \bm{p}_i(\bm{\lambda}_f,\bm{\lambda}_i; -\mu)   = -\begin{pmatrix}
    \sqrt{\mu - m} \sinh (\tau_i + \varphi) \\
     \sqrt{\mu + m} \sinh (\tau_i - \varphi)
\end{pmatrix},
\end{align}
where $y^{\mu}$ are the coordinates $(\tau_i,\tau_f, m,\varphi)$ and we set $E= -\mu$.

\section{Derivations of Terms in Bosonic Action \label{sec:DerBA}}

We collect here some derivations of various terms that appear in the bosonized effective action produced by the coadjoint orbit method, as described in Section \ref{sec:COMgeneral}.

\subsection{Weak field expansion}

\label{app:weakfield}

For some purposes, it is useful to have a compact expression for each of the terms in the weak field expansion of the bosonized action, which we derive below. To begin, recall that the general form of the bosonic action is as follows,
	\begin{align}
		S = S_H + S_{\rm WZW},
	\end{align}
with the above individual terms admitting the following expansion:
	\begin{align}
		\begin{split}
			S_{\rm WZW}& = \int \mathrm dt \, \langle f_0, - \dot \phi + \frac{1}{2} \{ \dot \phi, \phi\} - \frac{1}{3!} \{ \{ \dot\phi, \phi\} , \phi\} + \cdots \rangle \\
			S_H &=- \int \mathrm dt \, \langle f_0, h + \frac{1}{2} \{ \phi, \{ \phi, h\} \} + \frac{1}{3!} \{ \phi, \{ \phi, \{ \phi, h \} \} \} + \cdots \rangle.
		\end{split}
	\end{align}
Rearranging terms and making use of the antisymmetry of the Poisson bracket, the two terms can be re-expressed as
	\begin{align}
		\begin{split}
			S_{\rm WZW}& = \int \mathrm dt \, \langle f_0, - \dot \phi + \frac{1}{2} \{ \phi, -\dot \phi\} + \frac{1}{3!} \{ \phi, \{ \phi, - \dot \phi \} \} + \cdots \rangle\\
			S_H &=\int \mathrm dt \, \langle f_0,- h + \frac{1}{2} \{ \phi, -\{ \phi, h\} \} + \frac{1}{3!} \{ \phi, \{ \phi,- \{ \phi, h \} \} \} + \cdots \rangle.
		\end{split}
	\end{align}
Combining the two terms, and defining the covariant derivative $D_t \equiv -\partial_t + \{ h, \cdot \}$ and adjoint action $\mathrm{ad}_\phi \equiv \{ \phi, \cdot\}$, the total action can be expressed as
	\begin{align}
	\begin{split}
		S &= S_{\rm WZW} + S_H\\
		&=\int \mathrm dt \, \langle f_0,  - h - \dot \phi + \frac{1}{2} \{ \phi,  D_t \phi \} + \frac{1}{3!} \{ \phi, \{ \phi, D_t \phi \} \} + \cdots \rangle \\
		&= S^{(0)} + \sum_{n=2}^{\infty} \frac{1}{n!} \int \mathrm dt \, \langle f_0,  \text{ad}_\phi^{n-1} D_t \phi  \rangle.
	\end{split}
	\end{align}
 Assuming that boundary terms can be dropped, we can make use of the identity $	\langle A, \{ B, C \} \rangle = - \langle B , \{ A, C\} \rangle$, and thus we are free to write
	\begin{align}
		\begin{split}
			\langle f_0, \text{ad}_\phi^{n-1} D_t \phi \rangle 
			&=\langle \text{ad}_\phi^{n-2} D_t \phi, \{ f_0, \phi \} \rangle.
		\end{split}
	\end{align}
We can introduce a similar expression for the density operators, working in an abstract basis of densities $\rho$. We can express the image of the distribution as
	\begin{align}
		f_\phi &= \sum_{n=0}^\infty \frac{(-1)^n}{n!} \text{ad}^n_\phi f_0,
	\end{align}
so that the weak field expansion of the density operator can be presented as follows:
	\begin{align}
		\begin{split}
			\rho[\phi] &=  \langle f_\phi , \rho \rangle \\
			&=\rho^{(0)}[\phi]+\sum_{n=1}^{\infty} \frac{(-1)^n}{n!} \langle \rho, \text{ad}_\phi^{n-1} \{ \phi, f_0\} \rangle\\
			&=\rho^{(0)}[\phi]+\sum_{n=1}^\infty \rho^{(n)}[\phi].
		\end{split}
	\end{align}

\subsection{Cubic Hamiltonian}
Here we provide the details of the derivation of \eqref{eq:H3}.
The cubic term is given by
\begin{align}
S_H ^{(3)} = -\frac{1}{3!} \int \mathrm dt \braket{\{ f_0,\phi\},\{ \phi , \{ \phi, h\}\}},
\end{align}
where above we have used the relation $\braket{F,\{\phi, G\}} = -\braket{\{\phi, F\},G}$ for the pairing between $\mathfrak{g}$ and $\mathfrak{g}^*$.
Integrating by parts, we find that the integrand of the cubic Hamiltonian term simplifies:
\begin{align}
 &\braket{\{ \phi,f_0\},\{ \phi , \{ \phi, h\}\}} \notag \\
&=- \int\frac{\mathrm d\tau \mathrm dh \mathrm dm \mathrm d\varphi}{(2\pi)^2} \delta (h+ \mu) \frac{\partial \phi}{\partial \tau}
\bigg[ \left(\frac{\partial \phi}{\partial \tau}\frac{\partial^2 \phi}{\partial \tau \partial h}  - \frac{\partial \phi}{\partial h} \frac{\partial^2\phi }{\partial \tau^2}\right)+\left(  \frac{\partial \phi}{\partial \varphi} \frac{\partial^2 \phi}{\partial m \partial  \tau}-\frac{\partial \phi}{\partial m}\frac{\partial^2 \phi}{\partial \varphi \partial \tau} \right)\bigg] \notag \\
&= -\int\frac{\mathrm d\tau \mathrm dh \mathrm dm \mathrm d\varphi}{(2\pi)^2} \delta (h+ \mu)
\bigg[ \left(\frac{\partial \phi}{\partial \tau} \right)^2\frac{\partial^2 \phi}{\partial \tau \partial h}  - \frac{\partial \phi}{\partial h}\frac{\partial \phi}{\partial \tau} \frac{\partial^2\phi }{\partial \tau^2}\bigg] \notag \\
&= -\int\frac{\mathrm d\tau \mathrm dh \mathrm dm \mathrm d\varphi}{(2\pi)^2} \delta (h+ \mu)
\bigg[ \left(\frac{\partial \phi}{\partial \tau} \right)^2\frac{\partial^2 \phi}{\partial \tau \partial h}  - \frac{1}{2}\frac{\partial \phi}{\partial h}\frac{\partial}{\partial \tau} \left(\frac{\partial\phi }{\partial \tau} \right)^2\bigg] \notag \\
&= -\int\frac{\mathrm d\tau \mathrm dh \mathrm dm \mathrm d\varphi}{(2\pi)^2} \delta (h+ \mu)
\bigg[ \left(\frac{\partial \phi}{\partial \tau} \right)^2\frac{\partial^2 \phi}{\partial \tau \partial h}  + \frac{1}{2}\frac{\partial^2 \phi}{\partial \tau\partial h}\left(\frac{\partial\phi }{\partial \tau} \right)^2\bigg] \notag \\
&= - \frac{3}{2}\int\frac{\mathrm d\tau \mathrm dh \mathrm dm \mathrm d\varphi}{(2\pi)^2} \delta (h+ \mu)
\frac{1}{3}\frac{\partial}{\partial h} \left(\frac{\partial \phi}{\partial \tau} \right)^3 \notag \\
&=  \frac{1}{2}\int\frac{\mathrm d\tau \mathrm dh \mathrm dm \mathrm d\varphi}{(2\pi)^2} \delta '(h+ \mu)
 \left(\frac{\partial \phi}{\partial \tau} \right)^3 . \label{eq:h3detail}
\end{align}
Note that the $m$ and $\varphi$ derivative terms vanish after integrating by parts.
For $c=1$ string theory, we can carry out the same manipulations in order to simplify the corresponding cubic term. The Hamiltonian term in the noncritical M-theory can now be evaluated using the following manipulations:
\begin{align}
\begin{split}
S_H ^{(3)} &=  \frac{1}{2\cdot 3!} \int \mathrm dt \frac{\mathrm dh \mathrm d\tau  \mathrm dm \mathrm d\varphi}{(2\pi)^2}  \delta'(h+\mu)  \left(\frac{\partial \phi}{\partial \tau} \right)^3  \\
&= \frac{1}{2\cdot 3!} \int \mathrm dt \frac{p \mathrm dp \mathrm d\xi \mathrm d^2\boldsymbol\lambda}{(2\pi)^2}  \frac{1}{\frac{\partial h}{\partial p}} \frac{\partial }{\partial p}\left( \frac{1}{\frac{\partial h}{\partial p}} \delta(p-p_F)\right) (v_h \phi)^3  \\
&= \frac{1}{2\cdot 3!} \int \mathrm dt \frac{\mathrm  dp \mathrm d\xi \mathrm d^2\boldsymbol \lambda}{(2\pi)^2}  \frac{\partial }{\partial p}\left( \frac{1}{p} \delta(p-p_F)\right) (v_h \phi)^3  \\
&= -\frac{1}{2\cdot 3!} \int \mathrm dt \frac{\mathrm dp \mathrm d\xi \mathrm d^2 \boldsymbol \lambda}{(2\pi)^2} \left( \frac{1}{p} \delta(p-p_F)\right) \frac{\partial }{\partial p} (v_h \phi)^3  \\
&= -\frac{1}{2\cdot 3!} \int \mathrm dt \frac{ \mathrm dp \mathrm d\xi \mathrm d^2 \boldsymbol \lambda}{(2\pi)^2} \left( \frac{1}{p} \delta(p-p_F)\right) 3 (v_h \phi)^2 \frac{\partial}{\partial p} v_h \phi  \\
&= -\frac{3}{2\cdot 3!} \int \mathrm dt \frac{\mathrm  dp \mathrm d\xi \mathrm d^2 \boldsymbol \lambda}{(2\pi)^2} \left( \frac{1}{p} \delta(p-p_F)\right)  (v_h \phi)^2 \left([\frac{\partial}{\partial p}, v_h] \phi + v_h \frac{\partial \phi}{\partial p} \right)  \\
&= -\frac{3}{ 2 \cdot 3!} \int \mathrm dt \frac{\mathrm d^2 \boldsymbol \lambda \mathrm d\xi}{(2\pi)^2} \frac{1}{p_F} (p_F^2 R \phi)^2 \left( \boldsymbol n_\xi \cdot \partial_{\boldsymbol \lambda} \phi - \frac{\bm{F}\cdot \bm{t}_\xi}{p_F^2}\frac{\partial \phi}{\partial \xi} \right)  \\
&=  -\frac{3}{ 2 \cdot 3!} \int \mathrm dt \frac{\mathrm d^2\boldsymbol \lambda \mathrm d\xi}{(2\pi)^2} \frac{1}{p_F^2} (p_F^2 R \phi)^2 \left ( p_F^2 R \phi- 2\frac{p_F \boldsymbol t_\xi \cdot \partial_{\boldsymbol \lambda} p_F}{p_F^2}\frac{\partial \phi}{\partial \xi}\right)  \\
&=  -\frac{3}{ 2 \cdot 3!} \int \mathrm dt \frac{\mathrm d^2\boldsymbol \lambda \mathrm d\xi}{(2\pi)^2} \frac{1}{p_F^2} \left( \frac{\partial \phi}{\partial \tau} \right)^2 \left ( \frac{\partial \phi}{\partial \tau}- 2\frac{ \boldsymbol t_\xi \cdot \partial_{\boldsymbol \lambda} p_F}{p_F}\frac{\partial \phi}{\partial \xi} \right).
\end{split}
\label{eq:H3derivation}
\end{align}
Here
\begin{align}
v_h = \frac{\partial}{\partial \tau}  = \bm{p}\cdot \partial_{\bm{\lambda}} + \bm{F}\cdot \partial_{\bm{p}}, \qquad \bm{F} = -\partial_{\bm{\lambda}} V(\bm{\lambda}) = \bm{\lambda},
\end{align}
is a Hamiltonian vector field on the phase space generated by $h$ and $R$ is the Reeb vector field.
This completes the derivation of the final expression \eqref{eq:H3}.

\section{Power Counting for Eigenvalue Density Operators}

In this section, we show that the perturbative expansion of the bosonized noncritical M-theory action is controlled by inverse powers of the Fermi momentum $p_F$.\footnote{The Fermi momentum $p_F$ is equal to the Fermi velocity $v_{F}$ in our conventions, as we set the single particle mass equal to unity.} This justifies our truncation of the bosonized action at leading nontrivial (i.e., cubic) order when studying the correlation functions of the eigenvalue density operators discussed in Section \ref{bosonicdensitycorr}.

Let us recall from (\ref{Skexp}) and (\ref{rhokexp}) that we can write the bosonized action and density operators as
	\begin{align}
		\begin{split}
			S &= S^{(0)} + \sum_{k=2}^{\infty} S^{(k)} ,~~~~ S^{(k)} = \frac{1}{k!} \int \mathrm dt \, \langle \text{ad}_\phi^{k-2} D_t \phi , \{ f_0,\phi \} \rangle\\
			\rho[\phi]&= \rho^{(0)} + \sum_{k=1}^{\infty} \rho^{(k)}[\phi] ,~~~~ \rho^{(k)}[\phi] =\frac{1}{k!} \langle \text{ad}_\phi^{k-1}  \rho , \{ f_0,\phi\} \rangle,
		\end{split}
	\end{align}
where in the first line above we have defined $D_t \equiv -\partial_t + \{ h, \cdot\}$, with $h$ being the single particle Hamiltonian. Setting $f_0=\theta(-h-\mu)$ and working in action-angle coordinates, we may write
	\begin{align}
		\{ \phi, f_0\} = - \partial_{\tau} \phi \delta(h+ \mu ),
	\end{align}
where $\{\tau,h\} = 1$.
This implies that the $k$th order corrections can be expressed as follows:
	\begin{align}
		\begin{split}
			S^{(k)} &=- \frac{1}{k!} \int \mathrm dt \frac{\mathrm dvol_{FS}}{(2\pi)^2} \,\frac{\partial \phi}{\partial \tau}  \text{ad}_\phi^{k-2} \left(\frac{\partial \phi}{\partial t} + \frac{\partial \phi}{\partial \tau}\right) \\
			\rho^{(k)}[\phi]&= \frac{1}{k!} \int \frac{\mathrm dvol_{FS}}{(2\pi)^2} \,  \frac{\partial \phi}{\partial \tau}  \text{ad}_\phi^{k-1} \rho.
		\end{split}
	\end{align}
Now, we specialize to the basis of densities $\rho = \delta^2(\boldsymbol \lambda - \bar{\boldsymbol{\lambda}})$ and work in $p$-gauge, $\phi = \phi(t, \lambda_r,\theta, \xi)$ on the Fermi surface. Our goal is to determine the scaling of $S^{(k)}, \rho^{(k)}$ at large $\lambda_r$.

We will need a few results. First, observe that the rectilinear coordinate partial derivatives of a function $F = F(\lambda_r, \theta, \xi)$ can be expressed as follows in doubly-polar coordinates:
	\begin{align}
		\begin{split}
			\frac{\partial F}{\partial \lambda_1}  &= \cos \theta \frac{\partial F}{\partial \lambda_r} - \frac{\sin \theta}{\lambda_r} \frac{\partial F}{\partial  \theta},~~~~ \frac{\partial F}{\partial \lambda_2} = \sin \theta \frac{\partial F}{\partial \lambda_r} + \frac{\cos \theta}{\lambda_r} \frac{\partial F}{\partial \theta} \\
			\frac{\partial F}{\partial p_1}  &= - \frac{\sin \xi}{p_F} \frac{\partial F}{\partial \xi},~~~~ \frac{ \partial F}{\partial p_2} =\frac{\cos \xi}{p_F} \frac{\partial F}{\partial \xi}.
		\end{split}
	\end{align}
It follows that the Poisson bracket of a pair of functions $F(\lambda_r,\theta, \xi), F'(\lambda_r, \theta, \xi)$ is given by
	\begin{align}
		\begin{split}
			\{ F, F' \}    \equiv \frac{1}{\lambda_r p_F} B(F,F'),
        \end{split}
    \end{align}
where above
    \begin{align}
        \begin{split}
        B(F,F')&= \cos(\theta -\xi) \left( \frac{\partial F}{\partial \theta} \frac{\partial F'}{\partial \xi} - \frac{\partial F}{\partial \xi} \frac{\partial F'}{\partial \theta}\right)\\
        &~~+ \sin(\theta -\xi) \left( \frac{\partial F}{\partial \log \lambda_r} \frac{\partial F'}{\partial \xi} - \frac{\partial F}{\partial \xi} \frac{\partial F'}{\partial  \log \lambda_r}  \right).
		\end{split}
	\end{align}
At large $\lambda_r$, the Fermi momentum $p_F$ scales as $\lambda_r$ on the Fermi surface, so that
	\begin{align}
		\{ F, F' \} \sim \frac{1}{\lambda_r^2} ( \cdots ),~~~~ \lambda_r\to \infty.
	\end{align}
Second, observe that
	\begin{align}
		\{ F, p \} \equiv \frac{1}{\lambda_r} C(F), ~~~~C(F) = \cos(\theta -\xi) \frac{\partial F}{\partial \log \lambda_r}  - \sin(\theta-\xi) \frac{\partial F}{\partial \theta}.
	\end{align}
Finally, let us recursively define
	\begin{align}
		F_{k+1}(x) \equiv \frac{1}{\lambda_r} \left[ B(x, F_k) - k F_k C(x) \right].
	\end{align}
Using the above results, it is straightforward to show by induction that given a base function $F_0(\lambda_r,\theta,\xi)$, the $N$-times composition of the operator $\text{ad}_\phi$ acting on $F_0$ goes like
	\begin{align}
		\text{ad}_\phi^N F_0 = \frac{1}{p_F^N} F_N \sim \frac{1}{p_F^N \lambda_r^N} ( \cdots).
	\end{align}
The base case is trivial. The inductive step is to assume that $\text{ad}_\phi^{N-1} F_0 = F_{N-1}/ p_F^{N-1}$. Then
	\begin{align}
    \begin{split}
		\text{ad}_\phi^{N} F_0 &= \{ \phi, F_{N-1} / p_F^{N-1} \} \\
		&= \frac{1}{p_F^{N-1}} \{ \phi, F_{N-1}\} + F_{N-1} \{ \phi, 1/p_F^{N-1} \}  \\
		&=\frac{1}{p_F^{N-1}} \frac{1}{\lambda_r p_F} B(\phi , F_{N-1}) - \frac{N-1}{p_F^{N}} F_{N-1}  \{\phi, p \} \\
		&=\frac{1}{\lambda_r p_F^{N}} B(\phi, F_{N-1}) - \frac{N-1}{\lambda_r p_F^{N}} F_{N-1}C(\phi)\\
		&= \frac{1}{p_F^N} F_N.
    \end{split}
	\end{align}
The above results imply that at large $\lambda_r$ and ignoring the integration measure in $S^{(k)}$,
	\begin{align}
		S^{(k)} \sim \frac{1}{p_F^{2k-4}} ,~~~~ \rho^{(k)} \sim \frac{1}{p_F^{2k-2}}.
	\end{align}
This in particular justifies truncating the $p$-gauge action at cubic order as well as organizing the eigenvalue density correlation functions in terms of a $1/p_F$ expansion---see Figure \ref{hgaugegraph}.

\section{Miscellaneous  Results}

\subsection{Complex exponential integrals}

The following integrals appear when computing propagators in action-angle coordinates (see Section \ref{actionangle}):
    \begin{align}
        \int_{-\infty}^{-\mu} \mathrm dE\, e^{i E u } ,~~~~ \int_{-\mu}^\infty \mathrm dE\, e^{i E u}, ~~~~ u = -t + \tau.
    \end{align}
 There are two possible situations that we could encounter, depending on whether or not $t$ is real. For real time, we need to use a prescription to get the integrals to converge, and a standard approach is to use an $i \epsilon$ prescription, so that, e.g.,
		\begin{align}
			\begin{split}
			\int_{-\infty}^{-\mu} \mathrm dE \, e^{i E u}	&=\lim_{\epsilon \to 0^+} \int_{-\infty}^{-\mu} \mathrm dE \, e^{ (i u + \epsilon) E}
				=  e^{- i \mu u } \biggl[ PV \frac{1}{iu} +  \pi \delta(u) \biggr],\\
				\int_{-\mu}^{\infty} \mathrm dE \, e^{i E u}	&=\lim_{\epsilon \to 0^+} \int_{-\mu}^{\infty}\mathrm  dE \, e^{ (i u - \epsilon) E} = e^{-i \mu u } \biggl[ - PV\frac{1}{iu} +  \pi \delta(u) \biggr].
			\end{split}
		\end{align}
For imaginary time $t = -ix$, we have
    \begin{align}
        \begin{split}
        \label{Eint}
          \int_{-\mu}^{\infty} \mathrm dE \, e^{i E u} \theta(x)&= -\frac{e^{-i \mu u}}{ iu } \theta(x),~~~~   \int_{-\infty}^{-\mu} \mathrm dE \, e^{i Eu}\theta(-x) = \frac{e^{-i \mu u}}{i u} \theta(-x).
        \end{split}
    \end{align}

\subsection{Chiral boson propagator}

\label{app:bosonprop}
The quadratic bosonized action for noncritical M-theory was found to take the following form in action-angle coordinates $\tau, \varphi,m$:
    \begin{align}
		S^{(2)}[\phi] 
         =- \frac{1}{2} \int \mathrm dt \, \frac{\mathrm d \tau \mathrm d\varphi  \mathrm dm}{(2\pi)^2} \, \frac{\partial \phi}{\partial \tau} \left(\frac{\partial \phi}{\partial t} +  \frac{\partial \phi}{\partial \tau}\right).
    \end{align}
After integrating by parts, the quadratic action takes the following form:
\begin{align}
		\begin{split}
			S^{(2)} &\cong\int \mathrm dt \mathrm d\tau \mathrm d\varphi \mathrm dm \,\frac{1}{2} \phi    K  \phi,~~~~~    K =  \frac{1}{(2\pi)^2}\frac{\partial}{\partial \tau}  \left(
        \frac{\partial}{\partial t} + \frac{\partial}{\partial \tau} \right).
		\end{split}
	\end{align}
The two-dimensional part of the above action is the Floreanini-Jackiw-type action for a chiral boson \cite{Floreanini:1987as}. Note that since the single particle Hamiltonian $h$ is unbounded, the angle variable $\tau$ is non-compact, whereas $\varphi$ may be either compact or non-compact. Using the defining property $K G = \delta^4$, the time-ordered Green function defined with respect to $ K$ admits the following momentum space representation,
    \begin{align}
            \begin{split}
              G= -\delta(\varphi) \delta(m) \int \mathrm d\omega \mathrm dk \, \frac{e^{- i (\omega t + k \tau)} }{k (k+\omega) + i \epsilon},
            \end{split}
    \end{align}
where we have included an $i\epsilon$ to implement time-ordering before continuing to Euclidean time. After evaluating the integral over $\omega$ and analytically-continuing to Euclidean time $x = i t$, the above momentum space integral becomes
    \begin{align}
          2\pi \int \frac{\mathrm dk}{k} \, e^{i k u } [\theta(-x) \theta(-k)-\theta(x) \theta(k)],~~~~ u= i x + \tau.
    \end{align}
The above expression contains an IR divergence corresponding to the zero mode $k=0$. In order to eliminate this divergence, we compute a derivative of the above integral with respect to $u$  and use the identity (\ref{Eint}) to evaluate the resulting integral. As a result, we find that
    \begin{align}
      2\pi  \frac{\partial}{\partial u}  \int \frac{\mathrm dk}{k} \, e^{i k u} [ \theta(-x) \theta(-k) - \theta(x) \theta(k) ] =  \frac{1}{u}.
    \end{align}
Integrating with respect to $u$ and dropping the integration constant, we find that the chiral Euclidean propagator is given by
    \begin{align}
    \label{aaprop}
      \langle T (\phi(f) \phi(i)) \rangle =- 2\pi \delta(\Delta \varphi) \delta(\Delta m) \log \Delta u,
    \end{align}
where we have taken $\epsilon \to 0$, and $\Delta u \equiv u_f - u_i$, etc.

\section{More on the Resolvent \label{sec:resolvent2nd}}
In this section, we study the contribution to the stationary phase approximation of the noncritical M-theory resolvent coming from the second saddle point.

\begin{figure}[t]
\begin{center}
\includegraphics[width=6.5cm]{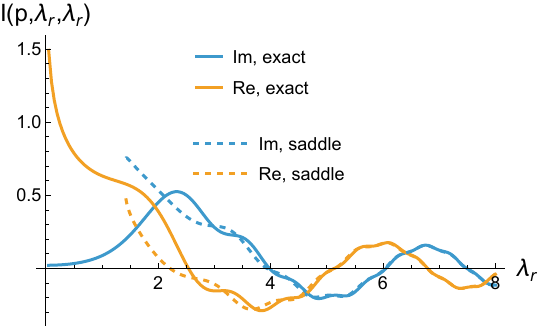}
\includegraphics[width=7.0cm]{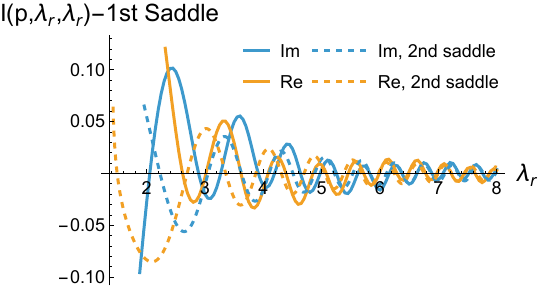}
\caption{
Comparison between the exact resolvent and the saddle-point approximation including both the first and second saddle contributions.
The left panel compares the exact resolvent with the two-saddle approximation.
The right panel isolates the second saddle contribution by subtracting the leading saddle contribution from the exact resolvent, and compares the remainder with the second saddle approximation.
The parameters are chosen as $p=0.001$, $\theta=\pi/20$ and $\mu=1$.
The agreement shows that the second saddle accurately reproduces the small rapid oscillations of the exact resolvent. }
\label{fig:2ndResolventp1}
\end{center}
\end{figure}
We can try to find another saddle for $\bm{\lambda}_ i \approx  \bm{\lambda}_f$ and $p \approx 0$.
The saddle-point equation in this regime is
\begin{align}
S'(u) \approx    \frac{1}{4} (\bm{\lambda}_i + \bm{\lambda}_f)^2   - (\mu- ip) \left (\frac{1}{1-u} + \frac{1}{1+u} \right)  + \frac{i}{u}.
\end{align}
Ignoring the last term assuming that both $ (\bm{\lambda}_i + \bm{\lambda}_f)^2 $ and $\mu$ are large, we obtain the solution
\begin{align}
u_1 = \pm \sqrt{1 - \frac{8(\mu-ip)}{(\bm{\lambda}_i + \bm{\lambda}_f)^2}}.
\end{align}
The branch cut is taken to be $\sqrt{\lambda_r e^{i\theta}} = \sqrt{\lambda_r} e^{i \frac{\theta}{2}} $.
The saddle point value  is
\begin{align}
S(u_1) &\approx \pm \frac{1}{4} \sqrt{(\bm{\lambda}_i + \bm{\lambda}_f)^2}\sqrt{(\bm{\lambda}_i + \bm{\lambda}_f)^2 - 8 (\mu-ip)}  \notag \\
& \quad -i (p + i \mu ) \log \left(1- \sqrt{1 - \frac{8(\mu-ip)}{(\bm{\lambda}_i + \bm{\lambda}_f)^2}} \right) + i \log \sqrt{1 - \frac{8(\mu-ip)}{(\bm{\lambda}_i + \bm{\lambda}_f)^2}} .
\end{align}
The plus sign saddle provides an exponentially small correction.
The one-loop determinant part becomes
\begin{align}
S''(u_1) & \approx - (\mu - ip) \left( \frac{1}{(1-u_1)^2} -  \frac{1}{(1 +u_1)^2} \right)  - i \frac{1}{u_1^2} \notag \\
&= - \frac{u_1 (\bm{\lambda}_i + \bm{\lambda}_f)^4}{16(\mu-ip)} -  \frac{i}{u_1^2},
\end{align}
and therefore the contribution to the resolvent takes the form
\begin{align}
&\frac{1}{2\pi }\sqrt{\frac{2\pi i}{{S''(u_1)}}} e^{i S(u_1) } \notag \\
& \approx \frac{1}{2\pi u_1} \sqrt{\frac{-32 \pi i (\mu -ip)}{u_1 (\bm{\lambda}_i + \bm{\lambda}_f)^4}} \left( \frac{ 1- u_1}{1+ u_1} \right)^{p + i \mu} \exp\bigg(\frac{i}{4} \sqrt{(\bm{\lambda}_i + \bm{\lambda}_f)^2}\sqrt{(\bm{\lambda}_i + \bm{\lambda}_f)^2 - 8 (\mu-ip)}  \bigg).
\end{align}
Including the contribution from the saddle  $u = u_1$ to \eqref{eq:RNMsaddle1}, the full resolvent is approximated as
\begin{align}
I(p,\bm{\lambda}_f,\bm{\lambda}_i) \approx    \frac{1}{2\pi}\sqrt{\frac{2\pi i}{{S''(u_0)}}} e^{i S(u_0) } + \frac{1}{2\pi }\sqrt{\frac{2\pi i}{{S''(u_1)}}} e^{i S(u_1) }.
\end{align}
We compare the exact resolvents and the saddle point approximations for them in Figure \ref{fig:2ndResolventp1}.
This second saddle reproduces the small rapid oscillations, similar to brane effects \cite{Saad:2019lba} in noncritical string theories.

\bibliography{NCMtheory}

@book{Dirac:1958qm,
  author    = {Dirac, P. A. M.},
  title     = {The Principles of Quantum Mechanics},
  edition   = {4th},
  series    = {The International Series of Monographs on Physics},
  publisher = {Clarendon Press},
  address   = {Oxford},
  year      = {1958}
}

@article{Coleman:1969sm,
    author = "Coleman, Sidney R. and Wess, J. and Zumino, Bruno",
    title = "{Structure of phenomenological Lagrangians. 1.}",
    doi = "10.1103/PhysRev.177.2239",
    journal = "Phys. Rev.",
    volume = "177",
    pages = "2239--2247",
    year = "1969"
}

@article{Callan:1969sn,
    author = "Callan, Jr., Curtis G. and Coleman, Sidney R. and Wess, J. and Zumino, Bruno",
    title = "{Structure of phenomenological Lagrangians. 2.}",
    doi = "10.1103/PhysRev.177.2247",
    journal = "Phys. Rev.",
    volume = "177",
    pages = "2247--2250",
    year = "1969"
}

@article{Jevicki:1979mb,
    author = "Jevicki, A. and Sakita, B.",
    editor = "Kikkawa, K. and Virasoro, M. and Wadia, S. R.",
    title = "{The Quantum Collective Field Method and Its Application to the Planar Limit}",
    reportNumber = "BROWN HET-397",
    doi = "10.1016/0550-3213(80)90046-2",
    journal = "Nucl. Phys. B",
    volume = "165",
    pages = "511",
    year = "1980"
}

@article{Sakita:1979gs,
    author = "Sakita, B.",
    editor = "Kikkawa, K. and Virasoro, M. and Wadia, S. R.",
    title = "{Field Theory of Strings as a Collective Field Theory of U($N$) Gauge Field}",
    reportNumber = "CCNY-HEP-79/8",
    doi = "10.1103/PhysRevD.21.1067",
    journal = "Phys. Rev. D",
    volume = "21",
    pages = "1067",
    year = "1980"
}

@article{Polyakov:1981rd,
    author = "Polyakov, Alexander M.",
    editor = "Khalatnikov, I. M. and Mineev, V. P.",
    title = "{Quantum Geometry of Bosonic Strings}",
    reportNumber = "Print-81-0351 (LANDAU INST)",
    doi = "10.1016/0370-2693(81)90743-7",
    journal = "Phys. Lett. B",
    volume = "103",
    pages = "207--210",
    year = "1981"
}

@article{Polyakov:1981re,
    author = "Polyakov, Alexander M.",
    editor = "Khalatnikov, I. M. and Mineev, V. P.",
    title = "{Quantum Geometry of Fermionic Strings}",
    reportNumber = "Print-81-0350 (LANDAU INST)",
    doi = "10.1016/0370-2693(81)90744-9",
    journal = "Phys. Lett. B",
    volume = "103",
    pages = "211--213",
    year = "1981"
}

@article{Witten:1983ar,
    author = "Witten, Edward",
    editor = "Stone, M.",
    title = "{Nonabelian Bosonization in Two-Dimensions}",
    reportNumber = "PRINT-83-0934 (PRINCETON)",
    doi = "10.1007/BF01215276",
    journal = "Commun. Math. Phys.",
    volume = "92",
    pages = "455--472",
    year = "1984"
}

@article{Floreanini:1987as,
    author = "Floreanini, R. and Jackiw, R.",
    title = "{Selfdual Fields as Charge Density Solitons}",
    journal = "Phys. Rev. Lett.",
    volume = "59",
    pages = "1873",
    year = "1987",
    doi = "10.1103/PhysRevLett.59.1873"
}

@article{Douglas:1989ve,
    author = "Douglas, Michael R. and Shenker, Stephen H.",
    editor = "Brezin, E. and Wadia, S. R.",
    title = "{Strings in Less Than One-Dimension}",
    reportNumber = "RU-89-34",
    doi = "10.1016/0550-3213(90)90522-F",
    journal = "Nucl. Phys. B",
    volume = "335",
    pages = "635",
    year = "1990"
}

@article{Das:1990kaa,
    author = "Das, Sumit R. and Jevicki, Antal",
    editor = "Brezin, E. and Wadia, S. R.",
    title = "{String Field Theory and Physical Interpretation of $D=1$ Strings}",
    reportNumber = "BROWN-HET-750, TIFR-TH-90-26",
    doi = "10.1142/S0217732390001888",
    journal = "Mod. Phys. Lett. A",
    volume = "5",
    pages = "1639--1650",
    year = "1990"
}

@article{Gross:1990md,
    author = "Gross, David J. and Klebanov, Igor R.",
    title = "{One-dimensional string theory on a circle}",
    journal = "Nucl. Phys. B",
    volume = "344",
    pages = "475--498",
    year = "1990",
    doi = "10.1016/0550-3213(90)90672-Z"
}

@article{Bershadsky:1990gs,
  author  = {Bershadsky, Michael and Klebanov, Igor R.},
  title   = {Genus one path integral in two-dimensional quantum gravity},
  journal = {Phys. Rev. Lett.},
  volume  = {65},
  pages   = {3088--3091},
  year    = {1990},
  doi     = {10.1103/PhysRevLett.65.3088}
}

@article{Sengupta:1990bt,
    author = "Sengupta, Anirvan M. and Wadia, Spenta R.",
    title = "{Excitations and interactions in d = 1 string theory}",
    reportNumber = "TIFR-TH-90-33",
    doi = "10.1142/S0217751X91000988",
    journal = "Int. J. Mod. Phys. A",
    volume = "6",
    pages = "1961--1984",
    year = "1991"
}

@article{Polchinski:1991uq,
    author = "Polchinski, Joseph",
    editor = "Brezin, E. and Wadia, S. R.",
    title = "{Classical limit of (1+1)-dimensional string theory}",
    reportNumber = "UTTG-06-91",
    doi = "10.1016/0550-3213(91)90559-G",
    journal = "Nucl. Phys. B",
    volume = "362",
    pages = "125--140",
    year = "1991"
}

@article{Gross:1990st,
    author = "Gross, David J. and Klebanov, Igor R.",
    title = "{Fermionic string field theory of c = 1 two-dimensional quantum gravity}",
    reportNumber = "PUPT-1198",
    doi = "10.1016/0550-3213(91)90103-5",
    journal = "Nucl. Phys. B",
    volume = "352",
    pages = "671--688",
    year = "1991"
}

@article{Moore:1991ir,
  author  = {Moore, Gregory W. and Seiberg, Nathan and Staudacher, Matthias},
  title   = {From loops to states in two-dimensional quantum gravity},
  journal = {Nucl. Phys. B},
  volume  = {362},
  pages   = {665--709},
  year    = {1991},
  doi     = {10.1016/0550-3213(91)90548-C}
}

@inproceedings{Klebanov:1991qa,
    author = "Klebanov, Igor R.",
    title = "{String theory in two-dimensions}",
    booktitle = "{Spring School on String Theory and Quantum Gravity (to be followed by Workshop)}",
    eprint = "hep-th/9108019",
    archivePrefix = "arXiv",
    reportNumber = "PUPT-1271",
    month = "7",
    year = "1991"
}

@article{Das:1991uta,
    author = "Das, Sumit R. and Dhar, Avinash and Mandal, Gautam and Wadia, Spenta R.",
    title = "{Bosonization of nonrelativistic fermions and W infinity algebra}",
    eprint = "hep-th/9111021",
    archivePrefix = "arXiv",
    reportNumber = "TIFR-TH-91-51, IASSNS-HEP-91-72",
    doi = "10.1142/S021773239200344X",
    journal = "Mod. Phys. Lett. A",
    volume = "7",
    pages = "71--84",
    year = "1992"
}

@article{DiFrancesco:1991daf,
    author = "Di Francesco, P. and Kutasov, D.",
    title = "{World sheet and space-time physics in two-dimensional (Super)string theory}",
    eprint = "hep-th/9109005",
    archivePrefix = "arXiv",
    reportNumber = "PUPT-1276",
    doi = "10.1016/0550-3213(92)90337-B",
    journal = "Nucl. Phys. B",
    volume = "375",
    pages = "119--170",
    year = "1992"
}

@article{Moore:1991zv,
  author  = {Moore, Gregory W.},
  title   = {Double scaled field theory at c = 1},
  journal = {Nucl. Phys. B},
  volume  = {368},
  pages   = {557--590},
  year    = {1992},
  doi     = {10.1016/0550-3213(92)90214-V}
}

@article{Moore:1991sf,
    author = "Moore, Gregory W. and Plesser, M. Ronen and Ramgoolam, Sanjaye",
    title = "{Exact S matrix for 2-D string theory}",
    eprint = "hep-th/9111035",
    archivePrefix = "arXiv",
    reportNumber = "YCTP-P35-91",
    doi = "10.1016/0550-3213(92)90020-C",
    journal = "Nucl. Phys. B",
    volume = "377",
    pages = "143--190",
    year = "1992"
}

@article{Dhar:1992hr,
    author = "Dhar, Avinash and Mandal, Gautam and Wadia, Spenta R.",
    title = "{Nonrelativistic fermions, coadjoint orbits of W(infinity) and string field theory at c = 1}",
    eprint = "hep-th/9207011",
    archivePrefix = "arXiv",
    reportNumber = "TIFR-TH-92-40",
    doi = "10.1142/S0217732392002512",
    journal = "Mod. Phys. Lett. A",
    volume = "7",
    pages = "3129--3146",
    year = "1992"
}

@inproceedings{Ginsparg:1993is,
    author = "Ginsparg, Paul H. and Moore, Gregory W.",
    title = "{Lectures on 2-D gravity and 2-D string theory}",
    booktitle = "{Theoretical Advanced Study Institute (TASI 92): From Black Holes and Strings to Particles}",
    eprint = "hep-th/9304011",
    archivePrefix = "arXiv",
    reportNumber = "YCTP-P23-92, LA-UR-92-3479",
    pages = "277--469",
    month = "10",
    year = "1993"
}

@article{Demeterfi:1993cm,
  author        = {Demeterfi, Kresimir and Klebanov, Igor R. and Rodrigues, Joao P.},
  title         = {The exact S matrix of the deformed c = 1 matrix model},
  journal       = {Phys. Rev. Lett.},
  volume        = {71},
  pages         = {3409--3412},
  year          = {1993},
  eprint        = {hep-th/9308036},
  archivePrefix = {arXiv},
  doi           = {10.1103/PhysRevLett.71.3409}
}

@inproceedings{Jevicki:1993qn,
    author = "Jevicki, Antal",
    title = "{Development in 2-d string theory}",
    booktitle = "{Workshop on String Theory, Gauge Theory and Quantum Gravity}",
    eprint = "hep-th/9309115",
    archivePrefix = "arXiv",
    reportNumber = "BROWN-HET-918, TA-502",
    doi = "10.1142/9789814447072_0004",
    month = "9",
    year = "1993"
}

@article{Jevicki:1993zg,
    author = "Jevicki, Antal and Yoneya, Tamiaki",
    title = "{A Deformed matrix model and the black hole background in two-dimensional string theory}",
    eprint = "hep-th/9305109",
    archivePrefix = "arXiv",
    reportNumber = "NSF-ITP-93-67, BROWN-HEP-904, UT-KOMABA-93-10",
    doi = "10.1016/0550-3213(94)90054-X",
    journal = "Nucl. Phys. B",
    volume = "411",
    pages = "64--96",
    year = "1994"
}

@article{Natsuume:1994sp,
    author = "Natsuume, Makoto and Polchinski, Joseph",
    title = "{Gravitational scattering in the c = 1 matrix model}",
    eprint = "hep-th/9402156",
    archivePrefix = "arXiv",
    reportNumber = "NSF-ITP-94-19",
    doi = "10.1016/0550-3213(94)90092-2",
    journal = "Nucl. Phys. B",
    volume = "424",
    pages = "137--154",
    year = "1994"
}

@inproceedings{Polchinski:1994mb,
    author = "Polchinski, Joseph",
    title = "{What is string theory?}",
    booktitle = "{NATO Advanced Study Institute: Les Houches Summer School, Session 62: Fluctuating Geometries in Statistical Mechanics and Field Theory}",
    eprint = "hep-th/9411028",
    archivePrefix = "arXiv",
    reportNumber = "NSF-ITP-94-97",
    month = "11",
    year = "1994"
}

@article{Hull:1994ys,
    author = "Hull, C. M. and Townsend, P. K.",
    title = "{Unity of superstring dualities}",
    eprint = "hep-th/9410167",
    archivePrefix = "arXiv",
    reportNumber = "QMW-94-30, DAMTP-R-94-33",
    doi = "10.1016/0550-3213(94)00559-W",
    journal = "Nucl. Phys. B",
    volume = "438",
    pages = "109--137",
    year = "1995"
}

@article{Townsend:1995kk,
    author = "Townsend, P. K.",
    title = "{The eleven-dimensional supermembrane revisited}",
    eprint = "hep-th/9501068",
    archivePrefix = "arXiv",
    reportNumber = "DAMTP-R-95-2",
    doi = "10.1016/0370-2693(95)00397-4",
    journal = "Phys. Lett. B",
    volume = "350",
    pages = "184--187",
    year = "1995"
}

@article{Witten:1995ex,
    author = "Witten, Edward",
    title = "{String theory dynamics in various dimensions}",
    eprint = "hep-th/9503124",
    archivePrefix = "arXiv",
    reportNumber = "IASSNS-HEP-95-18",
    doi = "10.1016/0550-3213(95)00158-O",
    journal = "Nucl. Phys. B",
    volume = "443",
    pages = "85--126",
    year = "1995"
}

@article{Dhar:1995nq,
    author = "Dhar, Avinash and Mandal, Gautam and Wadia, Spenta R.",
    title = "{Discrete state moduli of string theory from the C=1 matrix model}",
    eprint = "hep-th/9507041",
    archivePrefix = "arXiv",
    reportNumber = "CERN-TH-95-186, TIFR-TH-95-30",
    doi = "10.1016/0550-3213(95)00493-C",
    journal = "Nucl. Phys. B",
    volume = "454",
    pages = "541--560",
    year = "1995"
}

@article{Horava:1995qa,
    author = "Horava, Petr and Witten, Edward",
    title = "{Heterotic and Type I string dynamics from eleven dimensions}",
    eprint = "hep-th/9510209",
    archivePrefix = "arXiv",
    reportNumber = "IASSNS-HEP-95-86, PUPT-1571A",
    doi = "10.1016/0550-3213(95)00621-4",
    journal = "Nucl. Phys. B",
    volume = "460",
    pages = "506--524",
    year = "1996"
}

@article{Banks:1996vh,
    author = "Banks, T. and Fischler, W. and Shenker, S. H. and Susskind, L.",
    title = "{M theory as a matrix model: A Conjecture}",
    eprint = "hep-th/9610043",
    archivePrefix = "arXiv",
    doi = "10.1103/PhysRevD.55.5112",
    journal = "Phys. Rev. D",
    volume = "55",
    pages = "5112--5128",
    year = "1997"
}

@article{Maldacena:1997re,
    author = "Maldacena, Juan M.",
    title = "{The Large N limit of superconformal field theories and supergravity}",
    eprint = "hep-th/9711200",
    archivePrefix = "arXiv",
    doi = "10.1023/A:1026654312961",
    journal = "Adv. Theor. Math. Phys.",
    volume = "2",
    pages = "231--252",
    year = "1998"
}

@article{Itzhaki:1998dd,
    author = "Itzhaki, N. and Maldacena, Juan M. and Sonnenschein, J. and Yankielowicz, S.",
    title = "{Supergravity and the large N limit of theories with sixteen supercharges}",
    eprint = "hep-th/9802042",
    archivePrefix = "arXiv",
    doi = "10.1103/PhysRevD.58.046004",
    journal = "Phys. Rev. D",
    volume = "58",
    pages = "046004",
    year = "1998"
}

@article{Berenstein:2002jq,
    author = "Berenstein, David Eliecer and Maldacena, Juan Martin and Nastase, Horatiu Stefan",
    title = "{Strings in flat space and pp waves from N=4 superYang-Mills}",
    eprint = "hep-th/0202021",
    archivePrefix = "arXiv",
    doi = "10.1088/1126-6708/2002/04/013",
    journal = "JHEP",
    volume = "04",
    pages = "013",
    year = "2002"
}

@article{McGreevy:2003kb,
    author = "McGreevy, John and Verlinde, Herman L.",
    title = "{Strings from tachyons: The c=1 matrix reloaded}",
    eprint = "hep-th/0304224",
    archivePrefix = "arXiv",
    reportNumber = "PUPT-2083",
    doi = "10.1088/1126-6708/2003/12/054",
    journal = "JHEP",
    volume = "12",
    pages = "054",
    year = "2003"
}

@article{Klebanov:2003km,
    author = "Klebanov, Igor R. and Maldacena, Juan Martin and Seiberg, Nathan",
    title = "{D-brane decay in two-dimensional string theory}",
    eprint = "hep-th/0305159",
    archivePrefix = "arXiv",
    reportNumber = "PUPT-2085",
    doi = "10.1088/1126-6708/2003/07/045",
    journal = "JHEP",
    volume = "07",
    pages = "045",
    year = "2003"
}

@inproceedings{Douglas:2003up,
    author = "Douglas, Michael R. and Klebanov, Igor R. and Kutasov, D. and Maldacena, Juan Martin and Martinec, Emil John and Seiberg, N.",
    title = "{A New hat for the c=1 matrix model}",
    booktitle = "{From Fields to Strings: Circumnavigating Theoretical Physics: A Conference in Tribute to Ian Kogan}",
    eprint = "hep-th/0307195",
    archivePrefix = "arXiv",
    reportNumber = "PUPT-2090, EFI-03-35, RUNHETC-2003-23",
    pages = "1758--1827",
    month = "7",
    year = "2003"
}

@article{Takayanagi:2003sm,
    author = "Takayanagi, Tadashi and Toumbas, Nicolaos",
    title = "{A Matrix model dual of type 0B string theory in two-dimensions}",
    eprint = "hep-th/0307083",
    archivePrefix = "arXiv",
    doi = "10.1088/1126-6708/2003/07/064",
    journal = "JHEP",
    volume = "07",
    pages = "064",
    year = "2003"
}

@mastersthesis{Alexandrov:2003ut,
    author = "Alexandrov, Sergei",
    title = "{Matrix quantum mechanics and two-dimensional string theory in nontrivial backgrounds}",
    eprint = "hep-th/0311273",
    archivePrefix = "arXiv",
    type = "Other thesis",
    month = "9",
    year = "2003"
}

@article{Strominger:2003tm,
    author = "Strominger, Andrew",
    title = "{A Matrix model for AdS(2)}",
    eprint = "hep-th/0312194",
    archivePrefix = "arXiv",
    doi = "10.1088/1126-6708/2004/03/066",
    journal = "JHEP",
    volume = "03",
    pages = "066",
    year = "2004"
}

@article{Klebanov:2003wg,
    author = "Klebanov, Igor R. and Maldacena, Juan Martin and Seiberg, N.",
    title = "{Unitary and complex matrix models as 1-d type 0 strings}",
    eprint = "hep-th/0309168",
    archivePrefix = "arXiv",
    reportNumber = "PUPT-2094",
    doi = "10.1007/s00220-004-1183-7",
    journal = "Commun. Math. Phys.",
    volume = "252",
    pages = "275--323",
    year = "2004"
}

@article{ho2004isometry,
  title={Isometry of AdS 2 and the c= 1 matrix model},
  author={Ho, Pei-Ming},
  journal={Journal of High Energy Physics},
  volume={2004},
  number={05},
  pages={008--008},
  year={2004}
}

@article{Nakayama:2004vk,
    author = "Nakayama, Yu",
    title = "{Liouville field theory: A Decade after the revolution}",
    eprint = "hep-th/0402009",
    archivePrefix = "arXiv",
    reportNumber = "UT-04-02",
    doi = "10.1142/S0217751X04019500",
    journal = "Int. J. Mod. Phys. A",
    volume = "19",
    pages = "2771--2930",
    year = "2004"
}

@inproceedings{Martinec:2004td,
    author = "Martinec, Emil J.",
    title = "{Matrix models and 2D string theory}",
    booktitle = "{NATO Advanced Study Institute: Marie Curie Training Course: Applications of Random Matrices in Physics}",
    eprint = "hep-th/0410136",
    archivePrefix = "arXiv",
    reportNumber = "EFI-04-34",
    pages = "403--457",
    month = "10",
    year = "2004"
}

@article{Seiberg:2004ei,
    author = "Seiberg, Nathan and Shih, David",
    title = "{Flux vacua and branes of the minimal superstring}",
    eprint = "hep-th/0412315",
    archivePrefix = "arXiv",
    reportNumber = "PUPT-2148",
    doi = "10.1088/1126-6708/2005/01/055",
    journal = "JHEP",
    volume = "01",
    pages = "055",
    year = "2005"
}

@article{Marino:2004uf,
    author = "Marino, Marcos",
    title = "{Chern-Simons theory and topological strings}",
    eprint = "hep-th/0406005",
    archivePrefix = "arXiv",
    reportNumber = "CERN-PH-TH-2004-098",
    doi = "10.1103/RevModPhys.77.675",
    journal = "Rev. Mod. Phys.",
    volume = "77",
    pages = "675--720",
    year = "2005"
}

@article{Aharony:2005hm,
    author = "Aharony, Ofer and Patir, Assaf",
    title = "{The Conformal limit of the 0A matrix model and string theory on AdS(2)}",
    eprint = "hep-th/0509221",
    archivePrefix = "arXiv",
    reportNumber = "WIS-22-05-SEP-DPP",
    doi = "10.1088/1126-6708/2005/11/052",
    journal = "JHEP",
    volume = "11",
    pages = "052",
    year = "2005"
}

@article{Aganagic:2003qj,
    author = "Aganagic, Mina and Dijkgraaf, Robbert and Klemm, Albrecht and Marino, Marcos and Vafa, Cumrun",
    title = "{Topological strings and integrable hierarchies}",
    eprint = "hep-th/0312085",
    archivePrefix = "arXiv",
    reportNumber = "UW-PT-03-33, ITFA-2003-58, CERN-TH-2003-290, MAD-TH-03-5, HUTP-03-A083",
    doi = "10.1007/s00220-005-1448-9",
    journal = "Commun. Math. Phys.",
    volume = "261",
    pages = "451--516",
    year = "2006"
}

@article{Horava:2005wm,
    author = "Horava, Petr and Keeler, Cynthia A.",
    title = "{Thermodynamics of noncritical M-theory and the topological A-model}",
    eprint = "hep-th/0512325",
    archivePrefix = "arXiv",
    doi = "10.1016/j.nuclphysb.2006.02.039",
    journal = "Nucl. Phys. B",
    volume = "745",
    pages = "1--28",
    year = "2006"
}

@article{Horava:2005tt,
    author = "Horava, Petr and Keeler, Cynthia A.",
    title = "{Noncritical M-theory in 2+1 dimensions as a nonrelativistic Fermi liquid}",
    eprint = "hep-th/0508024",
    archivePrefix = "arXiv",
    doi = "10.1088/1126-6708/2007/07/059",
    journal = "JHEP",
    volume = "07",
    pages = "059",
    year = "2007"
}

@article{Horava:2007ds,
    author = "Horava, Petr and Keeler, Cynthia A.",
    title = "{Strings on AdS(2) and the High-Energy Limit of Noncritical M-Theory}",
    eprint = "0704.2230",
    archivePrefix = "arXiv",
    primaryClass = "hep-th",
    doi = "10.1088/1126-6708/2007/06/031",
    journal = "JHEP",
    volume = "06",
    pages = "031",
    year = "2007"
}

@book{Wen:2007joe,
    author = "Wen, Xiao-Gang",
    title = "{Quantum Field Theory of Many-Body Systems}",
    doi = "10.1093/acprof:oso/9780199227259.001.0001",
    isbn = "978-0-19-171301-9, 978-0-19-922725-9",
    publisher = "Oxford University Press",
    month = "9",
    year = "2007"
}

@article{Aharony:2008ug,
    author = "Aharony, Ofer and Bergman, Oren and Jafferis, Daniel Louis and Maldacena, Juan",
    title = "{N=6 superconformal Chern-Simons-matter theories, M2-branes and their gravity duals}",
    eprint = "0806.1218",
    archivePrefix = "arXiv",
    doi = "10.1088/1126-6708/2008/10/091",
    journal = "JHEP",
    volume = "10",
    pages = "091",
    year = "2008"
}

@book{Fradkin:2013sab,
    author = "Fradkin, Eduardo H.",
    title = "{Field Theories of Condensed Matter Physics}",
    isbn = "978-0-521-76444-5, 978-1-107-30214-3",
    publisher = "Cambridge Univ. Press",
    address = "Cambridge, UK",
    volume = "82",
    month = "2",
    year = "2013"
}

@article{Saad:2019lba,
    author = "Saad, Phil and Shenker, Stephen H. and Stanford, Douglas",
    title = "{JT gravity as a matrix integral}",
    eprint = "1903.11115",
    archivePrefix = "arXiv",
    primaryClass = "hep-th",
    month = "3",
    year = "2019"
}

@article{Balthazar:2017mxh,
    author = "Balthazar, Bruno and Rodriguez, Victor A. and Yin, Xi",
    title = "{The $c$ = 1 string theory S-matrix revisited}",
    eprint = "1705.07151",
    archivePrefix = "arXiv",
    primaryClass = "hep-th",
    doi = "10.1007/JHEP04(2019)145",
    journal = "JHEP",
    volume = "04",
    pages = "145",
    year = "2019"
}

@article{Sen:2019qqg,
  author        = {Sen, Ashoke},
  title         = {Fixing an ambiguity in two dimensional string theory using string field theory},
  journal       = {JHEP},
  volume        = {03},
  pages         = {005},
  year          = {2020},
  eprint        = {1908.02782},
  archivePrefix = {arXiv},
  doi           = {10.1007/JHEP03(2020)005}
}

@article{Sen:2020oqr,
  author        = {Sen, Ashoke},
  title         = {Divergent $\Longrightarrow$ complex amplitudes in two dimensional string theory},
  journal       = {JHEP},
  volume        = {02},
  pages         = {086},
  year          = {2021},
  eprint        = {2003.12076},
  archivePrefix = {arXiv},
  doi           = {10.1007/JHEP02(2021)086}
}

@article{Sen:2020eck,
  author        = {Sen, Ashoke},
  title         = {D-instantons, string field theory and two dimensional string theory},
  journal       = {JHEP},
  volume        = {11},
  pages         = {061},
  year          = {2021},
  eprint        = {2012.11624},
  archivePrefix = {arXiv},
  doi           = {10.1007/JHEP11(2021)061}
}

@article{Sen:2021qdk,
  author        = {Sen, Ashoke},
  title         = {Normalization of D-instanton amplitudes},
  journal       = {JHEP},
  volume        = {11},
  pages         = {077},
  year          = {2021},
  eprint        = {2101.08566},
  archivePrefix = {arXiv},
  doi           = {10.1007/JHEP11(2021)077}
}

@article{Delacretaz:2022ocm,
    author = "Delacretaz, Luca V. and Du, Yi-Hsien and Mehta, Umang and Son, Dam Thanh",
    title = "{Nonlinear bosonization of Fermi surfaces: The method of coadjoint orbits}",
    eprint = "2203.05004",
    archivePrefix = "arXiv",
    primaryClass = "cond-mat.str-el",
    reportNumber = "EFI 22-3",
    doi = "10.1103/PhysRevResearch.4.033131",
    journal = "Phys. Rev. Res.",
    volume = "4",
    number = "3",
    pages = "033131",
    year = "2022"
}

@article{Balthazar:2019rnh,
  author        = {Balthazar, Bruno and Rodriguez, Victor A. and Yin, Xi},
  title         = {ZZ instantons and the non-perturbative dual of c = 1 string theory},
  journal       = {JHEP},
  volume        = {05},
  pages         = {048},
  year          = {2023},
  eprint        = {1907.07688},
  archivePrefix = {arXiv},
  doi           = {10.1007/JHEP05(2023)048}
}

@article{Balthazar:2019ypi,
  author        = {Balthazar, Bruno and Rodriguez, Victor A. and Yin, Xi},
  title         = {Multi-instanton calculus in c = 1 string theory},
  journal       = {JHEP},
  volume        = {05},
  pages         = {050},
  year          = {2023},
  eprint        = {1912.07170},
  archivePrefix = {arXiv},
  doi           = {10.1007/JHEP05(2023)050}
}

@article{Alexandrov:2023fvb,
  author        = {Alexandrov, Sergei and Mahajan, Raghu and Sen, Ashoke},
  title         = {Instantons in sine-Liouville theory},
  journal       = {JHEP},
  volume        = {01},
  pages         = {141},
  year          = {2024},
  eprint        = {2311.04969},
  archivePrefix = {arXiv},
  doi           = {10.1007/JHEP01(2024)141}
}

@article{Lu:2023emm,
    author = "Lu, Da-Chuan and Wang, Juven and You, Yi-Zhuang",
    title = "{Definition and classification of Fermi surface anomalies}",
    eprint = "2302.12731",
    archivePrefix = "arXiv",
    primaryClass = "cond-mat.str-el",
    doi = "10.1103/PhysRevB.109.045123",
    journal = "Phys. Rev. B",
    volume = "109",
    number = "4",
    pages = "045123",
    year = "2024"
}

@article{Huang:2024uap,
    author = "Huang, Xiaoyang and Lucas, Andrew and Mehta, Umang and Qi, Marvin",
    title = "{Effective field theory for ersatz Fermi liquids}",
    eprint = "2402.14066",
    archivePrefix = "arXiv",
    primaryClass = "cond-mat.str-el",
    doi = "10.1103/PhysRevB.110.035102",
    journal = "Phys. Rev. B",
    volume = "110",
    number = "3",
    pages = "035102",
    year = "2024"
}

@article{Chen:2025jdv,
    author = "Chen, Sihan and Delacretaz, Luca V.",
    title = "{Quantizing bosonized Fermi surfaces}",
    eprint = "2510.07583",
    archivePrefix = "arXiv",
    primaryClass = "cond-mat.str-el",
    doi = "10.1103/zdtz-mdtn",
    journal = "Phys. Rev. B",
    volume = "113",
    number = "12",
    pages = "125150",
    year = "2026"
}

@article{Collier:2026pxi,
    author = "Collier, Scott and Eberhardt, Lorenz and Rodriguez, Victor A.",
    title = "{$c=1$ strings as a matrix integral}",
    eprint = "2604.06301",
    archivePrefix = "arXiv",
    primaryClass = "hep-th",
    month = "4",
    year = "2026"
}

@ARTICLE{2005cond.mat..5529H,
       author = {{Haldane}, F.~D.~M.},
        title = "{Luttinger's Theorem and Bosonization of the Fermi Surface}",
      journal = {arXiv e-prints},
         year = 2005,
        month = may,
          eid = {cond-mat/0505529},
        pages = {cond-mat/0505529},
          doi = {10.48550/arXiv.cond-mat/0505529},
archivePrefix = {arXiv},
       eprint = {cond-mat/0505529},
 primaryClass = {cond-mat.str-el},
       adsurl = {https://ui.adsabs.harvard.edu/abs/2005cond.mat..5529H}
}

@book{Gutzwiller:1990,
  author    = {Gutzwiller, Martin C.},
  title     = {Chaos in Classical and Quantum Mechanics},
  series    = {Interdisciplinary Applied Mathematics},
  volume    = {1},
  publisher = {Springer},
  address   = {New York},
  year      = {1990},
  doi       = {10.1007/978-1-4612-0983-6},
  isbn      = {978-0-387-97173-5}
}

\end{document}